\documentclass[12pt]{article}

\usepackage{graphicx}
\usepackage{natbib}
\usepackage{url} % not crucial - just used below for the URL 

\newcommand{\blind}{0}

\usepackage{amssymb}
\usepackage[tbtags]{amsmath}
\usepackage{algorithm}
\usepackage{algpseudocode}
\usepackage{subfigure}
\usepackage{multirow}
\usepackage{amsthm}
\usepackage{bbm}
\usepackage{enumerate}
\usepackage{tabu}
\usepackage{color}

\newtheorem{theorem}{Theorem}

\newtheorem{assumption}{Assumption}

\begin{document}

\def\spacingset#1{\renewcommand{\baselinestretch}%
{#1}\small\normalsize} \spacingset{1}

%%%%%%%%%%%%%%%%%%%%%%%%%%%%%%%%%%%%%%%%%%%%%%%%%%%%%%%%%%%%%%%%%%%%%%%%%%%%%%

\if0\blind
{
  \title{\bf Efficient Bayesian Synthetic Likelihood with\\ Whitening Transformations}
  \author{Jacob W. Priddle \footnote{Communicating Author: jacob.priddle@hdr.qut.edu.au} \hspace{.2cm}\\
    School of Mathematical Sciences, Queensland University of Technology\\
    Scott A. Sisson\\
    School of Mathematics and Statistics, UNSW Sydney\\
    David T. Frazier \\
    Department of Econometrics and Business Statistics, Monash University and \\
    Christopher Drovandi \\
    School of Mathematical Sciences, Queensland University of Technology}
  \maketitle
} \fi

\if1\blind
{
  \bigskip
  \bigskip
  \bigskip
  \begin{center}
    {\LARGE\bf Title}
\end{center}
  \medskip
} \fi

\bigskip
\begin{abstract}
Likelihood-free methods are an established approach for performing approximate Bayesian inference for models with intractable likelihood functions. However, they can be computationally demanding.
Bayesian synthetic likelihood (BSL) is a popular such method that approximates the likelihood function of the summary statistic  with a known, tractable distribution -- typically Gaussian -- and then performs statistical inference using standard likelihood-based techniques.
However, as the number of summary statistics grows, the number of model simulations required to accurately estimate the covariance matrix for this likelihood rapidly increases. This poses significant challenge for the application of BSL, especially in cases where model simulation is expensive. 
In this article we propose whitening BSL (wBSL) -- an efficient BSL method that uses approximate whitening transformations to decorrelate the summary statistics at each algorithm iteration. We show empirically that this can reduce the number of model simulations required to implement BSL by more than an order of magnitude, without much loss of accuracy. 
%There is an infinite number of whitening transformations possible, meaning a large contribution of this work involves finding the most suitable whitening transformation for BSL.
We explore a range of whitening procedures and demonstrate  the performance of wBSL on a range of simulated and real modelling scenarios from ecology and biology.
\end{abstract}

\noindent%
\vfill

\newpage
\spacingset{1} % DON'T change the spacing!
\section{Introduction}
\label{sec:Introduction}

Likelihood-free methods have become a well established tool over the past two decades for performing statistical inference in the presence of computationally intractable likelihood functions. Such intractability can arise through a desire to fit realistically complex models, or through the shear size of a dataset, rendering the straightforward application of standard likelihood-based procedures practically infeasible.
One popular and well studied likelihood-free approach is approximate Bayesian computation (ABC) \citep{sisson2018handbook}. ABC methods operate by repeated simulation of data under the model of interest, and then comparing observed and simulated data on the basis of summary statistics of these data under some kernel function. ABC methods are known to scale poorly to high-dimensional problems \citep{prangle2018summary,nott+ofs17}.
%, and rely on the availability of informative summary statistics, and an appropriate distance function. \\

Recently, Bayesian synthetic likelihood (BSL) \citep{price2018bayesian} has been gaining popularity as an alternative method to ABC for likelihood-free inference. BSL is the Bayesian extension of the synthetic likelihood approach of \cite{wood2010statistical}, which approximates the unknown likelihood function of the summary statistics with a known, tractable distribution, typically Gaussian. Compared to the non-parametric estimate of the likelihood function that is implied by ABC methods \citep{blum10,sisson+f18}, by
%both of which make the assumption of multivariate normality for the vector of summary statistics. By 
making a parametric assumption, BSL is able to scale better than ABC to high dimensional problems 
\cite[in both summary statistics and model parameters;][]{ong2018likelihood,nott+ofs17}, 
%(in both summary statistics and model parameters; \citet{ong2018likelihood,nott+ofs17}), 
and makes the usual ABC trade-off between the dimensionality and informativeness of the summary statistics much easier. 
%Furthermore, 
\cite{nott2019bayesian} show that an importance sampling BSL algorithm with the posterior as a proposal distribution is more computationally efficient than the corresponding ABC algorithm.

Despite the relative advantages and efficiencies of BSL, and recent work in this area \citep[e.g.][]{nott2019bayesian,an2019accelerating,ong+ntsd18}
there remain some key inefficiencies in the method. Most prominently, for a Gaussian synthetic likelihood the unknown mean and covariance matrix must be estimated by simulation for every proposed parameter within any inference algorithm. This is especially problematic when the dimension of the summary statistics is high, as a large number of model simulations are then required to produce an accurate estimate of the covariance matrix, or when simulation from the model itself is expensive.

A number of efficient
covariance matrix estimation techniques have been considered to reduce the needed number of model simulations in BSL. \cite{an2019accelerating} use the graphical lasso to provide a sparse estimate of the precision matrix. However, performance is inhibited when there is a low degree of sparsity in the covariance or inverse covariance matrix. \cite{ong2018likelihood} and \cite{nott2019bayesian} consider shrinkage estimation 
%(see \citealp{warton2008penalized}), 
to shrink the off-diagonal elements of the correlation matrix by a factor and leave the estimated variances (i.e.~the diagonals of the covariance matrix) unadjusted. However, in a number of empirical examples when there is significant correlation between summaries, these estimators result in poor BSL posterior approximations -- in particular, recovering the wrong dependence structure between parameters and over- or under-estimates of variances. \cite{nott2019bayesian} deliberately mis-specify the form of the covariance matrix (as diagonal or taking a factor form) to allow more shrinkage to be applied, and then use asymptotic results to correct the resulting posterior variances post-hoc.
%of inferences.
 \cite{everitt2017bootstrapped} consider an alternative method to reduce the number of model simulations in a bootstrapped version of synthetic likelihood.

In this article we consider the application of whitening transformations within BSL. Whitening is a linear transformation that maps a set of random variables into a new set of variables with an identity covariance matrix. In the context of BSL, we perform an approximate whitening transformation of the set of simulated summary statistics at each algorithm iteration. The transformation requires a whitening matrix which is based on a point estimate of the parameter that is supplied by the user \cite[following e.g.][]{luciani+sjft09}. The whitening transformation can be effective in decorrelating the summary statistics across important parts of the parameter space. 
In addition, because the resulting transformed summary statistics should be significantly less correlated, a greater amount of shrinkage can then be applied to the covariance estimator. Accordingly, the number of required model simulations can be substantially reduced without a detrimental effect on the accuracy of the resulting posterior approximation, relative to standard BSL.  We show that when estimating the summary statistic covariance matrix as a diagonal matrix (corresponding to complete shrinkage), the number of model simulations required to control the variance of the log synthetic likelihood scales linearly with the dimension of the summary statistic, as opposed to the standard synthetic likelihood estimator, for which we show that the number of simulations must grow quadratically with the summary statistic dimension. These results provide a strong motivation for our whitening method; we refer to the method of whitening transformation and covariance shrinkage within BSL as wBSL.
%
%% OK, but I don't think we really need to spell this out here.
%Because the covariance of the whitened summary statistics may not be close to the identity matrix at parameter values away from the point estimate. Thus, we advocate the use of Warton shrinkage with our whitening BSL method, since this estimator is effective when the covariance matrix is close to diagonal (rather than identity) as it does not shrink the variance estimates. 

Due to the rotational freedom of the whitening transformation, there is an infinite number of whitening transformation matrices available. We consider the five whitening transformations examined by \cite{kessy2018optimal} and find that the principal component analysis (PCA) based whitening transformation performs best within the BSL framework. We also empirically demonstrate that the whitening BSL posterior approximation is quite insensitive to the point at which the whitening matrix is initially estimated.

This article is structured as follows: Section \ref{sec:BSL} details BSL, its
 properties and practical recommendations, as well as background information on shrinkage covariance matrix estimation. Section \ref{sec:wBSL} describes the whitening transformations and introduces the wBSL algorithm.
We examine the performance of wBSL under controlled simulations in Section \ref{sec:examples}, in addition to two real world analyses in ecology and biology.
Section \ref{sec:investigate} explores the choice of whitening transformation in terms of the effectiveness of the transformation over the parameter space, and the sensitivity of the whitening procedure to the initial point estimate.
We conclude with a discussion.

%Section 2 details BSL, its properties and practical recommendations. Section 3 and 4 provide relevant background information on shrinkage estimation and the whitening transformation respectively. Our wBSL method is presented in Section 5. In Section 6 the new method is applied to five different examples. The first three examples are toy examples, and the last two are real-life examples from ecology and biology. In Section 7, we investigate the choice of whitening transformation in terms of the effectiveness of the transformation over the parameter space. The sensitivity of the PCA whitening procedure to the initial point estimate is investigated in Section 8. Final discussions, remarks and recommendations are presented in Section 9.

%%%%%%%%%%%%%%%%%%%%%%%%%%%%%%%%%%%%%%%%%%%
%%%%%%%%%%%%%%%%%%%%%%%%%%%%%%%%%%%%%%%%%%%
\section{Bayesian Synthetic Likelihood}
\label{sec:BSL}
%%%%%%%%%%%%%%%%%%%%%%%%%%%%%%%%%%%%%%%%%%%
%%%%%%%%%%%%%%%%%%%%%%%%%%%%%%%%%%%%%%%%%%%

Suppose we have developed a statistical model $p(\cdot|\boldsymbol{\theta})$ and are interested in learning the parameters $\boldsymbol{\theta}\in\boldsymbol{\Theta}$ for a given set of observed data $\boldsymbol{y} = (y_1,...,y_m)^\top$. The model may contain many parameters and hidden states, making it sufficiently complex so that a computationally tractable expression for the likelihood $p(\boldsymbol{y}|\boldsymbol{\theta})$ is unavailable. Bayesian synthetic likelihood (BSL) is a likelihood-free inference technique that permits an approximate Bayesian inference in this setting, but without direct evaluation of the intractable likelihood function \citep{wood2010statistical,price2018bayesian}. Like ABC methods, BSL relies on reducing $\boldsymbol{y}$ to a lower-dimensional set of informative summary statistics $\boldsymbol{s}_y = S(\boldsymbol{y})$, where $S(\cdot)$ is a summary statistic mapping function. BSL aims to target the partial posterior distribution
\begin{align*}
    p(\boldsymbol{\theta}|\boldsymbol{s_y})\propto p(\boldsymbol{\theta})p(\boldsymbol{s_y}|\boldsymbol{\theta}),
\end{align*}
where $p(\boldsymbol{\theta})$ is the prior for $\boldsymbol{\theta}$. Because $p(\boldsymbol{s_y}|\boldsymbol{\theta})$ will also likely be computationally intractable, BSL then makes the assumption that the summary statistic likelihood $p(\boldsymbol{s_y}|\boldsymbol{\theta})$ follows a convenient specified parametric form. Typically this will be a multivariate normal distribution, so that an auxiliary or synthetic approximation of the summary statistic likelihood is 
\begin{align*}
    p(\boldsymbol{s}_y|\boldsymbol{\theta}) \approx p_A(\boldsymbol{s}_y|\boldsymbol{\theta}) = \mathcal{N}(\boldsymbol{s}_y|\boldsymbol{\mu}(\boldsymbol{\theta}),\boldsymbol{\Sigma}(\boldsymbol{\theta})).
\end{align*}
The auxiliary parameters $\boldsymbol{\mu}(\boldsymbol{\theta})$ and $\boldsymbol{\Sigma}(\boldsymbol{\theta})$ are generally unknown (as a function of $\boldsymbol{\theta}$), but can be straightforwardly estimated by Monte Carlo simulation. Denote by $\boldsymbol{s}_{1:n} = (\boldsymbol{s}_1,...,\boldsymbol{s}_n)^\top$ the sequence of summary statistics for $n$ i.i.d.~simulated data sets $\boldsymbol{y}_{1:n} = (\boldsymbol{y}_1,...,\boldsymbol{y}_n)^\top$ such that $\boldsymbol{s}_i$ is the set of summary statistics for $\boldsymbol{y}_i\sim p(\cdot|\boldsymbol{\theta})$, and $\boldsymbol{s}_i = (s_1,...,s_{d})^\top$, where $d$ is the number of summary statistics. The parameters of the auxiliary likelihood can then be estimated by the sample statistics
\begin{align}
    \boldsymbol{\mu}_n(\boldsymbol{\theta}) &= \frac{1}{n}\sum_{i=1}^n\boldsymbol{s}_i,\label{eq:samplemean}\\
    \boldsymbol{\Sigma}_n(\boldsymbol{\theta}) &= \frac{1}{n-1}\sum_{i=1}^{n}(\boldsymbol{s}_i-\boldsymbol{\mu}_n(\boldsymbol{\theta}))(\boldsymbol{s}_i-\boldsymbol{\mu}_n(\boldsymbol{\theta}))^\top,\label{eq:samplecov}
\end{align}
and so the estimated auxiliary likelihood, as an explicit  function of $n$, becomes
\begin{align*}
    p_{A,n}(\boldsymbol{s}_y|\boldsymbol{\theta}) = \mathcal{N}(\boldsymbol{s}_y|\boldsymbol{\mu}_n(\boldsymbol{\theta}),\boldsymbol{\Sigma}_n(\boldsymbol{\theta})).
\end{align*}
We write $p_{A,n}(\boldsymbol{s_y}|\boldsymbol{\theta})$ to emphasise the dependence on $n$. In practice, the dependence on $n$ is weak \citep{price2018bayesian}, and if $n$ tends toward infinite with $m$ at any rate, then the effect of estimating $\boldsymbol{\mu}(\boldsymbol{\theta})$ and $\boldsymbol{\Sigma}(\boldsymbol{\theta})$ is asymptotically negligible \citep{nott2019bayesian}. As a result,  \cite{price2018bayesian} suggest choosing $n$ to maximise computational efficiency, with large $n$ providing expensive but precise likelihood estimates, compared to low $n$ producing fast but variable estimates.
% When a small value of $n$ is chosen, synthetic likelihood estimates are fast but highly variable, whereas when $n$ is high, synthetic likelihood estimates are slow and precise. 
 \cite{price2018bayesian}  recommend choosing a point of high posterior support for $\boldsymbol{\theta}$ and then tuning  $n$ so that the standard deviation of the $\log$ synthetic likelihood is roughly between $1$ and $2$ \cite[see also][]{doucet+pdk15}.

In practice, \eqref{eq:samplecov} is not the most efficient estimator of $\boldsymbol{\Sigma}(\boldsymbol{\theta})$, and several authors have adopted different strategies, including shrinkage, to improve on this within the BSL context \citep[e.g.][]{an2019accelerating,ong2018likelihood,ong+ntsd18,nott2019bayesian,everitt2017bootstrapped}.
With
%In this section, we provide some information on shrinkage estimation of covariance matrices. First, we reiterate our motivation for using 
shrinkage, the primary aim is to estimate the covariance matrix  $\boldsymbol{\Sigma}(\boldsymbol{\theta})$ with as few model simulations as possible such that the performance of a BSL sampler is efficient. As the number of model simulations approaches the number of summary statistics ($d$) from above, (\ref{eq:samplecov}) becomes increasingly close to singular -- with $n < d$ guaranteeing a singular estimate. 

One simple approach used by e.g.~\cite{ong2018likelihood, ong+ntsd18} makes use of
%\cite{warton2008penalized} use an approach based on 
ridge regularisation to avoid such instabilities \citep{warton2008penalized}. The standard ridge regulariser for the covariance matrix estimate is $\boldsymbol{\Sigma}_{\kappa} = \boldsymbol{\Sigma}_n + \kappa\boldsymbol{I}_d$, where $\kappa>0$ is the ridge parameter and $\boldsymbol{I}_d$ is the $d\times d$ identity matrix. When the variables are measured on different scales (as is usual for the summary statistics in BSL), \cite{warton2008penalized} derived a ridge estimator of the correlation matrix $\boldsymbol{R}$ using maximum penalised Gaussian likelihood estimation, with a  $\text{tr}(\boldsymbol{R}^{-1})$ penalty. For the estimated correlation matrix
\begin{align*}
    \hat{\boldsymbol{R}} = \boldsymbol{\Sigma}_d^{-1/2}\boldsymbol{\Sigma}_n\boldsymbol{\Sigma}_d^{-1/2}
\end{align*}
where $\boldsymbol{\Sigma}_d=\mbox{diag}\boldsymbol{\Sigma}_n$ is formed using the diagonals of $\boldsymbol{\Sigma}_n$, the ridge estimator is 
\begin{align}\label{eq:warton}
    \hat{\boldsymbol{R}}_{\gamma} = \gamma\hat{\boldsymbol{R}} + (1-\gamma)\boldsymbol{I}_d,
\end{align}
with $\gamma \in (0,1]$. 
%Upon inspection of (\ref{eq:warton}) is clear that 
The estimator $\hat{\boldsymbol{R}}_{\gamma}$ is always a valid correlation matrix with unit diagonals. The estimated covariance matrix is then
\begin{align}\label{eq:warton2}
    \boldsymbol{\Sigma}_{n,\gamma} = \boldsymbol{\Sigma}_d^{1/2}\hat{\boldsymbol{R}}_\gamma\boldsymbol{\Sigma}_d^{1/2}.
\end{align}
The smaller the value of $\gamma$, the closer $\boldsymbol{\Sigma}_{n,\gamma}$ comes to being a diagonal matrix. In the context of BSL, a smaller $\gamma$ reduces the variance of the synthetic likelihood estimator $\hat{p}_{A,n}(\boldsymbol{s}_y|\boldsymbol{\theta})$ for a given number of simulations, $n$. This implies that less model simulations are required to achieve the same acceptance rate (as a measure of sampler performance) within BSL. 

Any shrinkage estimator may be used within wBSL. However here we adopt the Warton estimator \eqref{eq:warton2} since it does not shrink the estimated variances of the transformed summary statistics, and we find that this is crucial for the accuracy of the best performing whitening transformation in wBSL (see Section \ref{sec:investigate}). It is also computationally trivial to calculate. We have also found (results not shown) that using the standard ridge shrinkage estimator with wBSL produces far less accurate posterior approximations. 
%An additional advantage of using the above estimator is that it is computationally trivial to calculate by multiplying the off-diagonals of $\hat{\boldsymbol{R}}$ by $\gamma$. 
Shrinkage on its own works well in cases where there is a low degree of correlation between summaries \citep{ong2018likelihood}, but performs poorly for small $\gamma$ when there is significant correlation between summaries (e.g.~Section \ref{sec:examples}, Figure \ref{fig:MA1}).

An understanding of the variance of the synthetic log-likelihood estimator provides insight into the computational efficiency of a BSL algorithm. To deduce the behaviour of the variance of the synthetic log-likelihood estimator, we consider the following assumptions. 

%In an ideal setting, there is no correlation between the summary statistics, and so the true covariance matrix $\boldsymbol{\Sigma}(\boldsymbol{\theta})$ is simply a diagonal matrix of variances. Under this assumption, Result 1 demonstrates how the variance of the log synthetic likelihood scales as $n$ and $d$ get large, given we estimate only a diagonal matrix of variances. In addition, Result 2 demonstrates how the variance of the log synthetic likelihood scales as $n$ and $d$ get large, when the full sample covariance matrix $\boldsymbol{\Sigma}_n(\boldsymbol{\theta})$ is used to estimate $\boldsymbol{\Sigma}(\boldsymbol{\theta})$. 

%Let $S(y)\in\mathbb{R}^{d}$ denote the observed summary statistic. Consider the standard BSL likelihood with  $\bar{\mu}_n(\theta)$ and $\Sigma_n(\theta)$ denoting the sample mean and variance of the simulated summary statistics. To deduce the behaviour of the variance we consider the following explicit assumptions. 
\begin{assumption}\label{ass:one}
For $i=1,\dots,n$, the simulated summaries $\boldsymbol{s}_i(\boldsymbol{\theta})$ are generated i.i.d.\ and satisfy	
$$\boldsymbol{s}_i(\boldsymbol{\theta})\sim_{}\mathcal{N}\left(\boldsymbol{\mu}(\boldsymbol{\theta}),\boldsymbol{\Sigma}(\boldsymbol{\theta})\right),$$ where $\sup_{\boldsymbol{\theta}\in\boldsymbol{\Theta}}\|\boldsymbol{\mu}(\boldsymbol{\theta})\|<\infty$ and $\boldsymbol{\Sigma}(\boldsymbol{\theta})$ is positive-definite for all $\boldsymbol{\theta}\in\boldsymbol{\Theta}$.	
\end{assumption}

In addition, consider a version of BSL that uses simulated summaries with a diagonal covariance structure, i.e., summaries that are uncorrelated across the $d$ dimensions. Denote these simulated summaries as $\boldsymbol{\zeta}_i(\boldsymbol{\theta})$. We maintain the following assumption about these simulated summaries. 
\begin{assumption}\label{ass:two}
For $i=1,\dots, n$, the simulated summaries $\boldsymbol{\zeta}_i(\boldsymbol{\theta})$ are generated i.i.d.\ and satisfy	
$$
\boldsymbol{\zeta}_i(\boldsymbol{\theta})\sim \mathcal{N}\left(\boldsymbol{\zeta}(\boldsymbol{\theta}),\boldsymbol{\Omega}(\boldsymbol{\theta})\right),\text{ where } \boldsymbol{\Omega}(\boldsymbol{\theta}):=\mathrm{diag}(\omega_{11}(\boldsymbol{\theta}),\dots,\omega_{dd}(\boldsymbol{\theta})),
$$where $$\sup_{\boldsymbol{\theta}\in\boldsymbol{\Theta}}\|\boldsymbol{\zeta}(\boldsymbol{\theta})\|<\infty,\mbox{ and }\max_{j\leq d}\sup_{\boldsymbol{\theta}\in\boldsymbol{\Theta}}\omega_{jj}(\boldsymbol{\theta})<\infty,\;\min_{j\leq d}\inf_{\boldsymbol{\theta}\in\boldsymbol{\Theta}}\omega_{jj}(\boldsymbol{\theta})>0.$$ 	
\end{assumption}
Let $\boldsymbol{\zeta}_n(\boldsymbol{\theta})$ denote the sample mean of these uncorrelated summary statistics and let $\boldsymbol{\Omega}_{n}(\boldsymbol{\theta})$ denote the sample variance. Throughout the remainder we explicitly consider that $\boldsymbol{\Omega}_{n}(\boldsymbol{\theta})$ is a diagonal matrix. 

Denote the BSL likelihood based on the uncorrelated summaries as $$p_{A,n,w}\left(\boldsymbol{s}_{\boldsymbol{y}}|\boldsymbol{\theta}\right)=\mathcal{N}(\boldsymbol{s}_{\boldsymbol{y}}|\boldsymbol{\zeta}_n(\boldsymbol{\theta}),\boldsymbol{\Omega}_{n}(\boldsymbol{\theta})),$$ and recall the standard BSL likelihood is given by:
$$p_{A,n}\left(\boldsymbol{s}_{\boldsymbol{y}}|\boldsymbol{\theta}\right)=\mathcal{N}(\boldsymbol{s}_{\boldsymbol{y}}|\boldsymbol{\mu}_n(\boldsymbol{\theta}),\boldsymbol{\Sigma}_{n}(\boldsymbol{\theta})).$$

For two real sequences $a_n,b_n$, we say that $b_n=\mathcal{O}(a_n)$ if the sequence $|b_n/a_n|$ is bounded. Theorem 1 demonstrates how the variance of the synthetic log-likelihood scales under each of the above assumptions.

\begin{theorem} For $d$ and $n$ large, but finite, with $n>d+4$, under Assumption \ref{ass:one},
\begin{flalign*}
\text{\em Var}\left[\log\left(p_{A,n}\left(\boldsymbol{s}_{\boldsymbol{y}}|\boldsymbol{\theta}\right) 
\right)\right]&=\mathcal{O}\left(\frac{d^2\cdot n^2}{(n-d)^3}\right),
\end{flalign*}however, under Assumption \ref{ass:two},
\begin{flalign*}
\text{\em Var}\left[\log\left(p_{A,n,w}\left(\boldsymbol{s}_{\boldsymbol{y}}|\boldsymbol{\theta}\right) 
\right)\right]&=\mathcal{O}\left(\frac{d\cdot n^2}{(n-d)^3}\right).
\end{flalign*}Therefore, for $d$ and $n$ large,
$$
\text{\em Var}\left[\log\left(p_{A,n,w}\left(\boldsymbol{s}_{\boldsymbol{y}}|\boldsymbol{\theta}\right) 
\right)\right] \leq \text{\em Var}\left[\log\left(p_{A,n}\left(\boldsymbol{s}_{\boldsymbol{y}}|\boldsymbol{\theta}\right) 
\right)\right].
$$
\end{theorem}	

Proof of Theorem 1 is given in Appendix A. Theorem 1 demonstrates that for a given value of $n$, the variance of $p_{A,n,w}$ is less than or equal to the variance of $p_{A,n}$. Moreover, we can deduce how $n$ must scale as $d$ increases in order to control the variance of the synthetic log-likelihood. For the synthetic likelihood with uncorrelated summaries, using Theorem 1 and letting $n \propto d$, we have that $\text{Var}\left[\log p_{A,n,w} (\boldsymbol{s}_{\boldsymbol{y}}|\boldsymbol{\theta})\right] = \mathcal{O}(1)$. In this case, $n$ must scale linearly with $d$ to control the variance of the synthetic log-likelihood. On the other hand, using Theorem 1, and letting $n \propto d^2$, we have that for the standard synthetic likelihood estimator with a full covariance matrix, $\text{Var}\left[\log p_{A,n} (\boldsymbol{s}_{\boldsymbol{y}}|\boldsymbol{\theta})\right] = \mathcal{O}(1)$. Therefore, when using $\boldsymbol{\Sigma}_n(\boldsymbol{\theta})$ to estimate $\boldsymbol{\Sigma}(\boldsymbol{\theta})$, $n$ must increase quadratically with $d$ to control the variance of the synthetic log-likelihood. Thus, if we had access to an uncorrelated set of summary statistics, this would greatly mitigate the curse of dimensionality with respect to the dimension of the summary statistic in BSL. However, in practice, a set of uncorrelated summary statistics that are informative about the model parameters is not usually available. In Section \ref{sec:wBSL} we propose a method that approximately decorrelates a set of correlated summary statistics, allowing us to achieve the computational gains associated with using an uncorrelated set of summary statistics (when $\gamma = 0$ in the shrinkage covariance estimator of (\ref{eq:warton})).

%Proofs of the above results are given in Appendix A of the supplementary materials. Given the results in Result 1 and Result 2, we can deduce how $n$ must scale as $d$ increases in order to control the variance of the log synthetic likelihood, depending on the form of the estimated covariance matrix. For the synthetic likelihood with a diagonal covariance matrix, using Result 1 and letting $n \propto d$, we have that $\text{Var}\left[\log\hat{p}_d (\boldsymbol{s}_{\boldsymbol{y}}|\boldsymbol{\theta})\right] = \mathcal{O}(1)$. In this case, $n$ must scale linearly with $d$ to control the variance of the log synthetic likelihood. On the other hand, using Result 2, and letting $n \propto d^2$, we have that for the synthetic likelihood estimator with a full covariance matrix, $\text{Var}\left[\log\hat{p}_d (\boldsymbol{s}_{\boldsymbol{y}}|\boldsymbol{\theta})\right] = \mathcal{O}(1)$. Therefore, when using $\Sigma_n(\boldsymbol{\theta})$ to estimate $\Sigma(\boldsymbol{\theta})$, $n$ must increase quadratically with $d$ to control the variance of the log synthetic likelihood.

%%%%%%%%%%%%%%%%%%%%%%%%%%%%%%%%%%%%%%%%%%%
%%%%%%%%%%%%%%%%%%%%%%%%%%%%%%%%%%%%%%%%%%%
\section{Whitening Bayesian synthetic likelihood (wBSL)}
\label{sec:wBSL}
%%%%%%%%%%%%%%%%%%%%%%%%%%%%%%%%%%%%%%%%%%%
%%%%%%%%%%%%%%%%%%%%%%%%%%%%%%%%%%%%%%%%%%%
%\textcolor{red}{[Please check the maths of this section]}
In order to reduce shrinkage estimation induced error within BSL, and thereby also increase the efficiency of the method, we propose the use of a whitening transformation \cite[e.g.][]{kessy2018optimal} to decorrelate the summary statistics at each iteration of the BSL algorithm. Whitening, also known as sphering, is a linear transformation commonly employed in data preprocessing to produce a decorrelated set of data with unit variance \citep[e.g.][]{bacus1976whitening}. Specifically, a whitening transformation converts the random vector $\boldsymbol{s}$ of arbitrary distribution with covariance matrix $\text{Var}(\boldsymbol{s}) = \boldsymbol{\Sigma}$, into a new set of variables 
% note that s is a column vector
\begin{align}
	\tilde{\boldsymbol{s}} = \boldsymbol{Ws} 
\end{align}
for some $d\times d$ whitening matrix $\boldsymbol{W}$, such that the covariance $\text{Var}(\tilde{\boldsymbol{s}}) = \boldsymbol{I}_d$ is the identity matrix of dimension $d$. 

Of course, in the context of BSL, $\boldsymbol{\Sigma}$ depends on $\boldsymbol{\theta}$ and thus, to achieve exactly $\text{Var}(\tilde{\boldsymbol{s}}) = \boldsymbol{I}_d$ at every parameter value, the whitening matrix must also vary as a function of $\boldsymbol{\theta}$. However, for wBSL, we hold $\boldsymbol{W}$ constant so that the transformed summary statistic likelihood
\begin{align*}
    g(\tilde{\boldsymbol{s}}|\boldsymbol{\theta}) 
    %&
    = \frac{p(\boldsymbol{W}^{-1}\tilde{\boldsymbol{s}}|\boldsymbol{\theta})}{|\text{det}(\boldsymbol{W}^\top)|}%\\
    %&
    =  \frac{p(\boldsymbol{W}^{-1}\boldsymbol{Ws}|\boldsymbol{\theta})}{|\text{det}(\boldsymbol{W}^\top)|}%\\
    %&
    \propto p(\boldsymbol{s}|\boldsymbol{\theta}),
\end{align*}
which ensures the posterior distribution conditional on the summary statistic remains unchanged with the whitening transformation. Therefore, for parameter values away from where $\boldsymbol{W}$ is estimated, the transformation is not exact, so that $\text{Var}(\tilde{\boldsymbol{s}}) \approx \boldsymbol{I}_d$. By approximately decorrelating summary statistics, the covariance shrinkage estimator may be applied with a greater amount of shrinkage. Heavier shrinkage permits fewer model simulations.

The only requirement for the whitening matrix $\boldsymbol{W}$ is that it must satisfy $\mbox{Var}(\tilde{\boldsymbol{s}}) = \mbox{Var}(\boldsymbol{Ws}) = \boldsymbol{W\Sigma W}^\top = \boldsymbol{I}_d$. Hence, as $\boldsymbol{W\Sigma} \boldsymbol{W}^\top \boldsymbol{W} = \boldsymbol{W}$, this is satisfied if 
\begin{align}\label{eq:white1}
    \boldsymbol{W}^\top \boldsymbol{W} = \boldsymbol{\Sigma}^{-1}.
\end{align}
Due to the rotational freedom, there are infinitely many whitening matrices that satisfy (\ref{eq:white1}), each resulting in uncorrelated but differing sets of variables $\tilde{\boldsymbol{s}}$. 

%At this stage, it is unclear which whitening transformation would be the most effective within a BSL algorithm. 
The most suitable whitening matrix $\boldsymbol{W}$ for wBSL is the one that most effectively decorrelates those summary statistics generated under the model, for parameter values that reside in regions with non-negligible posterior density. This would minimise posterior approximation errors caused by shrinkage estimation of the transformed summary statistic covariance matrix $\mbox{Var}(\tilde{\boldsymbol{s}})$, and thereby produce the most accurate inference. Here we consider the five natural whitening procedures outlined by \cite{kessy2018optimal}: zero-phase component analysis  whitening (ZCA), ZCA correlation whitening (ZCA-cor), principal component analysis  whitening (PCA), PCA correlation whitening (PCA-cor) and Cholesky whitening. Each transform arises naturally by either optimising some criteria with respect to the cross-covariance $\boldsymbol{\Phi} = \mbox{Cov}(\tilde{\boldsymbol{s}},\boldsymbol{s})$, the cross-correlation $\boldsymbol{\Psi} = \mbox{Cor}(\tilde{\boldsymbol{s}},\boldsymbol{s})$, or by satisfying some symmetry constraint. 
%Here we only provide a brief outline of each transform, for more details such as formal proofs, we refer the reader to \cite{kessy2018optimal}.\\
Each transform is described briefly below \cite[see e.g.][for further details]{kessy2018optimal}.

%We will 
The transformations make use of various matrix decompositions. Specifically, the covariance matrix may be decomposed as $\boldsymbol{\Sigma} = \boldsymbol{V}^{1/2}\boldsymbol{PV}^{1/2}$, where  $\boldsymbol{P}$ is the correlation matrix  and $\boldsymbol{V}$ is the diagonal matrix of variances. 
The eigendecomposition of the covariance matrix is $\boldsymbol{\Sigma} = \boldsymbol{U\Lambda U}^\top$, where $\boldsymbol{U}$ is the matrix of eigenvectors and $\boldsymbol{\Lambda}$ the diagonal matrix of eigenvalues,  and the eigendecomposition of the correlation matrix is $\boldsymbol{P} = \boldsymbol{G\Xi G}^\top$, where $\boldsymbol{G}$ is the eigenvector matrix  and $\boldsymbol{\Xi}$ is the diagonal matrix of eigenvalues. Finally, the Cholesky decomposition of the precision matrix is $\boldsymbol{\Sigma}^{-1} = \boldsymbol{LL}^\top$, where $\boldsymbol{L}$ is a unique lower triangular matrix.

%The first whitening procedure is the 
ZCA or ZCA-Mahalanobis whitening aims to produce a transformed set of data that remains maximally similar to the original data. This is achieved by minimising the squared distance between the original and transformed data
%, so that
\begin{align*}
    \mathbb{E}\left[(\tilde{\boldsymbol{s}}-\boldsymbol{s})^\top(\tilde{\boldsymbol{s}}-\tilde{\boldsymbol{s}})\right] &= \text{tr}(\boldsymbol{I}_d) - 2\mathbb{E}\left[(\tilde{\boldsymbol{s}}\boldsymbol{s})\right]+\text{tr}(\boldsymbol{\Sigma})\\
    &= d - 2\text{tr}(\boldsymbol{\Phi})+\text{tr}(\boldsymbol{V}),
\end{align*}
or equivalently maximising the average cross-covariance $\text{tr}(\boldsymbol{\Phi})$. The resulting whitening matrix is $\boldsymbol{W}_{\text{ZCA}} = \boldsymbol{\Sigma}^{-1/2}$. 
ZCA-cor whitening is the scale invariant analogue of ZCA whitening, 
where the objective is to minimise the distance between the variables on a standardised scale, so that 
\begin{align*}
    \mathbb{E}\left[(\tilde{\boldsymbol{s}}-\boldsymbol{V}^{-1/2}\boldsymbol{s})^\top(\tilde{\boldsymbol{s}}-\boldsymbol{V}^{-1/2}\boldsymbol{s})\right] = 2d - 2\text{tr}(\boldsymbol{\Psi}).
\end{align*}
The resulting whitening matrix, $\boldsymbol{W}_{\text{ZCA-corr}} = \boldsymbol{P}^{-1/2}\boldsymbol{V}^{-1/2}$, minimises the average cross-correlation $\text{tr}(\boldsymbol{\Psi})$.
%
%\cite{kessy2018optimal} also consider PCA whitening and PCA-corr whitening. It can be shown that 
PCA whitening and PCA-corr whitening are equivalent to maximising the compression with respect to the cross-covariance
\begin{align*}
    (\phi_1,...,\phi_{d})^\top = \text{diag}(\boldsymbol{\Phi\Phi}^\top)
\end{align*}
with $\phi_i \geq \phi_{i+1}$ and cross-correlation
\begin{align*}
    (\psi_1,...,\psi_{d})^\top = \text{diag}(\boldsymbol{\Psi\Psi}^\top)
\end{align*}
respectively. PCA whitening results in $\boldsymbol{W}_{\text{PCA}} = \boldsymbol{\Lambda}^{-1/2}\boldsymbol{U}^\top$ for the whitening matrix, whereas PCA-cor whitening results in $\boldsymbol{W}_{\text{PCA-cor}} = \boldsymbol{\Xi}^{-1/2}\boldsymbol{G}^\top\boldsymbol{V}^{-1/2}$.
%
%The last form of whitening considered is Cholesky whitening. 
Finally, Cholesky whitening is directly based on the Cholesky decomposition of the precision matrix and results in $W_{\text{Cholesky}} = \boldsymbol{L}^\top$. In the examples later we test all the whitening approaches in the context of BSL and find the PCA variants to be most effective.

%\subsection{wBSL}

The original BSL algorithm \citep{price2018bayesian} was presented in the form of a Metropolis-Hastings Markov chain Monte Carlo (MCMC) sampler, and so we present wBSL similarly. Of course, the (w)BSL procedure is Monte Carlo algorithm agnostic, and so alternative posterior simulation samplers (such as sequential Monte Carlo) are straightforward to construct.
The full MCMC-based wBSL procedure is outlined in Algorithm \ref{alg:BSLWHITE}.

In wBSL the unknown mean and covariance matrix of the $\tilde{\boldsymbol{s}}$ are estimated by simulation. First, the $n\times d$ matrix of summary statistics $\boldsymbol{s}_{1:n}$ is transformed according to $\tilde{\boldsymbol{s}}_{1:n} = \boldsymbol{s}_{1:n}\boldsymbol{W}^\top$.
The whitening matrix $\boldsymbol{W}$ is estimated prior to implementing the MCMC sampler using $n_\text{cov}>n$ simulations 
$\boldsymbol{x}_{1:n_{\text{cov}}}\sim p(\cdot|\boldsymbol{\theta}^0)$, given some parameter value
 %many off-line simulations at a point estimate 
 $\boldsymbol{\theta}^0$ located in a region of high posterior density.  
 This is not an uncommon procedure within ABC \citep[e.g.][]{luciani+sjft09}, and any suitable method can be used to find an appropriate $\boldsymbol{\theta}^0$, such as prior information,
 % from previous studies, expert opinion, a 
a pilot ABC analysis with a large kernel scale parameter \citep{fearnhead2012constructing} or a fast likelihood-free optimisation method \citep[e.g.][]{gutmann2016bayesian}.

Given an approximate whitening transformation, we use \eqref{eq:warton2} to estimate the covariance matrix of the transformed summary statistics, $\tilde{\boldsymbol{s}}_{1:n}$. Ideally, this covariance matrix  is approximately diagonal, meaning that the off-diagonal elements are close to zero. In this case, the whitened summary statistics in wBSL permit a large amount of shrinkage (a low value of $\gamma$) to be used, and a correspondingly large reduction in $n$ compared to standard BSL. That is, shrinkage covariance estimation can be much more effective when a whitening transformation is applied. 
%The full wBSL procedure is outlined in Algorithm \ref{alg:BSLWHITE}; note for notational convenience we drop the dependence on $\boldsymbol{\theta}$ for $\tilde{\boldsymbol{\mu}}_n(\boldsymbol{\theta})$ and $\tilde{\boldsymbol{\Sigma}}_n(\boldsymbol{\theta})$.\\

%The target distribution of this is NOT DEFINED YET.
% \textcolor{red}{[I made several ``corrections" here -- please check.]}

\begin{algorithm}[h!]
\caption{{\bf MCMC wBSL} \label{alg:BSLWHITE} 
\vspace{2mm}
\newline
{\em Inputs:} 
An initial value of the chain with non-negligible posterior support $\boldsymbol{\theta}^0$; the level of shrinkage $\gamma$; the number of model simulations $n$; the number of model simulations $n_{\text{cov}}$ to estimate $\boldsymbol{W}$; the model $p(\cdot|\boldsymbol{\theta})$; the prior $p(\boldsymbol{\theta})$; the observed data $\boldsymbol{y}$; 
the MCMC proposal distribution $q(\cdot|\boldsymbol{\theta})$; the number of chain iterations $T$.
\vspace{2mm}
\newline
{\em Outputs:} MCMC samples $\boldsymbol{\theta}^0,\ldots,\boldsymbol{\theta}^T$ from the wBSL posterior approximation.
%, $p_{A,n,\gamma}(\boldsymbol{\theta}|\tilde{\boldsymbol{s}}_{\boldsymbol{y}})$
}
\begin{algorithmic}[1]
\State Generate $\boldsymbol{x}_{1:n_{\text{cov}}}\stackrel{\text{iid}}{\sim} p(\cdot|\boldsymbol{\theta}^0)$.
\State Compute $\boldsymbol{s_y}$, $\boldsymbol{s}_{1:n_{\text{cov}}}$ and whitening matrix $\boldsymbol{W}$.
\State Compute whitened statistics $\tilde{\boldsymbol{s}}_{\boldsymbol{y}} = \boldsymbol{Ws}_{\boldsymbol{y}}$ and $\tilde{\boldsymbol{s}}_{1:n_{\text{cov}}} = \boldsymbol{s}_{1:n_{\text{cov}}}\boldsymbol{W}^\top$.
\State Set $\tilde{\boldsymbol{\Sigma}}_{n,\gamma}^0 = 
%\tilde{\boldsymbol{{\Sigma}}}^0_{n_{\text{cov}},\gamma}=
\boldsymbol{I}_d$ and compute $\tilde{\boldsymbol{\mu}}^0_n=\tilde{\boldsymbol{\mu}}^0_{n_{\text{cov}}}$ using \eqref{eq:samplemean}.
\For{$t=1$ to $T$}
\State Draw candidate parameter $\boldsymbol{\theta^*}\sim q(\cdot|\boldsymbol{\theta}^{i-1})$ from proposal distribution.
\State Generate $\boldsymbol{x}_{1:n}^*\stackrel{\text{iid}}{\sim} p(\cdot|\boldsymbol{\theta}^*)$.
\State Compute $\boldsymbol{s}_{1:n}^*$.
\State Compute $\tilde{\boldsymbol{s}}_{1:n}^* = \boldsymbol{s}_{1:n}^*\boldsymbol{W}^\top$.
\State Compute $\tilde{\boldsymbol{\mu}}_n^*$ via \eqref{eq:samplemean} and $\tilde{\boldsymbol{\Sigma}}_n^*$ via \eqref{eq:samplecov} using  $\tilde{\boldsymbol{s}}^*_{1:n}$.
\Comment{Ideally $\tilde{\boldsymbol{\Sigma}}_n^* \approx \text{diagonal}$}
\State Compute $\tilde{\boldsymbol{\Sigma}}_{n,\gamma}^*$ using \eqref{eq:warton2}.
\State Calculate $r = \frac{\mathcal{N}(\tilde{\boldsymbol{s}}_y|\tilde{\boldsymbol{\mu}}_n^*,\tilde{\boldsymbol{\Sigma}}_{n,\gamma}^*)p(\boldsymbol{\theta}^*)q(\boldsymbol{\theta}^{t-1}|\boldsymbol{\theta}^*)}{\mathcal{N}(\tilde{\boldsymbol{s}}_y|\tilde{\boldsymbol{\mu}}_n^{t-1},\tilde{\boldsymbol{\Sigma}}_{n,\gamma}^{t-1})p(\boldsymbol{\theta}^{t-1})q(\boldsymbol{\theta}^{*}|\boldsymbol{\theta}^{t-1})}$.
\If{$\mathcal{U}(0,1) < r$} 
\State  Set $\boldsymbol{\theta}^t = \boldsymbol{\theta}^*$, $\tilde{\boldsymbol{\Sigma}}_{n,\gamma}^t = \tilde{\boldsymbol{\Sigma}}_{n,\gamma}^*$ and $\boldsymbol{\mu}^t_n = \boldsymbol{\mu}^*_n$.
\Else
\State Set $\boldsymbol{\theta}^t = \boldsymbol{\theta}^{t-1}$, $\tilde{\boldsymbol{\Sigma}}_{n,\gamma}^t = \tilde{\boldsymbol{\Sigma}}_{n,\gamma}^{t-1}$ and $\boldsymbol{\mu}^t_n = \boldsymbol{\mu}^{t-1}_n$.
\EndIf
\EndFor
\end{algorithmic}
\end{algorithm}

%%%%%%%%%%%%%%%%%%%%%%%%%%%%%%%%%%%%%%%%%%%
%%%%%%%%%%%%%%%%%%%%%%%%%%%%%%%%%%%%%%%%%%%
\section{Examples}
\label{sec:examples}
%%%%%%%%%%%%%%%%%%%%%%%%%%%%%%%%%%%%%%%%%%%
%%%%%%%%%%%%%%%%%%%%%%%%%%%%%%%%%%%%%%%%%%%

We examine the performance of wBSL for 
%a variety of examples. The first 
three 
%are toy examples 
models with simulated data where the covariance of the summary statistics depends explicitly on the parameters. 
This is realistic in practice.
%making them interesting test examples for the whitening method. 
We also consider two real data analyses from ecology and biology. For the first four models we compare wBSL with each of the five whitening transformations and Warton shrinkage by itself, 
%and compare the results 
to either standard BSL or to the true posterior (where known). 
%Further, we investigate the potential power of wBSL for improving the efficiency of BSL by testing a wide range of shrinkage and number of model simulations combinations. \\

To find an appropriate combination of $\gamma$ and $n$, we estimate the number of model simulations required to maximise the computational efficiency of standard BSL,
which we define as wBSL but with no whitening transformation or shrinkage covariance estimation (i.e.~$\gamma=1$).
Following  \cite{price2018bayesian}, this is the value for $n$
% Following the recommendation of \cite{price2018bayesian}, to maximise BSL's computational efficiency, we tune $n$ 
 such that the estimate of the log synthetic likelihood at $\boldsymbol{\theta}^0$ has a standard deviation in the range  $[1,2]$. 
 %roughly between 1 and 2. 
 We then fix  $n$ to achieve a $50\%$, $80\%$ and $90\%$ reduction in the number of model simulations at each sampler iteration compared to standard BSL, and tune the value of $\gamma$ to similarly constrain the log-likelihood variance. 
 We also consider complete shrinkage ($\gamma=0$) so that the covariance matrix of $\tilde{\boldsymbol{s}}_{1:n}$ is forced to be diagonal, and again choose $n$ to constrain the variance of the log-likelihood. This latter setting represents the most computationally efficient wBSL algorithm (lowest $n$), but is potentially the least accurate.
 
% We also tune $n$ so that $\gamma = 0$ (complete shrinkage), so that the covariance matrix of the transformed summaries is forced to be diagonal at each iteration, representing the most computationally efficient wBSL algorithm possible but least accurate. \\

For each method and analysis we use a Gaussian random walk MCMC proposal distribution with covariance set to be roughly equal to the (approximate) posterior covariance.
%For each example, we use a Gaussian proposal distribution, with the covariance set to be roughly the (approximate) posterior covariance. This allows a fair comparison between methods, since each method provides a different approximation of the target posterior.\\
%
To quantify the accuracy of each method, we use the total variation distance between two probability 
%distributions with 
density functions $f_1(\boldsymbol{\theta})$ and $f_2(\boldsymbol{\theta})$ given by
	$\text{tv}(f_1,f_2) = \frac{1}{2}\int
	%_{\Theta}
	|f_1(\boldsymbol{\theta})-f_2(\boldsymbol{\theta})|d\boldsymbol{\theta}$.
The distance is estimated using kernel density estimation from the (approximate) posterior samples and by numerical integration over a grid of carefully chosen parameter values. For models with more than two parameters, we present all pairwise results.

The first two examples are of an AR(1) model and a normal model, and full details of each are provided in Appendix B. The findings from both of these models are similar to the other examples.

An implementation of wBSL is available in the BSL package in R \citep{an2019bsl}, which is available at https://github.com/ziwenan/BSL.
%%%%%%%%%%%%%%%%%%%%%%%%%%%%%%%%%%%%%%%%%%%
%%%%%%%%%%%%%%%%%%%%%%%%%%%%%%%%%%%%%%%%%%%
\subsection{An MA(2) model}
%%%%%%%%%%%%%%%%%%%%%%%%%%%%%%%%%%%%%%%%%%%
%%%%%%%%%%%%%%%%%%%%%%%%%%%%%%%%%%%%%%%%%%%

%\textcolor{red}{[How are you handling $x_1$ and $x_2$ here? Need to tweak the below to account for this]}\\
The MA$(2)$ model represents a univariate series of temporally dependent observations  as
%. The model assumes the last $q$ error terms $w_{t-1},...,w_{t-q}$ are combined linearly to form the observed data,
\begin{align*}
    x_t = w_t + \theta_1 w_{t-1} + \theta_2 w_{t-2}
    %x_t = w_t + \theta_1 w_{t-1} + \theta_2 w_{t-2} +\cdots+ \theta_q w_{t-q}
    \quad\mbox{where}\quad
    w_i\sim\mathcal{N}(0,\sigma^2), i=-1,0,1,\ldots,T_0,
\end{align*}
for $t=1,\ldots,T_0$,
%It is common to assume that $w_{t-1},...,w_{t-q}$ is a Gaussian white noise series such that $w_i \sim \mathcal{N}(0,\sigma^2)$ for $i=t-1,...,t-q$. 
%Given that the likelihood of the moving average model is quick to compute for moderately sized time series, moving average models serve as useful toy examples for the verification of likelihood-free methodologies. \\
%
%Here we consider the case where  $q=2$ 
%so that each observation is of the form: 
%\begin{align*}
%    x_t = w_t + \theta_1 w_{t-1} + \theta_2 w_{t-2}
%\end{align*}
%and that $w_i\sim \mathcal{N}(0,1)$ for $i = t,t-1$ and $t-2$, 
%which results in t
and has parameter constraints $-1<\theta_2<1$, $\theta_1+\theta_2>-1$ and $\theta_1-\theta_2<1$. Defining $\gamma(h) = \text{Cov}(x_t,x_{t-h})$, then the likelihood is Gaussian with zero mean vector and covariance matrix constructed from $\gamma(0) = 1+\theta_1^2 + \theta_2^2$, $\gamma(1) = \theta_1 + \theta_1\theta_2$, $\gamma(2) = \theta_2$ and $\gamma(h)=0$ for $h>2$. 
We generate 200 observations from the MA$(2)$ process with $\boldsymbol{\theta}_{\text{true}} = (\theta_1,\theta_2)^\top = (0.6,0.2)^\top$ and fixed $\sigma^2=1$, and specify the full observed dataset as  summary statistics.
Under this setting, the summary statistics are exactly multivariate normal distributed and so standard BSL should perform well in terms of posterior approximation accuracy.
% are taken to be the full data set. Given this, and the multivariate normal likelihood, the multivariate normality assumption for the summary statistics is satiated and standard BSL should perform well in terms of posterior accuracy.\\
%
We compare the results of wBSL and Warton shrinkage by itself to the output of a standard Metropolis-Hastings sampler using the known likelihood. We find that $n=10\,000$ simulations are efficient for standard BSL, and we use $n_{\text{cov}} = 20\,000$ model simulations at $\boldsymbol{\theta}^0=\boldsymbol{\theta}_{\text{true}}$ to accurately estimate $\boldsymbol{W}$. We use $T = 200\,000$ MCMC sampler iterations and a uniform prior over the parameter support.
% for each method.\\

\begin{figure}[h!]
\centering
\begin{subfigure}
\centering\includegraphics[width = 15cm]{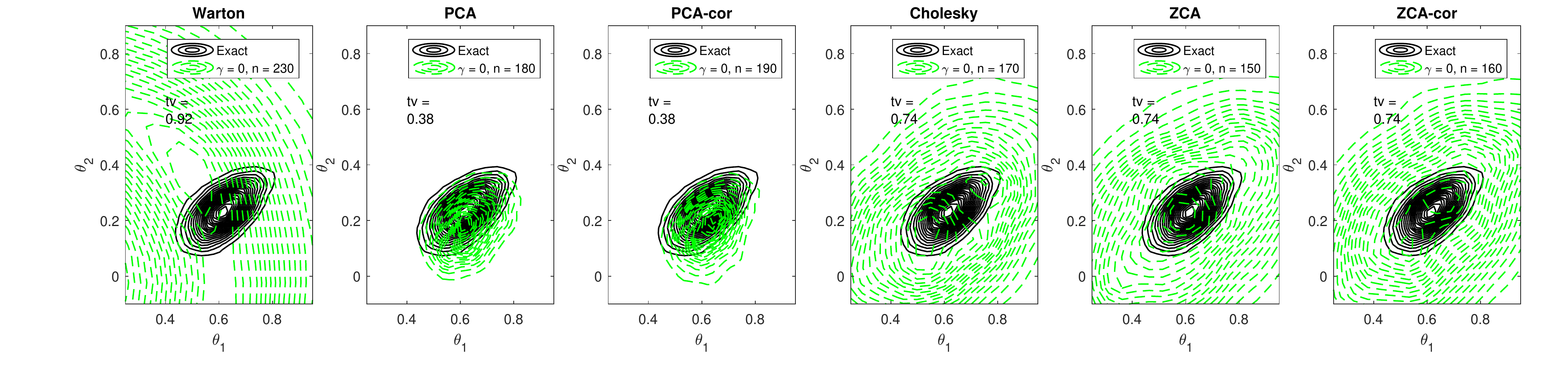}
\end{subfigure}
\begin{subfigure}
\centering\includegraphics[width = 15cm]{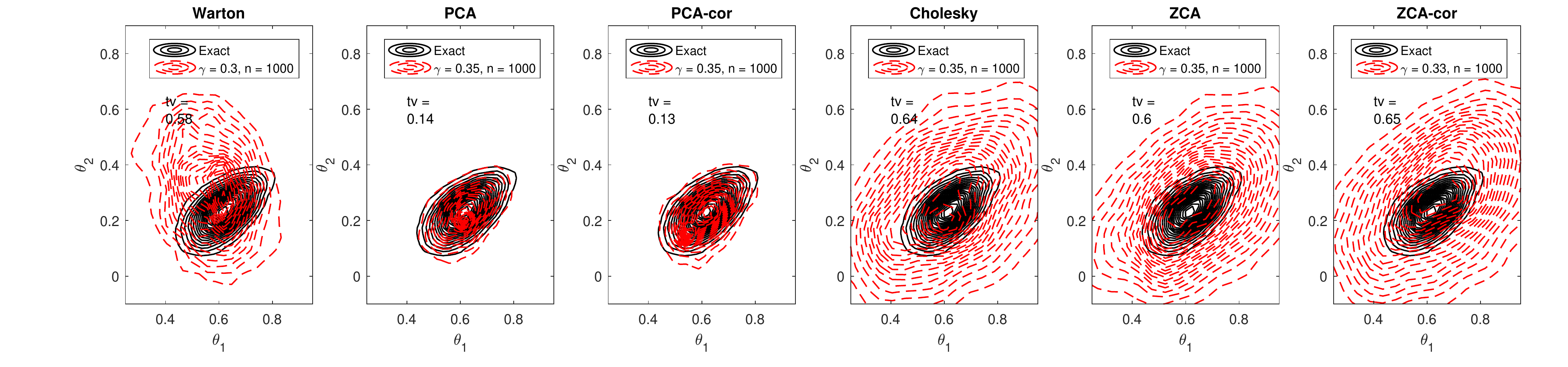}

\end{subfigure}
\begin{subfigure}
\centering\includegraphics[width = 15cm]{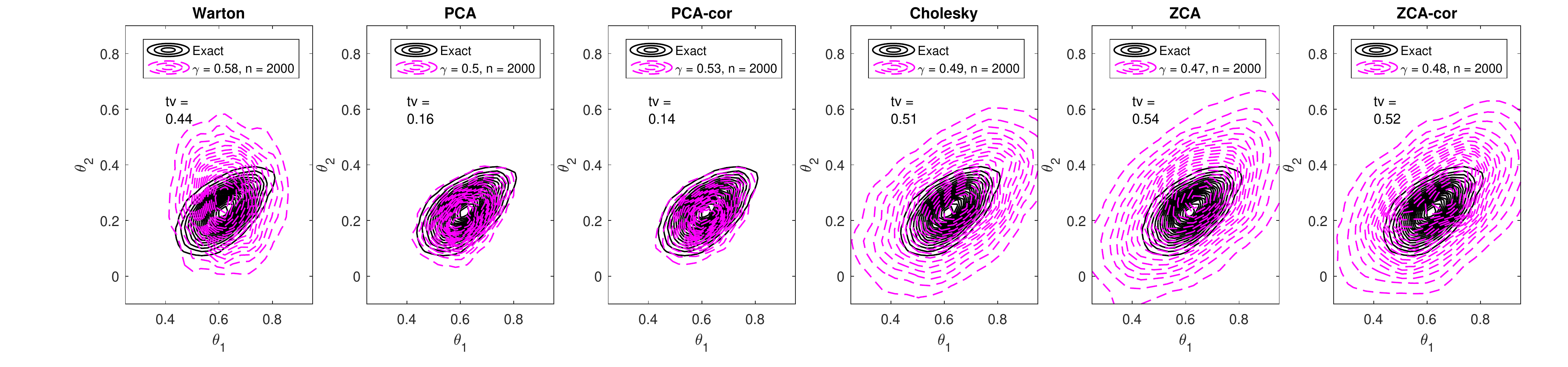}
\end{subfigure}

\begin{subfigure}
\centering\includegraphics[width = 15cm]{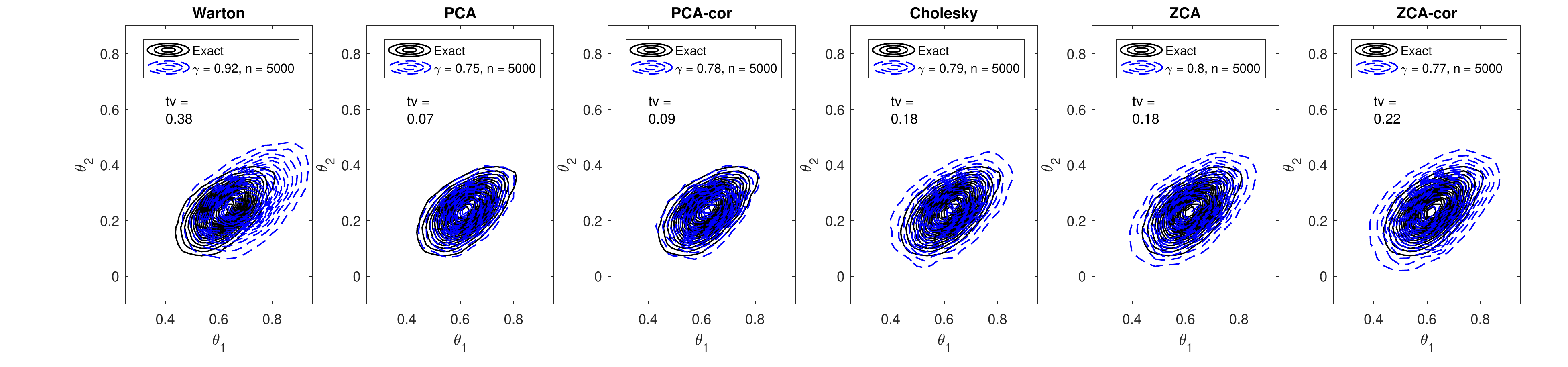}\end{subfigure}
\caption{\small
%MA(2) example bivariate 
Contour plots of the wBSL posterior approximations for the MA(2) model. 
Columns denote (left to right) Warton shrinkage alone, and the whitening methods PCA, PCA-cor, Cholesky, ZCA and ZCA-cor. Rows correspond to complete shrinkage ($\gamma=0$; top row) and 
 $90\%$, $80\%$ and $50\%$ reductions in the number of model simulations (rows 2--4).
 tv denotes total variation distance between approximate and true bivariate distributions.
 }
\label{fig:MA1}
\end{figure}

%We use bivariate c
Contour plots of the estimated joint posterior distribution under each method are shown in Figure \ref{fig:MA1}.
 It is evident that when the number of model simulations for estimating the synthetic likelihood is less than $n=5\,000$, using Warton shrinkage alone (leftmost column) fails to recover an accurate posterior approximation. This is likely
 %-- a likely result of 
 due to significant dependence between the summary statistics.
As the level of shrinkage is increased (i.e.~$\gamma$ is reduced), the estimated posterior variances and dependence structure become increasingly poor. 
 
% On the other hand, 
 In contrast, all forms of whitening produce accurate dependence structures. PCA and PCA-cor whitening are the only procedures that consistently provide accurate estimates of the variance for varying $n$: ZCA, ZCA-cor and Cholesky whitening all have inflated variances for smaller $n$ to roughly the same extent. Note that for  $n=5\,000$ model simulations (a 50\% computational reduction compared to standard BSL), all whitening methods produce reasonably accurate posterior distributions. However,  PCA-based results are reasonably accurate for all levels of shrinkage. This is  impressive as for complete shrinkage ($\gamma=0$) the number of model simulations ($n=180$) is reduced by two orders of magnitude for wBSL compared to BSL ($n=10\,000$ in BSL).

%Since PCA and PCA-cor whitening appear to be the most effective, it is natural to consider models with highly correlated summary statistics. As such, we turn to the AR(1) model.

%%%%%%%%%%%%%%%%%%%%%%%%%%%%%%%%%%%%%%%%%%%
%%%%%%%%%%%%%%%%%%%%%%%%%%%%%%%%%%%%%%%%%%%
\subsection{Movement models for Fowler's toads}
%%%%%%%%%%%%%%%%%%%%%%%%%%%%%%%%%%%%%%%%%%%
%%%%%%%%%%%%%%%%%%%%%%%%%%%%%%%%%%%%%%%%%%%

Understanding the movement behaviour of native and invasive species is an important topic in ecology \citep{lindstrom+bsps13}.
\cite{marchand2017stochastic} consider three
%In this example, we consider the 
individual-based movement models 
%(IBM) 
for a species of Fowler's toads (\textit{Anaxyrus fowleri}), 
%by \cite{marchand2017stochastic}. The study is 
motivated by a desire to understand the link between small scale movements and larger phenomena, such as home ranges, dispersal and migrations at seasonal, annual or life-time scales. 
%
%\cite{marchand2017stochastic} devise 3 movement models (random return, nearest return and 
%distance-based return probability), each 
%assuming the toads are nocturnal -- 
In particular, the random-return  model  assumes that  toads take refuge during the day and forage throughout the night,
% simulating an overnight displacement, and differs in the returning behaviour.
%
%The random return model first 
generating a net overnight displacement $\Delta x_n$ from a Levy alpha-stable distribution, $S(\alpha,\xi)$, with stability parameter 
%$\alpha$, such that 
$0\leq\alpha\leq2$ and scale parameter $\xi>0$. Toads are assumed to return only at the end of the nighttime foraging path, with constant probability, $p_0$. The refuge site is determined random from any of the previous refuge sites, with previously visited sites given a higher weighting.

Previously \cite{marchand2017stochastic} and \cite{an2018robust} used ABC and synthetic likelihood, respectively, for inference for this model.
%for inference and find that the random return model performs best, and following the analysis of \cite{an2018robust}, we consider this model for the application of synthetic likelihood. \\
%
Following \cite{marchand2017stochastic} we consider synthetically generated data for  $n_t = 66$ toads  recorded at least once per night (active foraging) and once per day (resting in refuge) over $n_d = 63$ days,
with $\boldsymbol{\theta}_{\text{true}} = (\alpha,\xi,p_0)^\top = (1.7, 35, 0.6)^\top$.
%, following the results in \cite{marchand2017stochastic}. 
We also specify uniform priors $\alpha\sim\mathcal{U}(1,2)$,  $\xi\sim\mathcal{U}(0,100)$ and  $p_0\sim\mathcal{U}(0,0.9)$. 
% are used for $\alpha$, $\gamma$ and $p_0$ respectively.\\
%
%The location of $n_t = 66$ toads  is recorded at least once per night (active foraging) and once per day (resting in refuge) over $n_d = 63$ days via radio tracking in Ontario, Canada. This information is stored in an observation matrix $\boldsymbol{Y}$ of dimension $n_t\times n_d$.\\
%
The distance moved distribution for each toad at time lags of 1, 2, 4 and 8 days was found, and the log of the differences in the $0,0.1,...,1$ quantiles, the number of absolute displacements less than 10m, and the median of the absolute displacements greater than 10m are used as summary statistics (48 in total).
As before, $n_{cov}=20\,000$ model simulations drawn at $\boldsymbol{\theta}^0=\boldsymbol{\theta}_{\text{true}}$ are used to estimate $\boldsymbol{W}$, and implement standard BSL and wBSL samplers for $T=100\,000$ MCMC iterations. BSL was found to perform efficiently for this setting with $n=500$ model simulations per iteration.

%All five whitening procedures are applied for inference, along with standard BSL and Warton shrinkage by itself. It was found that $n=500$ model simulations are needed at each iteration for standard BSL. For each run, we use $T = 100000$ iterations of MCMC. In wBSL, $n_{\text{cov}} = 20000$ model simulations are used to estimate $\boldsymbol{W}$.\\

The resulting estimated bivariate marginal distributions for wBSL with PCA whitening and Warton shrinkage by itself are shown in Figure \ref{fig:Toad2}; the results for wBSL with the remaining whitening methods are shown in Appendix C. 
Here Warton shrinkage by itself performs very poorly compared to standard BSL, producing both biased estimates and significantly underestimating marginal variances, unless large numbers of samples ($n=250$) are used.

In contrast, wBSL performs well compared to standard BSL. There are smaller differences between each of the whitening methods than seen in the previous examples,
most likely due to a lack of sensitivity of the covariance matrix of the summary statistics to the model parameters.
%. We suspect that in this example the covariance is not very sensitive to parameters, relative to the previous toy examples.  
PCA-based whitening appears to perform the best, followed by ZCA-based whitening and Choleksy whitening the worst performing. However, for all whitening types, the posterior approximations with complete shrinkage ($\gamma = 0$) provide reasonable approximations to the standard BSL posterior.  In this case, the number of model simulations is reduced by an order of magnitude from standard BSL ($n=500$)  to wBSL ($n\leq 44$).

\begin{figure}[h!]
\centering
\begin{subfigure}
\centering\includegraphics[width = 15cm]{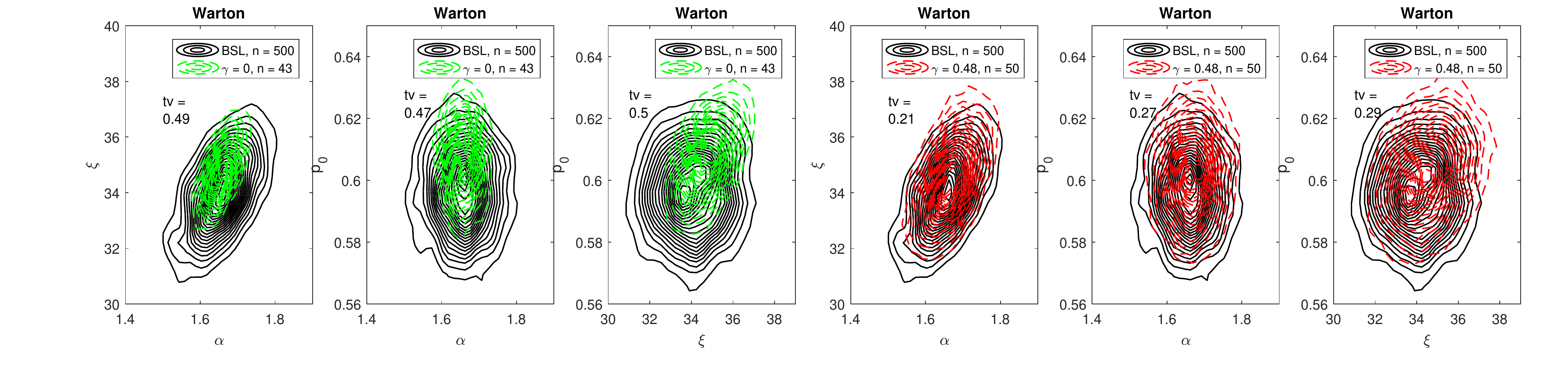}
\end{subfigure}
\begin{subfigure}
\centering\includegraphics[width = 15cm]{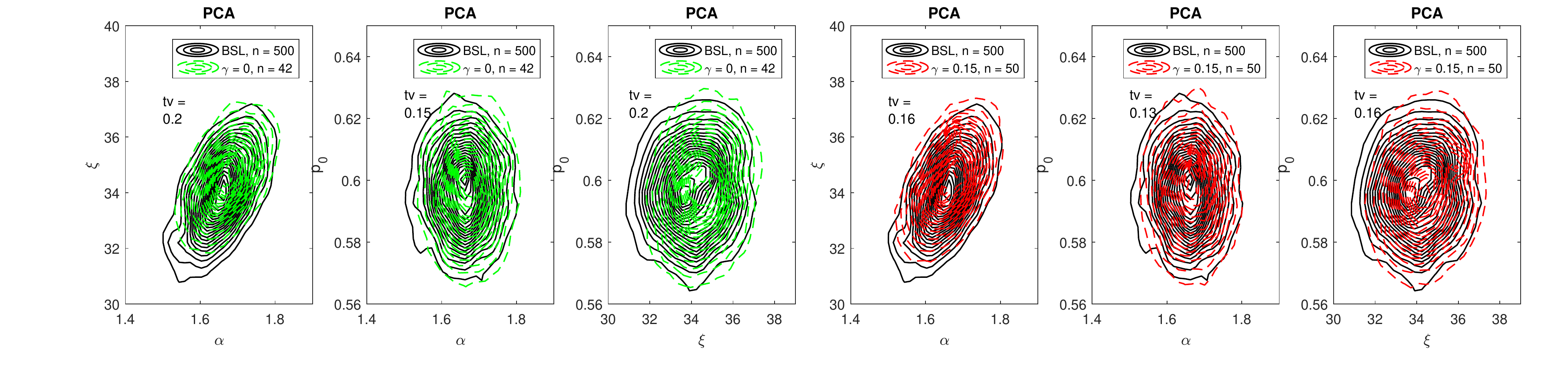}
\end{subfigure}
\begin{subfigure}
\centering\includegraphics[width = 15cm]{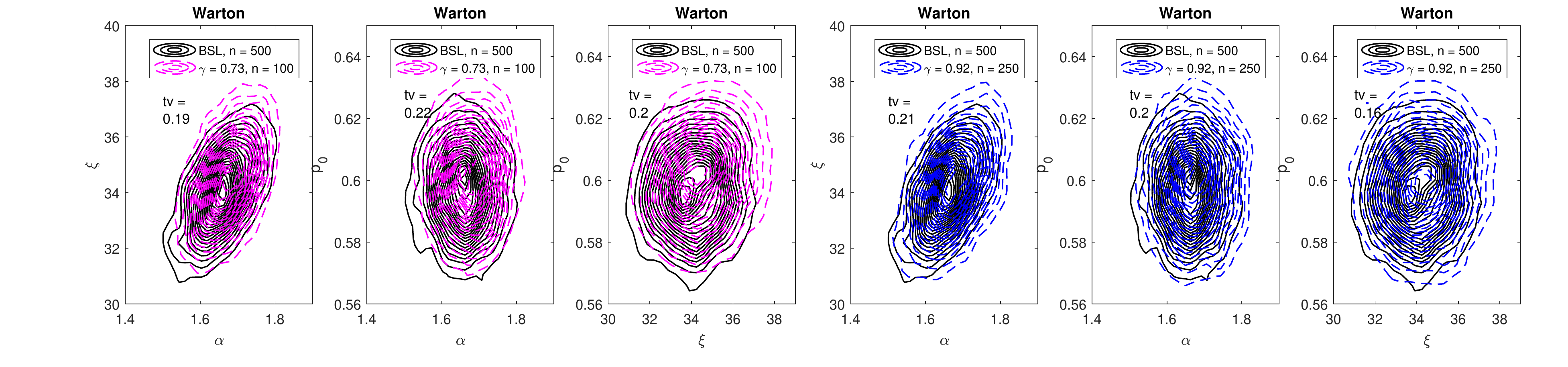}
\end{subfigure}
\begin{subfigure}
\centering\includegraphics[width = 15cm]{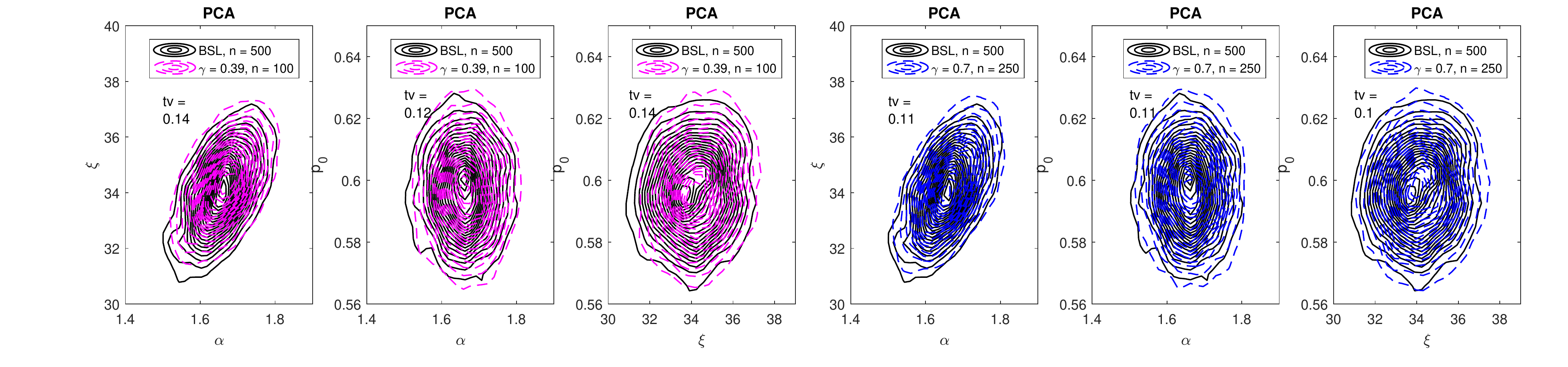}
\end{subfigure}\\
\caption{\small
%Toad example approximate posterior distributions. From top left to bottom right, the level of shrinkage and number of model simulations increases. Rows 1 and 3 are Warton shrinkage by itself, and rows 2 and 4 are PCA wBSL.
%
Contour plots of the bivariate margins of the synthetic likelihood posterior approximations for the toad displacement random-return  model. 
Solid lines denote BSL ($n=500$) estimates.
Rows denote Warton shrinkage alone (rows 1, 3) and the PCA whitening method (rows 2,4).
Results correspond to complete shrinkage ($\gamma=0$) and 
 $90\%$, $80\%$ and $50\%$ reductions in the number of model simulations.
 tv denotes total variation distance between approximate and `true' (BSL) bivariate marginal distributions.
%
%Columns denote (left to right) Warton shrinkage alone, and the whitening methods PCA, PCA-cor, Cholesky, ZCA and ZCA-cor. Rows correspond to complete shrinkage ($\gamma=0$; top row) and 
% $90\%$, $80\%$ and $50\%$ reductions in the number of model simulations (rows 2--4).
}
\label{fig:Toad2}
\end{figure}

%\textcolor{red}{[I wonder if we should make Figure 4 the same layout as all of the other figures (and include all whitening methods) by just showing the $(\alpha, \gamma)$ posterior, and putting the other 2 bivariate margin plots in the Supporting Information. The layout here has just changed completely from the previous examples.]}

%%%%%%%%%%%%%%%%%%%%%%%%%%%%%%%%%%%%%%%%%%%
%%%%%%%%%%%%%%%%%%%%%%%%%%%%%%%%%%%%%%%%%%%
\subsection{Collective cell spreading}
%%%%%%%%%%%%%%%%%%%%%%%%%%%%%%%%%%%%%%%%%%%
%%%%%%%%%%%%%%%%%%%%%%%%%%%%%%%%%%%%%%%%%%%

Central to the understanding of many biological phenomena, such as tissue repair \citep{shaw2009wound} and cancer \citep{friedl2003tumour}, is an understanding of collective cell behaviour. Mathematical models are a flexible tool for gaining insight into the movement, proliferation and interactions between cells on a cell-to-cell level (e.g.~\citealp{vo2015quantifying} \citealp{johnston2014interpreting}). An appealing approach is the continuous time, continuous space stochastic individual-based model of \cite{Binny2016}. Using ABC methods \cite{browning2018inferring} calibrate this model to experimental results obtained by a cell proliferation assay experiment.

%The individual-based model used in \cite{browning2018inferring} 
The model assumes that cells
%, or agents, 
are uniformly sized discs with diameter $\sigma = 24\mu m$ and location $\boldsymbol{x}_n = (x_1, x_2)^\top$ for $n = 1,...,N(t)$ cells. Two events occur: proliferation and movement, each evolving according to a Poisson process with intrinsic parameters $p>0$ and $m>0$, respectively. The rates of the $n^{\text{th}}$ cell, $P_n$ and $M_n$ depend on the crowding of neighbouring cells as determined by a Gaussian kernel 
%$w^{(\cdot)}(r)$ 
$w(r)$ 
given separation distance $r\geq0$. \cite{browning2018inferring} assume that that the net proliferation and movement rates reduce to zero under maximum hexagonal cell packing. Upon proliferation events, the location of the daughter cell is simulated from a bivariate normal distribution, $\mathcal{N}(\boldsymbol{x}_n, \sigma^2\boldsymbol{I}_2)$. For movement events, the preferred direction of movement is in the direction away from regions of high cell density $-\nabla B(\boldsymbol{x}_n)$, according to the crowding surface $B(\boldsymbol{x})$, with closeness governed by a Gaussian kernel and repulsive strength parameter $\gamma_b \geq0$. The movement distance is $\sigma$, which is equal to the cell diameter. The parameters to be inferred are $\boldsymbol{\theta} = (m,p,\gamma_b)^\top$.

We follow the results from \cite{browning2018inferring}'s PC-3 prostate cancer cell line in their cell proliferation assay experiment, for which images are taken every 12 hours for a total duration of 36 hours. 
We generated simulated data under this setting with $\boldsymbol{\theta}_{\text{true}} = (1,0.04,5)^\top$.
%In the implementation of ABC rejection in \cite{browning2018inferring}, 6 summary statistics are used -- the number of cells at each time $N(t)$, and a measure of the spatial structure $\mathcal{P}(t)$ given by the ratio of observed pairs of cells separated by a distance less than $R = 50\mu m$, to the expected number of pairs of cells separated by a distance less than $R\mu m$.\\
%
%Here we consider the application of synthetic likelihood to a data set simulated from the model of \cite{Binny2016} and \cite{browning2018inferring}. In this example we only consider PCA whitening due to its good performance in previous examples. A brief description of the model is also provided, but we refer the reader to \cite{Binny2016} and \cite{browning2018inferring} for more details. \\
%
In our BSL implementation we use 21 summary statistics. At 12, 24 and 36 hours, we record the number of cells,  Ripley's $K$ function evaluated at $r = 25, 50$ and $100\mu m$ and Ripley's $J$ function evaluated at $r = 10, 20$ and $40\mu m$
%. The $K$ and $J$ functions are useful descriptive statistics for spatial point processes; we refer the reader to 
\citep[see e.g.][for a discussion of these]{baddeley2007spatial}. 
Priors are specified as $p\sim \mathcal{U}(0,10)$, $m\sim \mathcal{U} (0,0.1)$ and $\gamma_b\sim\mathcal{U} (0,20)$, and it is found that $n = 150$ model simulations are required to implement standard BSL efficiently.
%to estimate the log SL at some $\boldsymbol{\theta}$ with high posterior support with standard deviation between 1 and 2 for standard BSL. 
%
%We tune $\gamma$ to allow for half, a fifth and a tenth of the number of model simulations to standard BSL in the same way, and find that appropriate levels of shrinkage are $\gamma = 0.8$, $\gamma = 0.5$ and $\gamma = 0$, respectively. 
We use $n_{\text{cov}} = 300$ model simulations to estimate the whitening matrix $\boldsymbol{W}$ given $\boldsymbol{\theta}^0=\boldsymbol{\theta}_{\text{true}}$ in a region of high posterior support, and a total of $T = 100\,000$ MCMC sampler iterations.

\begin{figure}[h!]
\centering
\begin{subfigure}
\centering\includegraphics[width = 15cm]{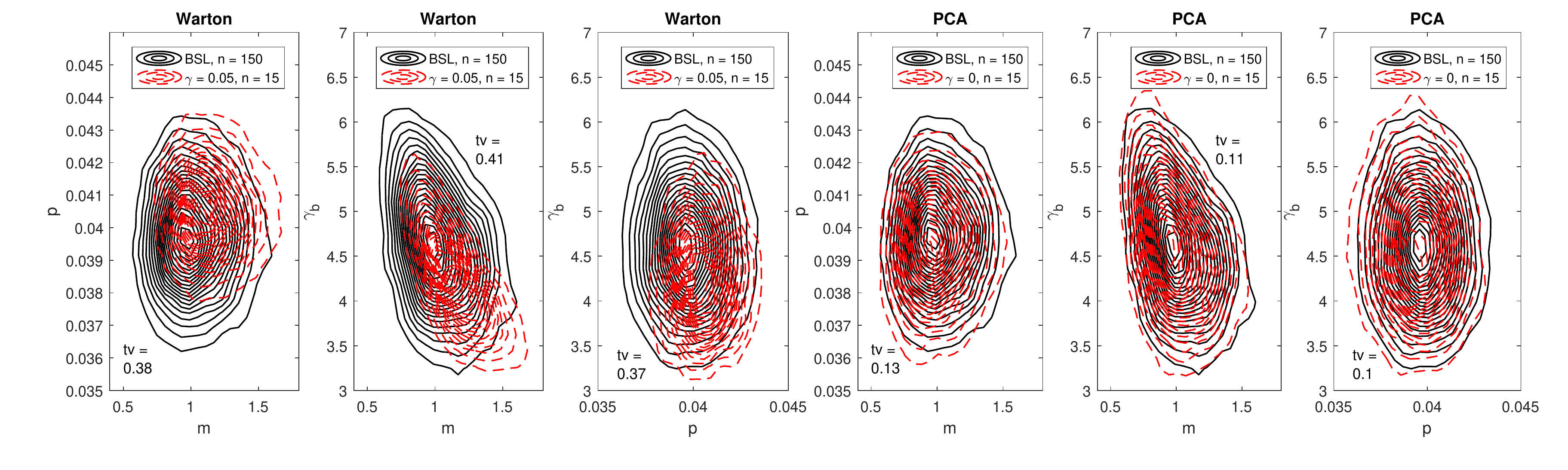}
\end{subfigure}
\begin{subfigure}
\centering\includegraphics[width = 15cm]{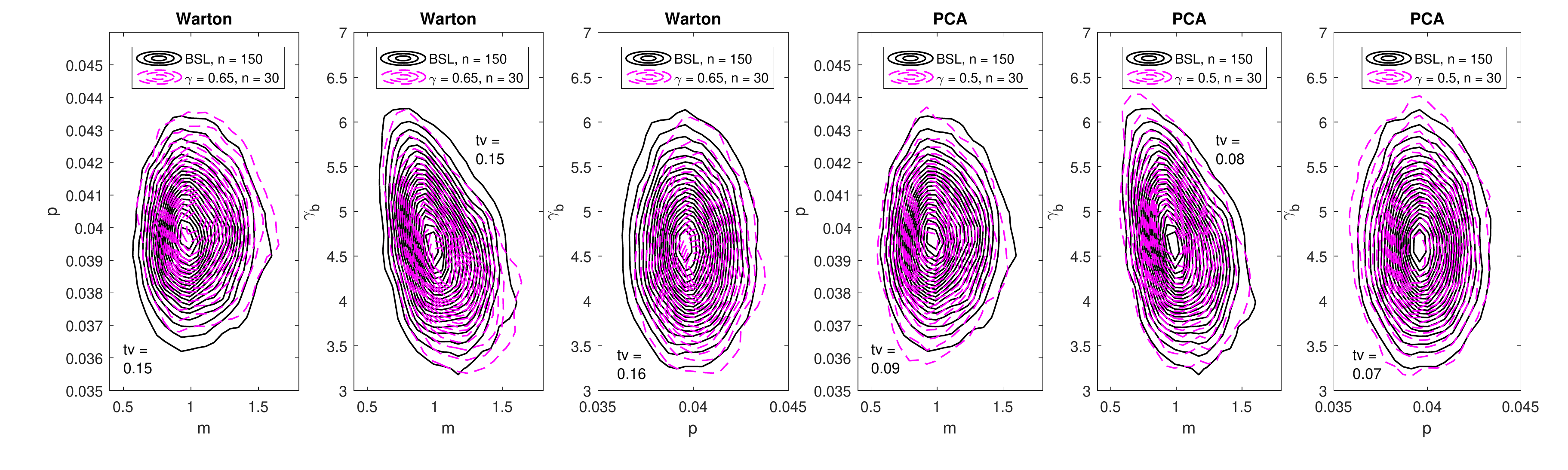}

\end{subfigure}
\begin{subfigure}
\centering\includegraphics[width = 15cm]{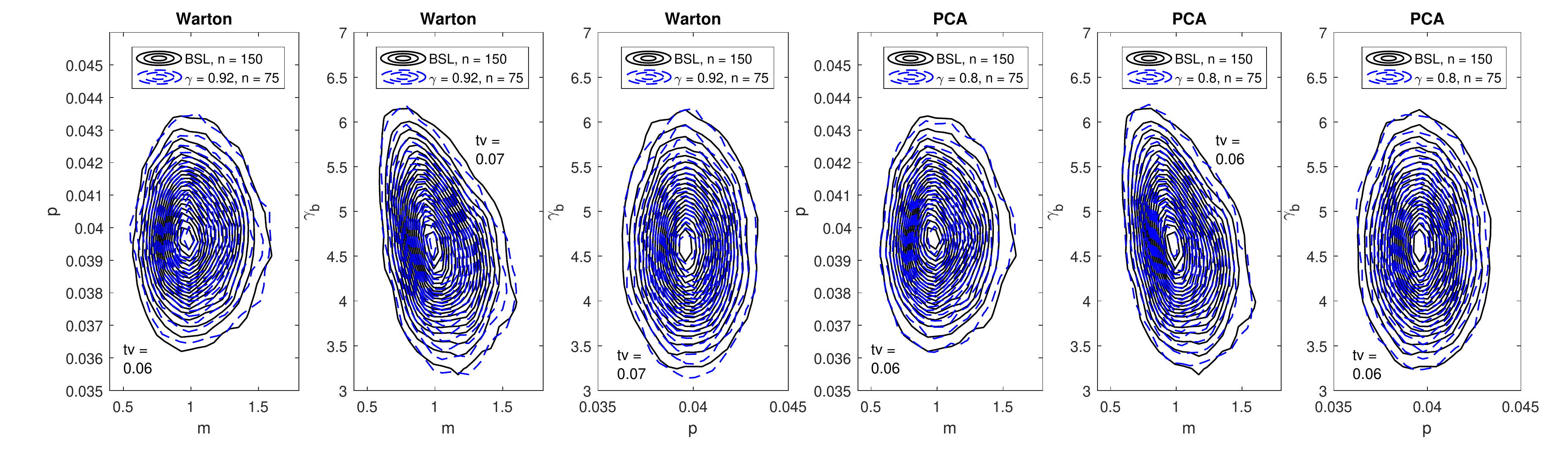}
\end{subfigure}

\caption{\small
%Collective cell spreading example results for wBSL using PCA whitening (right three columns) and Warton shrinkage by itself (left three columns). The top row shows the results for $n = 15$, the second row is for $n = 30$ and the bottom row is $n=75$. All results are compared to standard BSL, which uses $n=150$ model simulations. 
%
Contour plots of the bivariate margins of the synthetic likelihood posterior approximations for the collective cell spreading model. 
Solid lines denote BSL ($n=150$) estimates.
Left columns denote Warton shrinkage alone and right columns denote PCA whitening wBSL.
Results correspond to complete shrinkage ($\gamma=0$) and 
 $90\%$, $80\%$ and $50\%$ reductions in the number of model simulations.
 tv denotes total variation distance between approximate and `true' (BSL) bivariate marginal distributions.
 %Other wBSL whitening method results are provided in the Supporting Information.
}
\label{fig:lfc1}
\end{figure}

The resulting estimated bivariate posterior approximations are shown in Figure \ref{fig:lfc1}, for both PCA whitening wBSL (as the best performing wBSL method) and Warton shrinkage alone. 
The shrinkage-only posterior approximations are close to the BSL posterior approximations for $n=75$ (bottom row) and $n=30$ (middle row) model simulations (compared to $n=150$ for standard BSL). However, both posterior location and variance are much less accurate for $n=15$ model simulations ($\gamma=0$). In contrast, PCA wBSL performs very well for all levels of shrinkage. This order of magnitude performance gain is a particularly significant result, as simulating data under this model is very is computationally expensive.

%%%%%%%%%%%%%%%%%%%%%%%%%%%%%%%%%%%%%%%%%%%
%%%%%%%%%%%%%%%%%%%%%%%%%%%%%%%%%%%%%%%%%%%
\section{Whitening method choice and sensitivity}
\label{sec:investigate}
%%%%%%%%%%%%%%%%%%%%%%%%%%%%%%%%%%%%%%%%%%%
%%%%%%%%%%%%%%%%%%%%%%%%%%%%%%%%%%%%%%%%%%%

%%%%%%%%%%%%%%%%%%%%%%%%%%%%%%%%%%%%%%%%%%%
%%%%%%%%%%%%%%%%%%%%%%%%%%%%%%%%%%%%%%%%%%%
\subsection{Choice of Whitening Method}
%%%%%%%%%%%%%%%%%%%%%%%%%%%%%%%%%%%%%%%%%%%
%%%%%%%%%%%%%%%%%%%%%%%%%%%%%%%%%%%%%%%%%%%

The empirical results in Section \ref{sec:examples} suggest that PCA-based whitening methods provide the most accurate posterior approximation. 
%We now explore why this may be the case.
% by considering the MA$(2)$ and AR$(1)$ examples from earlier. 
Recall that for covariance shrinkage to be effective, the whitening transformation should decorrelate the summary statistics so that their covariance matrix is close to diagonal for parameter values that reside in regions with non-negligible posterior density.
%parts of the parameter space with non-negligible posterior support. 
We explore how well this has been achieved for the MA(2) and AR(1) models considered earlier.

For each model we compute the whitening matrix $\boldsymbol{W}_{\text{true}}$ using the known analytical covariance $\boldsymbol{\Sigma}(\boldsymbol{\theta}_{\text{true}})$ at the true parameter value $\boldsymbol{\theta}_{\text{true}}$. We then 
%use $\boldsymbol{W}$ and the analytical covariance $\boldsymbol{\Sigma}$ at MCMC parameter samples from the %true posterior to 
compute the covariances of the transformed summary statistics $\tilde{\boldsymbol{\Sigma}}(\boldsymbol{\theta}) = \boldsymbol{W}_{\text{true}}\boldsymbol{\Sigma}(\boldsymbol{\theta})\boldsymbol{W}_{\text{true}}^\top$ where $\boldsymbol{\Sigma}(\boldsymbol{\theta})$ is the known analytical covariance matrix for values of $\boldsymbol{\theta}$ drawn from the true posterior.
%, with \textcolor{red}{$n=XXX$.}
%
 We compute the difference between the upper triangular portion of $\tilde{\boldsymbol{\Sigma}}(\boldsymbol{\theta})$, both including and excluding the diagonal, from the identity $\boldsymbol{I}_d$ and zero matrix $\boldsymbol{0}_{d\times d}$ respectively, and then calculate the $L_1$ matrix norm. 
 %We also repeat the same process over a grid of values that covers the entire parameter space. 
These matrix norm deviations quantify the location and magnitude of the lack of effectiveness of the whitening transformation, and consequently where this deviation may have a direct effect on the wBSL posterior approximation. 
  By using the known analytical covariances there is no Monte Carlo error in these results. 
% We compare the error between whitening methods. \\

\begin{figure}[h!]
\centering
\begin{subfigure}
\centering\includegraphics[width = 5cm]{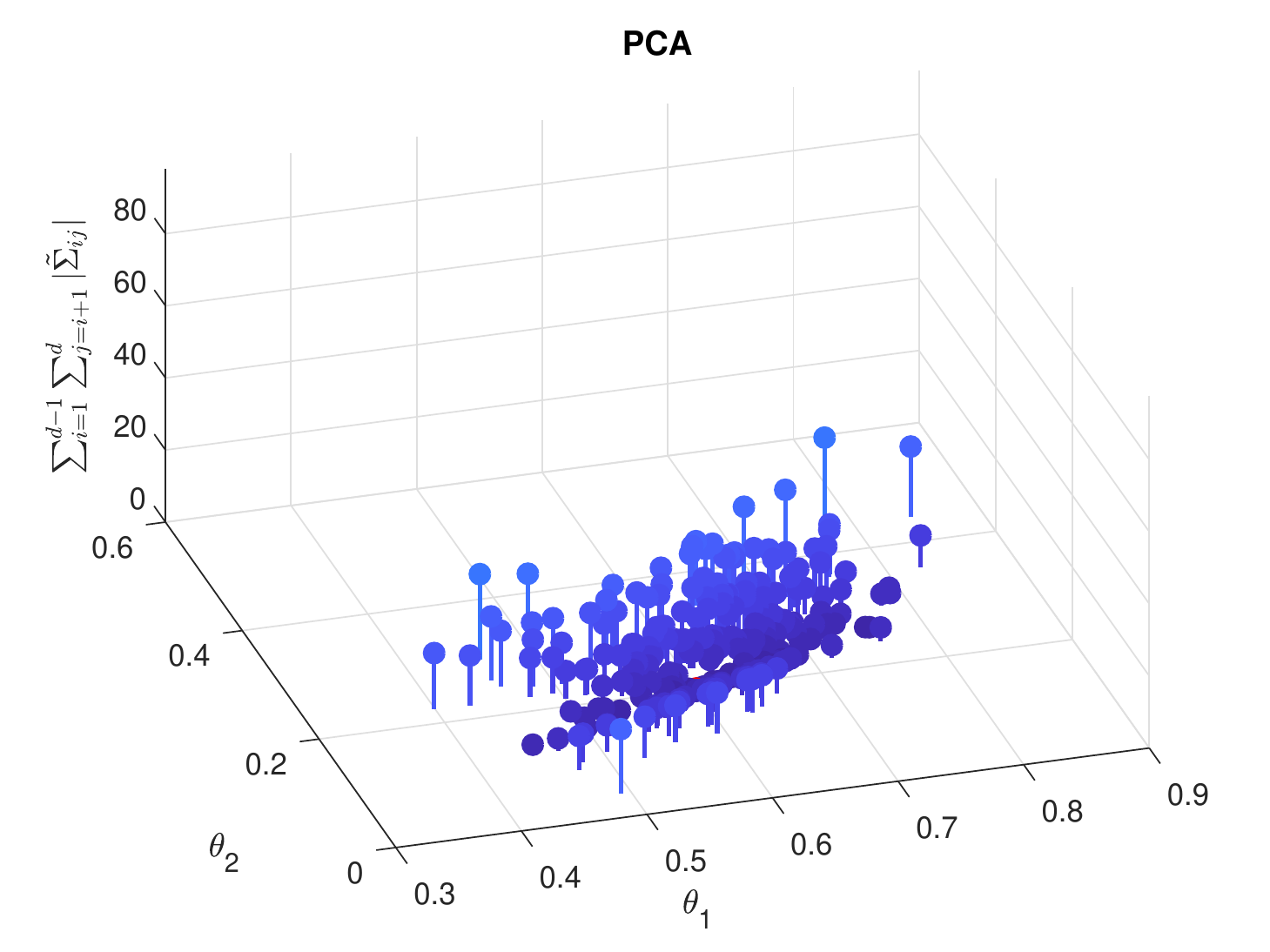}
\end{subfigure}
\begin{subfigure}
\centering\includegraphics[width = 5cm]{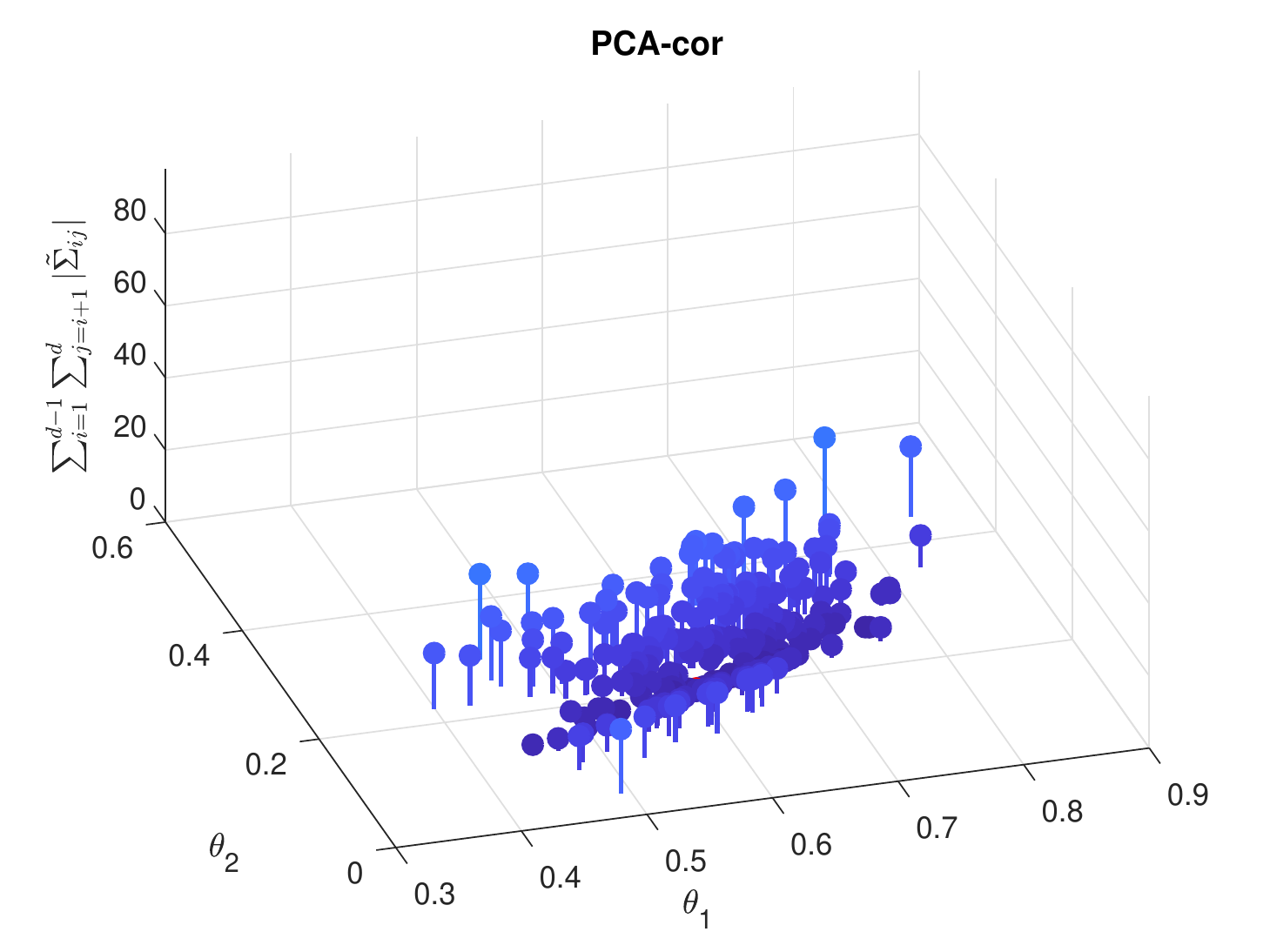}
\end{subfigure}
\begin{subfigure}
\centering\includegraphics[width = 5cm]{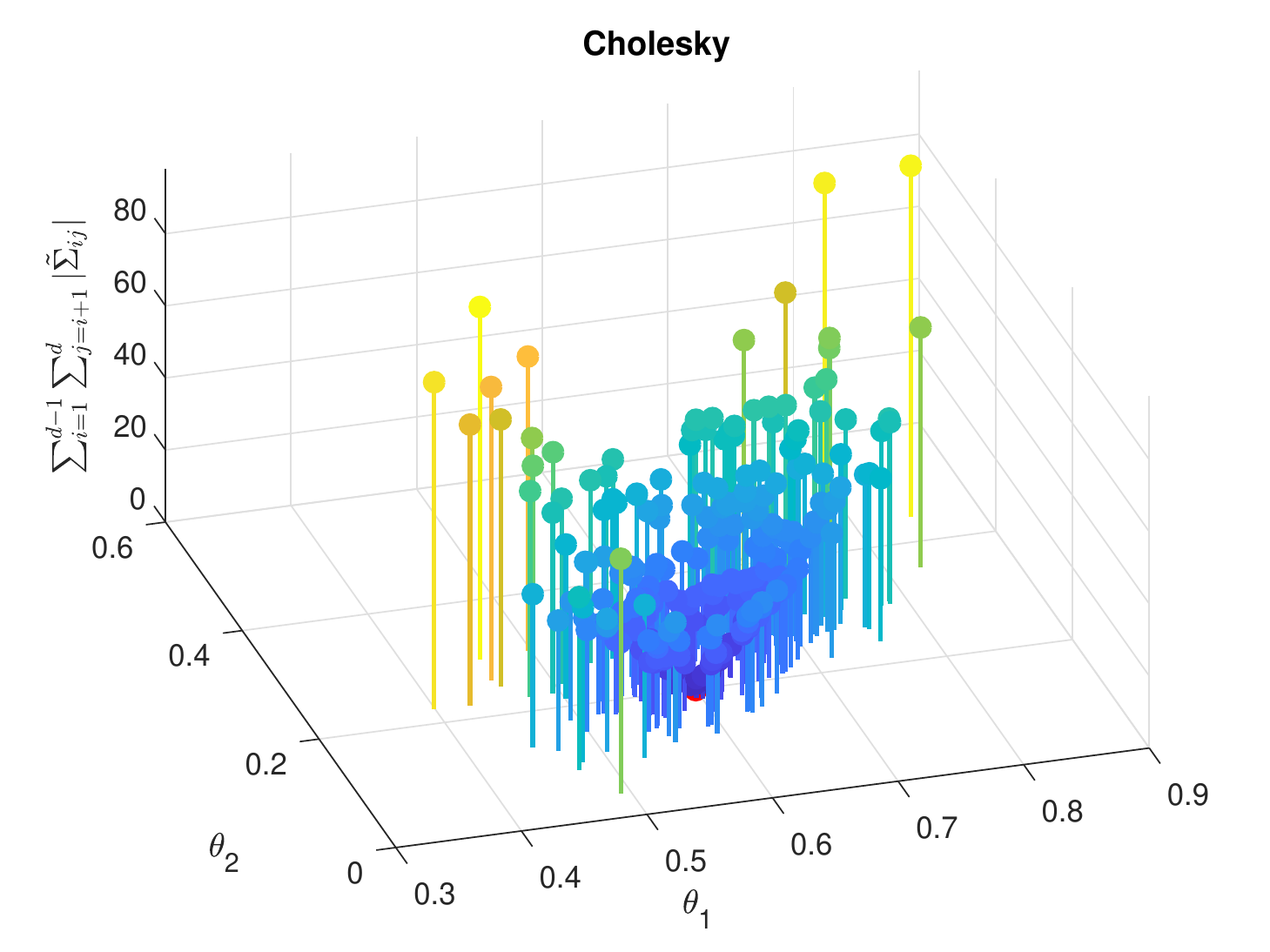}
\end{subfigure}
\begin{subfigure}
\centering\includegraphics[width = 5cm]{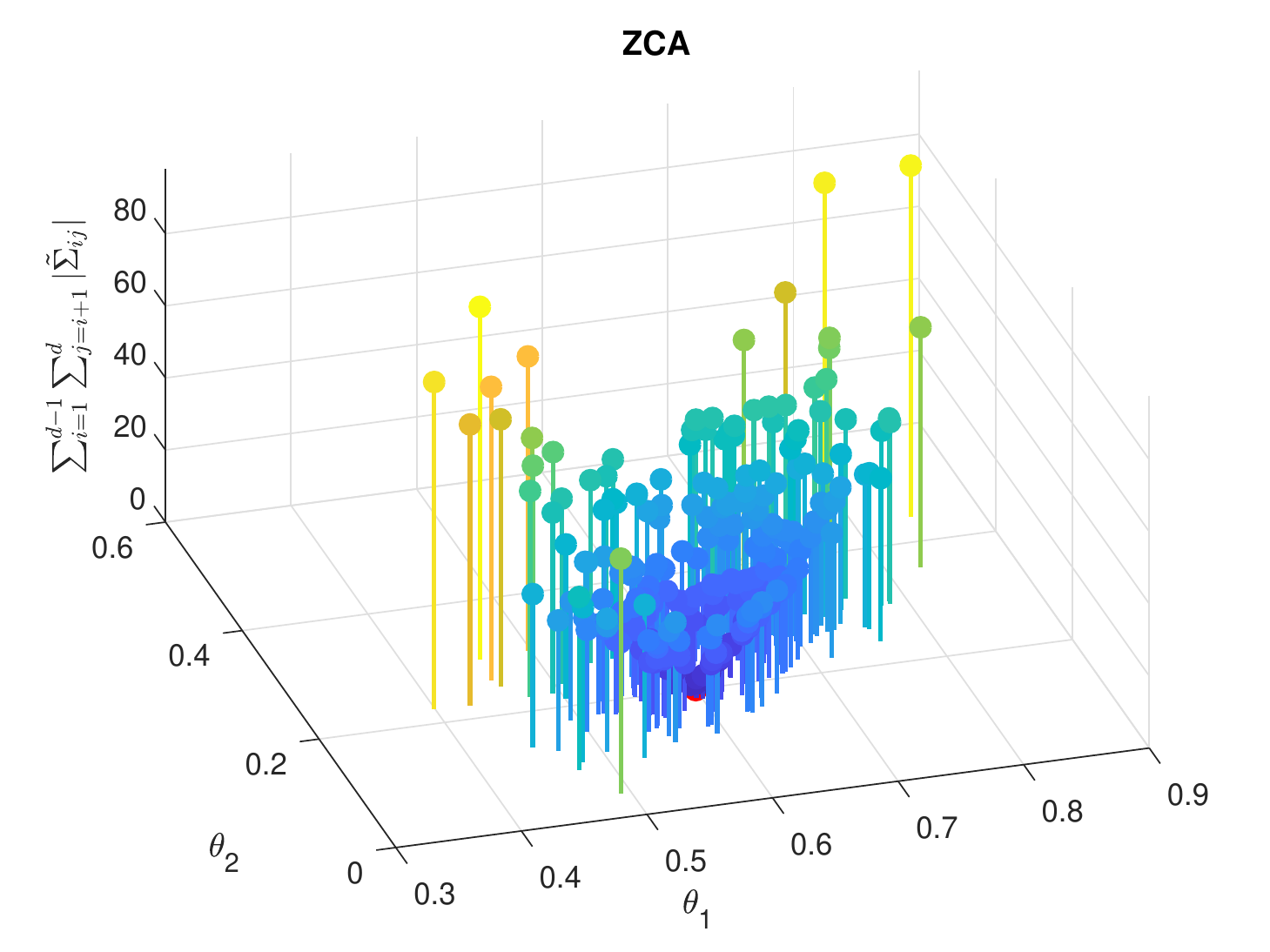}
\end{subfigure}
\begin{subfigure}
\centering\includegraphics[width = 5cm]{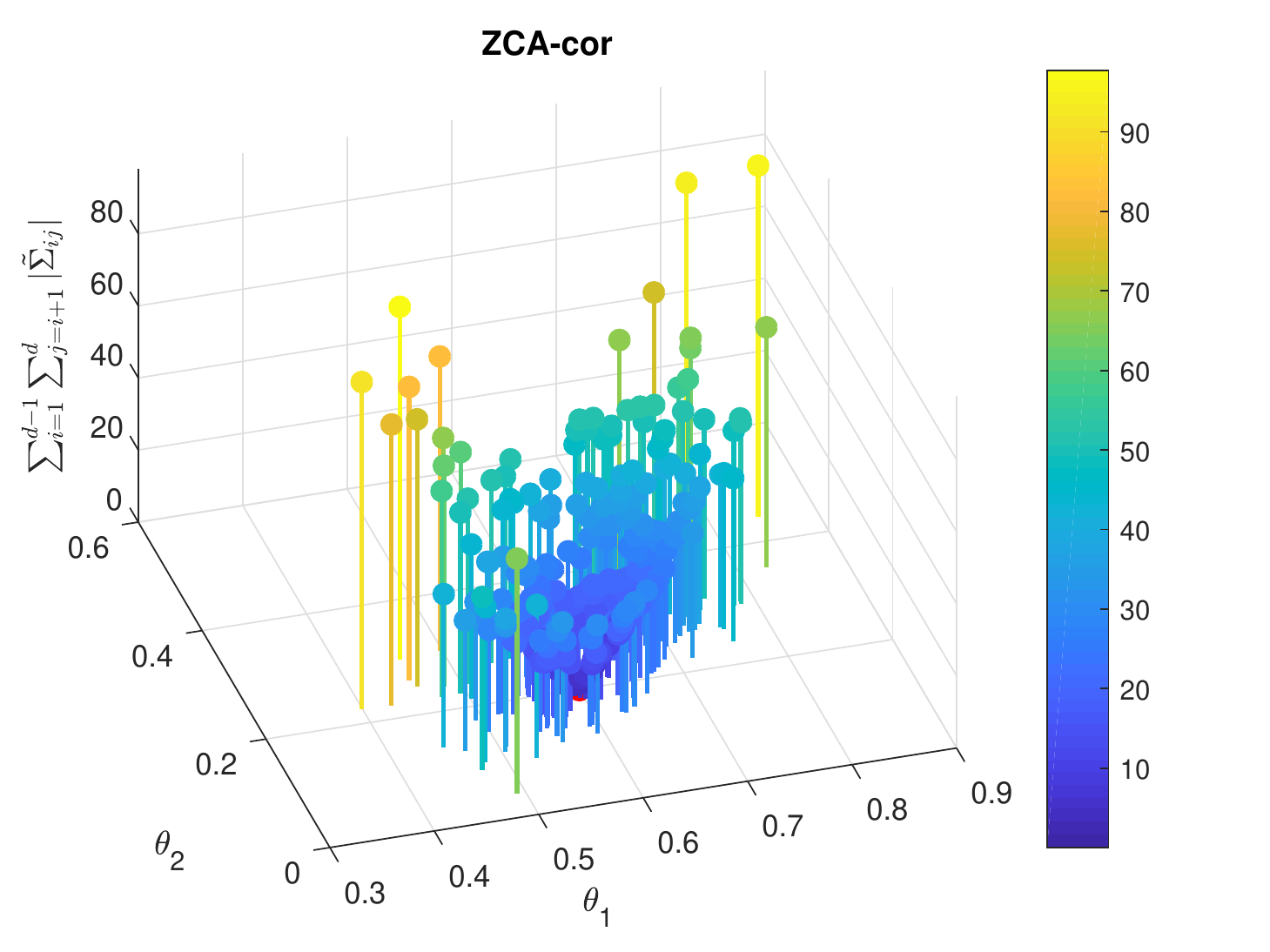}
\end{subfigure}
\caption{\small 
$L_1$ matrix norm deviation of the upper-triangular (excluding diagonals) elements of $\tilde{\boldsymbol{\Sigma}}(\boldsymbol{\theta})$ from the zero matrix $\boldsymbol{0}_{d\times d}$, for each of the whitening methods, under the MA(2) model.
Values of $\boldsymbol{\theta}$ are drawn from the true posterior distribution.
%Covariance deviation (without variances) from the zero matrix for whitened summary statistics over parameter values sampled from the MCMC posterior for the MA$(2)$ example. Deviation is measured using the $L_1$ matrix norm. 
Bar height and colour indicate the magnitude of the deviation. The true parameter value $\boldsymbol{\theta}_{\text{true}} = (\theta_{1,\text{true}},\theta_{2,\text{true}})^\top = (0.6,0.2)^\top$ is shown by the red dot (where visible).}
\label{fig:choiceofwhiteningsamplesMA}
\end{figure}

As might be expected, for both the MA$(2)$ and AR$(1)$ models, as $\boldsymbol{\theta}$ moves further away from $\boldsymbol{\theta}_{\text{true}}$, the deviation of $\tilde{\boldsymbol{\Sigma}}(\boldsymbol{\theta})$ from the identity matrix increases for each whitening method (see Appendix C). Here, PCA-based whitening has slightly lower deviations than the other whitening methods.
However, 
the differences between the different whitening methods become much clearer
when considering the off-diagonal deviations only (see Figure \ref{fig:choiceofwhiteningsamplesMA} and Appendix C).
% there is a much clearer discrepancy between PCA-based whitening and the other methods. 
%
Relative to the other whitening transformations, for PCA-based whitening, the covariance deviation (excluding variances) does not increase as rapidly as $\boldsymbol{\theta}$ moves  away from $\boldsymbol{\theta}_{\text{true}}$. This suggests that PCA-based whitening should be the most effective at decorrelating summary statistics 
%for parameter values that reside in regions with non-negligible posterior support
within the BSL algorithm, and that PCA and PCA-cor whitening in wBSL should provide posterior approximations closest to standard BSL. This is aligned with the results in Section \ref{sec:examples}.

The results also demonstrate why coupling the Warton shrinkage with the whitening (particularly the PCA-based whitening) is so effective; in the Warton shrinkage estimator, the variances are always re-estimated from the model simulations while only the correlations are shrunk.  Therefore it is only necessary for the whitening transformation to generate covariance matrices close to diagonal away from the point estimate, rather than being close to the identity, a stronger requirement.

The same conclusions can also be drawn based on the $L_1$ norm of the off diagonal elements of $\tilde{\boldsymbol{\Sigma}}(\boldsymbol{\theta})$ when $\boldsymbol{\theta}$ is taken over the entire parameter space (see Appendix C). 
It is clear that correlations between the transformed summary statistics are far less sensitive to changing $\boldsymbol{\theta}$ for PCA-based whitening compared to ZCA-based or Cholesky whitening. For PCA-based whitening the deviation surface is almost flat over $\boldsymbol{\theta}$, in contrast to the clear bowl-shaped surface for the other whitening methods.

%Here, it is very clear that correlations between transformed statistics are far less sensitive for PCA and PCA-cor whitening compared to ZCA, ZCA-cor and Cholesky whitening. For PCA and PCA-cor the surface is almost flat, even on the boundaries, compared to the other methods where the surface is roughly bowl shaped.\\ 

%%%%%%%%%%%%%%%%%%%%%%%%%%%%%%%%%%%%%%%%%%%
%%%%%%%%%%%%%%%%%%%%%%%%%%%%%%%%%%%%%%%%%%%
\subsection{Sensitivity to the value of $\boldsymbol{\theta}^0$}
%%%%%%%%%%%%%%%%%%%%%%%%%%%%%%%%%%%%%%%%%%%
%%%%%%%%%%%%%%%%%%%%%%%%%%%%%%%%%%%%%%%%%%%

A necessary step in implementing wBSL is estimation of the whitening matrix $\boldsymbol{W}$ before implementing the Monte Carlo sampler. This requires specification of two quantities: a parameter vector $\boldsymbol{\theta}^0$ believed to lie in region of high posterior probability, under which the summary statistics $\boldsymbol{s}_{1:n_{\text{cov}}}$ are generated and $\boldsymbol{\Sigma}$ is estimated, and the number of these statistics $n_{\text{cov}}$. Because  $\boldsymbol{W}$ is only estimated once at the start of the wBSL algorithm, it will take up a small fraction of the overall computational budget, and so $n_{\text{cov}}$ can be sufficiently large to estimate $\boldsymbol{\Sigma}$ well. In the analyses in Section \ref{sec:examples} we used $n_{\text{cov}}=20\,000$ for the first four examples.

Once $\boldsymbol{W}$ has been estimated then $\boldsymbol{\theta}^0$ is a natural candidate from which to initialise the MCMC sampler (or other Monte Carlo algorithm wBSL variant). In the examples in Section \ref{sec:examples} we used the true parameter value from which the observed dataset was generated $\boldsymbol{\theta}^0=\boldsymbol{\theta}_{\text{true}}$, however in practice little information about the true posterior may be available. 
Within the ABC literature a pilot analysis is commonly used to identify the region of high posterior density before performing a full analysis \citep{fearnhead2012constructing,fan+ns13} and similar ideas could be adopted here. However there would still be uncertainty regarding the best choice of $\boldsymbol{\theta}^0$ within this region.
Accordingly interest is in understanding the sensitivity of the wBSL posterior approximation to the choice for $\boldsymbol{\theta}^0$. Here we re-examine the MA(2) and AR(1) models and focus on PCA whitening as the best performing whitening procedure. The results for the other whitening methods are provided in Appendix C.

%before running the main loop of MCMC. This requires specification of a point estimate to perform $n_{\text{cov}}$ model simulations at, then computation of the whitening matrix, $\boldsymbol{W}$. This point estimate would then be used to initialise the MCMC chain. In practice, little information surrounding the true posterior may be available, so it is of interest to investigate the sensitivity of the wBSL posterior to the initial point estimate. Here we reconsider the MA$(2)$ and AR$(1)$ examples as candidates for such experiments. The results for PCA whitening are shown in this section, with the results for the other whitening types provided in Appendix C.\\

For the MA$(2)$ process, the parameters $\theta_1$ and $\theta_2$ are subject to the constraints $\theta_1 + \theta_2>-1$, $\theta_1 - \theta_2 < 1$ and $-1<\theta_2<1$. We choose five  different initial parameter configurations. The first three are the vertices of these boundaries: $\boldsymbol{\theta}^0 = (-2,1)^\top$, $(0,-1)^\top$ and $(2,1)^\top$, which are likely the worst possible choices of the point estimate. The other two values are $\boldsymbol{\theta}^0 = (c_1,c_2)^{\top}$ and $(c_3,c_4)^{\top}$ such that $0.75=p(\theta_1|\boldsymbol{s}_y < c_1) = p(\theta_2|\boldsymbol{s}_y<c_2)$ and $0.99=p(\theta_1|\boldsymbol{s}_y < c_3) = p(\theta_2|\boldsymbol{s}_y<c_4)$. 
%
%\textcolor{red}{[Unclear - there are 2 parameters. What quantiles are you referring to?]}
 We draw samples from the wBSL posterior approximation  using $T = 100\,000$ iterations, with a  burn-in of $1000$ iterations, and $n_{\text{cov}} = 20\,000$ model simulations.
%to estimate the whitening matrix, and the same priors as in Section 6.1. A burn-in of $1000$ iterations is discarded for each run. 

\begin{figure}[h!]
\centering
\begin{subfigure}
\centering\includegraphics[width = 11cm]{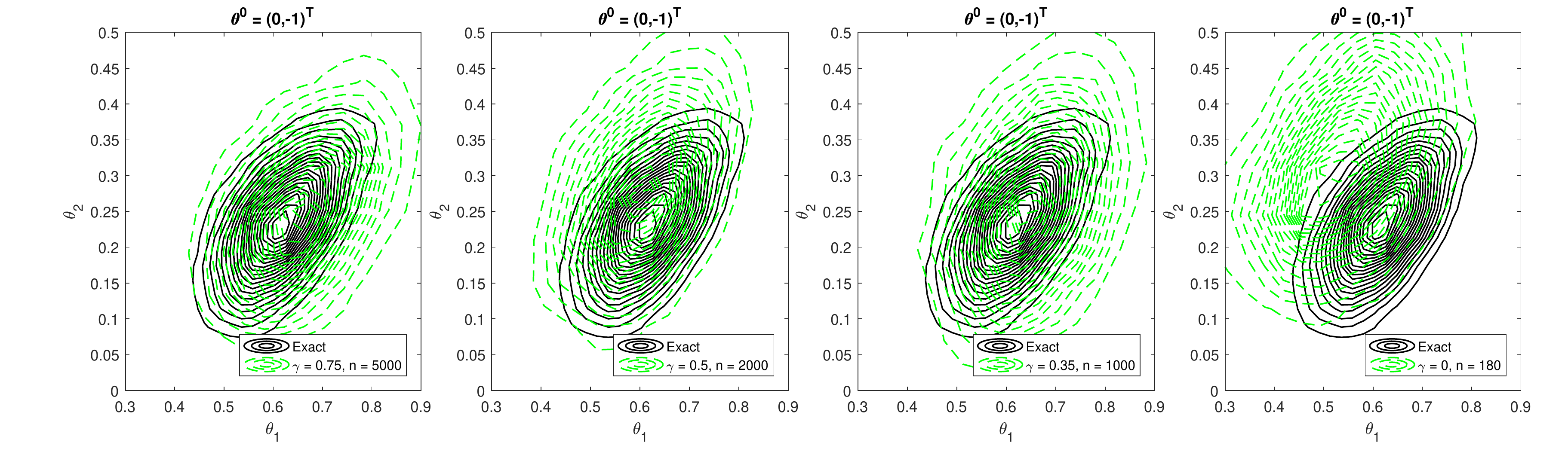}
\end{subfigure}

\begin{subfigure}
\centering\includegraphics[width = 11cm]{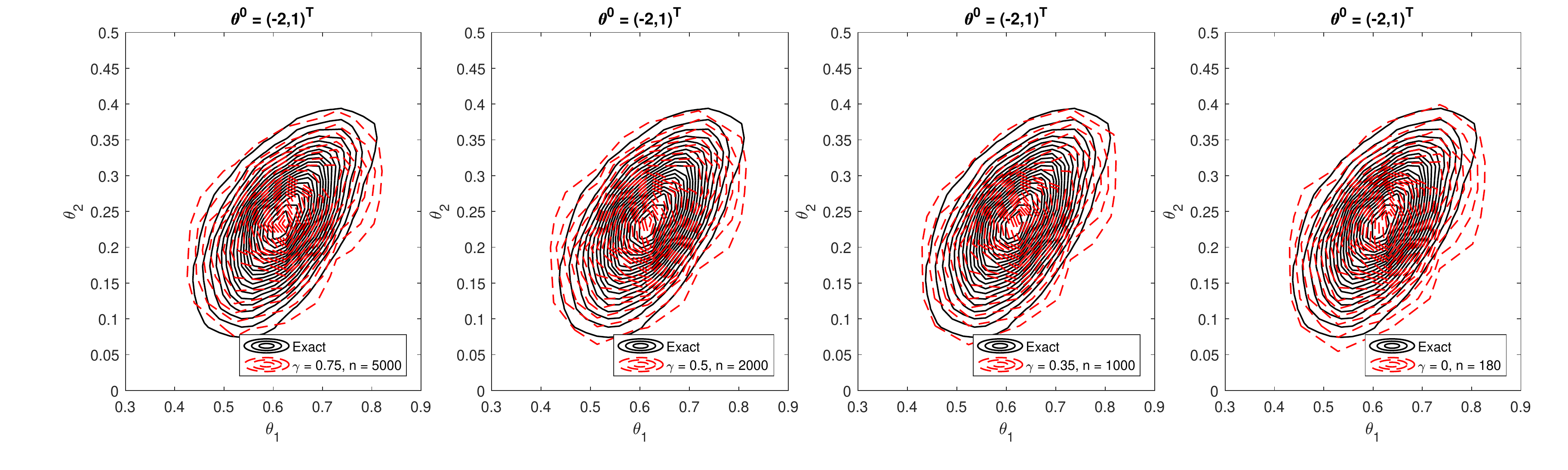}
\end{subfigure}

\begin{subfigure}
\centering\includegraphics[width = 11cm]{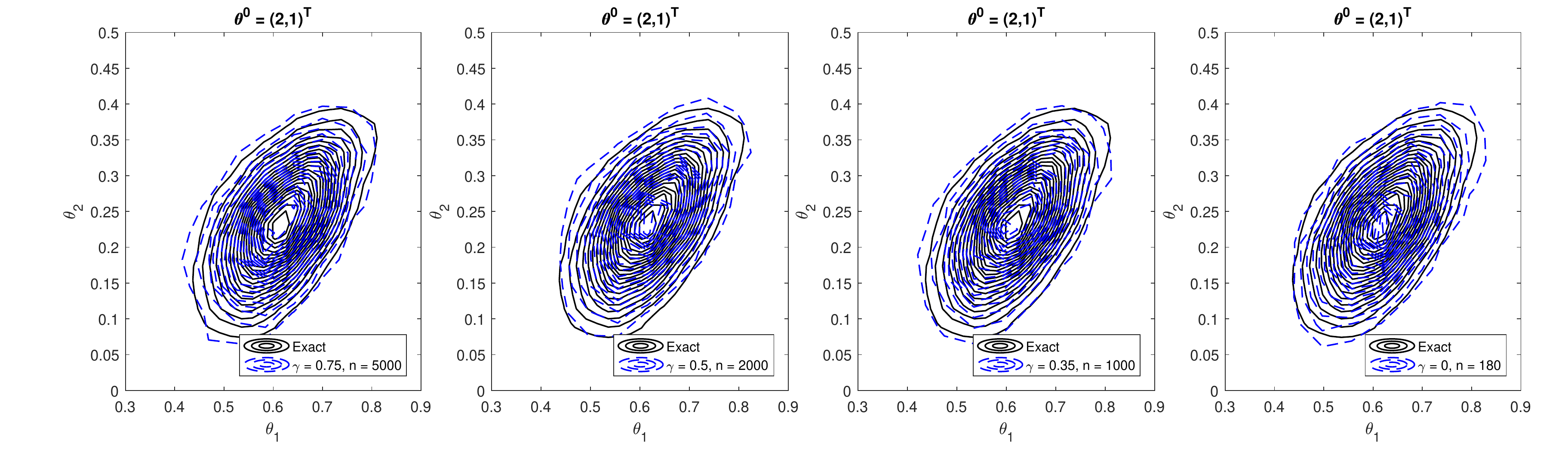}
\end{subfigure}

\begin{subfigure}
\centering\includegraphics[width = 11cm]{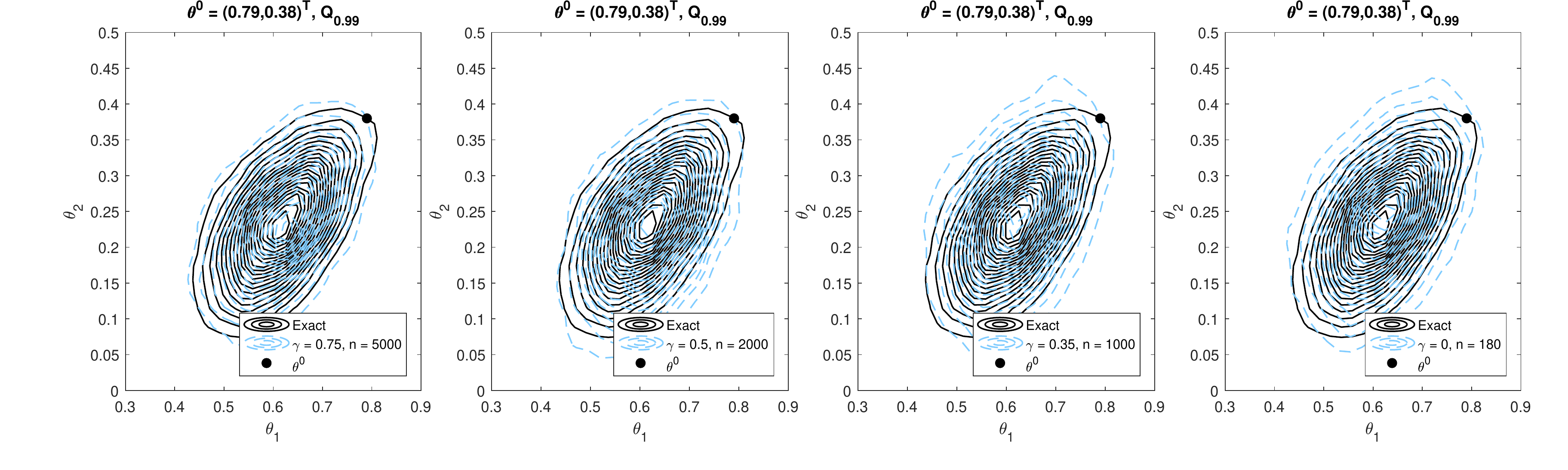}
\end{subfigure}

\begin{subfigure}
\centering\includegraphics[width = 11cm]{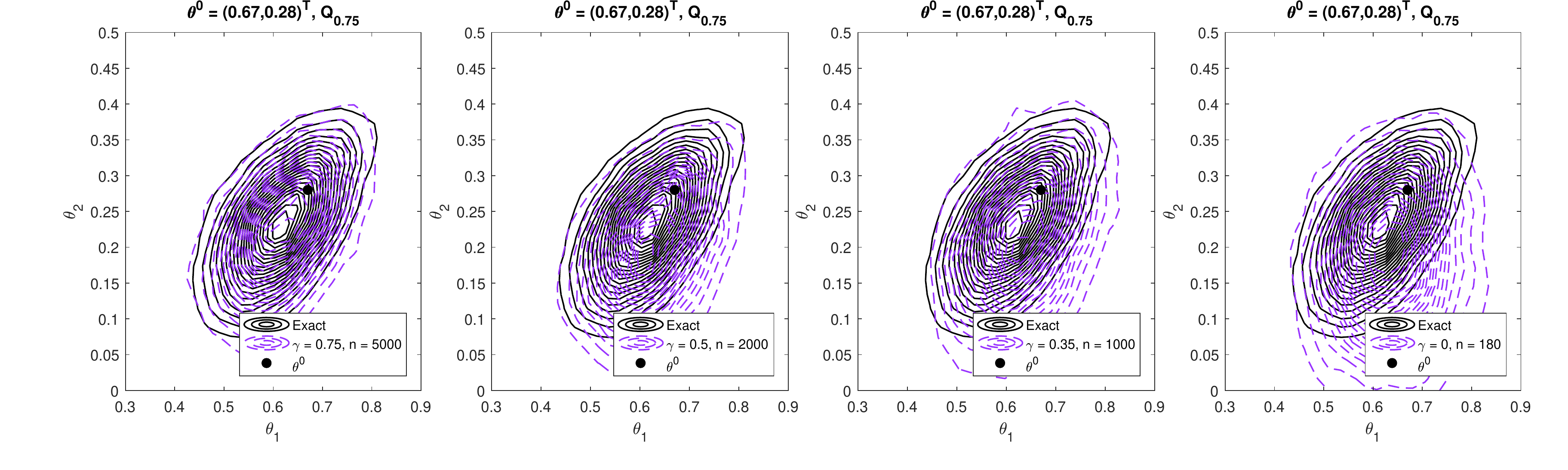}
\end{subfigure}

\caption{\small
Sensitivity of the wBSL posterior approximation with PCA whitening (dashed lines) to the point estimate for $\boldsymbol{\theta}^0$ for the MA(2) model. Rows correspond to different point estimates for $\boldsymbol{\theta}^0$. Columns correspond to different $\gamma,n$ combinations, with increasing levels of shrinkage (decreasing $\gamma$ and $n$) from left to right.
Estimated true posterior is shown by solid lines, $\boldsymbol{\theta}^0$ is illustrated by a dot.
}
\label{fig:sensMA2}
\end{figure}

The resulting posterior approximations are illustrated in Figure \ref{fig:sensMA2}, which demonstrate that the wBSL posterior has some robustness to the value of $\boldsymbol{\theta}^0$. 
When the point estimate was chosen on the boundary, there are mixed results. For $\boldsymbol{\theta}^0 = (0,-1)^\top$ (top row), the posterior variances are inflated, and more so for greater shrinkage (lower $\gamma,n$). 
However, when $\boldsymbol{\theta}^0 = (-2,1)^\top$ or $\boldsymbol{\theta}^0 = (2,1)^\top$ (rows 2 and 3), the posterior is recovered with high accuracy.  Interestingly, the wBSL posterior initialised at the ${0.99}$ true posterior quantiles (row 4) performs better than the posterior initialised at the ${0.75}$ quantiles (row 5), yet both appear less accurate than when initialising at $\boldsymbol{\theta}^0 = (-2,1)^\top$ or $\boldsymbol{\theta}^0 = (2,1)^\top$ (rows 2 and 3), which are on the boundary of the parameter support. We would expect point estimates with high posterior support to perform better overall, since the whitening transform would likely be more effective at decorrelating the simulated summary statistics over the region of the parameter space closer to $\boldsymbol{\theta}^0$.
%with non-negligible posterior support. 
The empirical results support this to some extent. % \textcolor{red}{[This last sentence is a generous conclusion.]}

% \textcolor{red}{[Are you sure that the $Q_{0.75}$ results are correct here? Worth checking? As this slightly undermines your arguments.]}

Details of the AR$(1)$ example are given in Appendix C. We observe similar results to the MA$(2)$ example. When the point estimate is close to regions of high posterior density the wBSL posterior approximation is accurate for all levels of shrinkage. When the point estimate is very far from the high posterior density region, then the wBSL approximation becomes poorer.

For both the MA$(2)$ and AR$(1)$ analyses, PCA-cor whitening wBSL performed similarly to PCA whitening under the same settings, whereas for ZCA-based whitening and Cholesky whitening there was a greater sensitivity to the initial point estimate (see Appendix C).
%(see Appendix). ZCA, ZCA-cor and Cholesky whitening all showed a much greater sensitivity to the initial point estimate (see Appendix).

%%%%%%%%%%%%%%%%%%%%%%%%%%%%%%%%%%%%%%%%%%%
%%%%%%%%%%%%%%%%%%%%%%%%%%%%%%%%%%%%%%%%%%%
\section{Discussion}
\label{sec:discussion}
%%%%%%%%%%%%%%%%%%%%%%%%%%%%%%%%%%%%%%%%%%%
%%%%%%%%%%%%%%%%%%%%%%%%%%%%%%%%%%%%%%%%%%%

%The main contribution of this paper is the integration of whitening transformations within Bayesian synthetic likelihood. 
In this article we have examined the integration of whitening transformations within the Bayesian synthetic likelihood framework.
In combination with shrinkage covariance estimation, the number of model simulations required to estimate the synthetic likelihood function can be drastically reduced compared to standard BSL methods, enabling the efficient implementation of BSL with high dimensional and highly correlated summary statistics. 
In particular, we obtained orders of magnitude computational gains over standard BSL in all analyses considered, with little detrimental impact on accuracy.

%% SCOTT: I took the below sentence out as you can just apply "whitening" to decorrelate ABC summary statistics. And in fact people have been doing this for many years E.g. Luciani et al (2009) is just one example (and we now mention that in this paper).
%This reduces the necessity of finding low-dimensional summary statistics that are informative about model parameters, as is required for competing methods such as ABC. \\

We examined five different whitening transformations: PCA, PCA-cor, ZCA, ZCA-cor and Choleksy whitening, on both simplified  and real-world models. In all cases, we empirically demonstrated that PCA-based whitening outperformed the other whitening methods, and produced transformations that were less sensitive to changes in the model parameter $\boldsymbol{\theta}$  in the sense that the covariance of the transformed summary statistics was closer to being diagonal. 
%
%. It was found that the PCA and PCA-cor transformations are more accurate over a range of parameter values than the other methods, in the sense that the covariance of the transformed summary statistics is closer to a diagonal matrix. 
%We attribute this to the respective sensitivities of the different whitening matrices to the changing covariance matrix over a range of parameter values. 
%It was also found in a number of examples 
We found that the PCA-based wBSL posterior approximation was fairly robust to the initial parameter point estimate $\boldsymbol{\theta}^0$ used to compute the whitening matrix $\boldsymbol{W}$. Although there was some variability in our results for the MA(2) model, we suggest using initial parameter estimates approximately located in regions of high posterior density in order to produce more accurate posterior approximations.
%However, the results did show some variability, with c
%While there was some variability, ases where 
%points further away from the true posterior produced more accurate inferences than points with higher posterior %support. 
Overall, since PCA and PCA-cor whitening produced similar results, we recommend PCA whitening as the standard choice of whitening for wBSL, since it is slightly more computationally efficient to implement.

In practice, we recommend that the user choose an appropriate number of model simulations ($n$) given their computational budget and then tune the corresponding shrinkage level ($\gamma$). Following the recommendation of \cite{price2018bayesian}, we suggest tuning the shrinkage level so that the estimated log synthetic likelihood has a standard deviation between 1 and 2. This should produce a good trade-off between computational and statistical efficiency. Of course, wBSL can produce more accurate results with more model simulations (less shrinkage). %However, for the examples considered in this paper, at least an order of magnitude reduction in the number of simulations was achieved with little detrimental effect on the accuracy relative to standard BSL.\\

It would be of future interest to investigate the applicability of whitening transformations in various extensions of BSL.  For example, \cite{an2018robust} develop a semi-parametric synthetic likelihood, which is more robust to departures from normality.  Further, \cite{frazier2019robust} develop synthetic likelihood methods that are more robust when there is incompatibility between the model and observed summary statistic.  An alternative extension could involve a method for automatically finding a whitening transformation that minimises the loss of accuracy compared to standard BSL.

%%%%%%%%%%%%%%%%%%%%%%%%%%%%%%%%%%%%%%%%%%%
%%%%%%%%%%%%%%%%%%%%%%%%%%%%%%%%%%%%%%%%%%%
\section*{Acknowledgements}
%%%%%%%%%%%%%%%%%%%%%%%%%%%%%%%%%%%%%%%%%%%
%%%%%%%%%%%%%%%%%%%%%%%%%%%%%%%%%%%%%%%%%%%

The authors would like to thank Ziwen An for providing some code for the toad  and the collective cell spreading models, and the High Performance Computing group at QUT for their computational resources. JWP was supported by a QUT Master of Philosophy Award. SAS was supported by the Australian Research Council under the Future Fellowship scheme (FT170100079). CD and SAS are also supported by the
ARC Centre of Excellence in Mathematical and Statistical Frontiers (ACEMS; CE140100049). CD is grateful to ACEMS for providing funding to visit SAS at UNSW where discussions on this research took place. We thank Ian Turner for some discussions on relevant linear algebra.
% grateful to ACEMS for providing funding to visit UNSW, where some discussions for this research took place.

\bibliographystyle{apalike}
\bibliography{Bibliography-MM-MC}

\begin{thebibliography}{}

\bibitem[An et~al., 2019a]{an2018robust}
An, Z., Nott, D.~J., and Drovandi, C. (2019a).
\newblock Robust {B}ayesian synthetic likelihood via a semi-parametric
  approach.
\newblock {\em Statistics and Computing}.

\bibitem[An et~al., 2019b]{an2019bsl}
An, Z., South, L.~F., and Drovandi, C. (2019b).
\newblock {BSL}: An {R} package for efficient parameter estimation for
  simulation-based models via {B}ayesian synthetic likelihood.
\newblock {\em arXiv preprint arXiv:1907.10940}.

\bibitem[An et~al., 2019c]{an2019accelerating}
An, Z., South, L.~F., Nott, D.~J., and Drovandi, C.~C. (2019c).
\newblock Accelerating {B}ayesian synthetic likelihood with the graphical
  lasso.
\newblock {\em Journal of Computational and Graphical Statistics},
  28(2):471--475.

\bibitem[Bacus, 1976]{bacus1976whitening}
Bacus, J.~W. (1976).
\newblock A whitening transformation for two-color blood cell images.
\newblock {\em Pattern Recognition}, 8(1):53--60.

\bibitem[Baddeley et~al., 2007]{baddeley2007spatial}
Baddeley, A., B{\'a}r{\'a}ny, I., and Schneider, R. (2007).
\newblock Spatial point processes and their applications.
\newblock {\em Stochastic Geometry: Lectures Given at the CIME Summer School
  Held in Martina Franca, Italy, September 13--18, 2004}, pages 1--75.

\bibitem[Binny et~al., 2016]{Binny2016}
Binny, R.~N., James, A., and Plank, M.~J. (2016).
\newblock Collective cell behaviour with neighbour-dependent proliferation,
  death and directional bias.
\newblock {\em Bulletin of Mathematical Biology}, 78(11):2277--2301.

\bibitem[Bishop, 2006]{bishop2006pattern}
Bishop, C.~M. (2006).
\newblock {\em Pattern recognition and machine learning}.
\newblock Springer.

\bibitem[Blum, 2010]{blum10}
Blum, M. G.~B. (2010).
\newblock Approximate {Bayesian} computation: a non-parametric perspective.
\newblock {\em Journal of the American Statistical Association}, 105:1178 --
  1187.

\bibitem[Browning et~al., 2018]{browning2018inferring}
Browning, A.~P., McCue, S.~W., Binny, R.~N., Plank, M.~J., Shah, E.~T., and
  Simpson, M.~J. (2018).
\newblock Inferring parameters for a lattice-free model of cell migration and
  proliferation using experimental data.
\newblock {\em Journal of Theoretical Biology}, 437:251--260.

\bibitem[Doucet et~al., 2015]{doucet+pdk15}
Doucet, A., Pitt, M., Deligiannidis, M.~K., and Kohn, R. (2015).
\newblock Efficient implementation of {Markov chain Monte Carlo} when using an
  unbiased likelihood estimator.
\newblock {\em Biometrika}, 102:295--313.

\bibitem[Everitt, 2017]{everitt2017bootstrapped}
Everitt, R.~G. (2017).
\newblock Bootstrapped synthetic likelihood.
\newblock {\em arXiv preprint arXiv:1711.05825}.

\bibitem[Fan et~al., 2013]{fan+ns13}
Fan, Y., Nott, D.~J., and Sisson, S.~A. (2013).
\newblock Approximate {B}ayesian computation via regression density estimation.
\newblock {\em Stat}, 2(1):34--48.

\bibitem[Fearnhead and Prangle, 2012]{fearnhead2012constructing}
Fearnhead, P. and Prangle, D. (2012).
\newblock Constructing summary statistics for approximate bayesian computation:
  semi-automatic approximate bayesian computation.
\newblock {\em Journal of the Royal Statistical Society: Series B (Statistical
  Methodology)}, 74(3):419--474.

\bibitem[Frazier and Drovandi, 2019]{frazier2019robust}
Frazier, D.~T. and Drovandi, C. (2019).
\newblock Robust approximate bayesian inference with synthetic likelihood.
\newblock {\em arXiv preprint arXiv:1904.04551}.

\bibitem[Frazier et~al., 2019]{nott2019bayesian}
Frazier, D.~T., Nott, D.~J., Drovandi, C., and Kohn, R. (2019).
\newblock Bayesian inference using synthetic likelihood: asymptotics and
  adjustments.
\newblock {\em arXiv preprint arXiv:1902.04827}.

\bibitem[Friedl and Wolf, 2003]{friedl2003tumour}
Friedl, P. and Wolf, K. (2003).
\newblock Tumour-cell invasion and migration: diversity and escape mechanisms.
\newblock {\em Nature Reviews Cancer}, 3(5):362.

\bibitem[Fujikoshi et~al., 2011]{fujikoshi2011multivariate}
Fujikoshi, Y., Ulyanov, V.~V., and Shimizu, R. (2011).
\newblock {\em Multivariate statistics: High-dimensional and large-sample
  approximations}, volume 760.
\newblock John Wiley \& Sons.

\bibitem[Gutmann and Corander, 2016]{gutmann2016bayesian}
Gutmann, M.~U. and Corander, J. (2016).
\newblock Bayesian optimization for likelihood-free inference of
  simulator-based statistical models.
\newblock {\em The Journal of Machine Learning Research}, 17(1):4256--4302.

\bibitem[Johnston et~al., 2014]{johnston2014interpreting}
Johnston, S.~T., Simpson, M.~J., McElwain, D.~S., Binder, B.~J., and Ross,
  J.~V. (2014).
\newblock Interpreting scratch assays using pair density dynamics and
  approximate {B}ayesian computation.
\newblock {\em Open Biology}, 4(9):140097.

\bibitem[Kessy et~al., 2018]{kessy2018optimal}
Kessy, A., Lewin, A., and Strimmer, K. (2018).
\newblock Optimal whitening and decorrelation.
\newblock {\em The American Statistician}, 72(4):309--314.

\bibitem[Lindstrom et~al., 2013]{lindstrom+bsps13}
Lindstrom, T., Brown, G.~P., Sisson, S.~A., Phillips, B.~L., and Shine, R.
  (2013).
\newblock Rapid shifts in dispersal behaviour on an expanding range edge.
\newblock {\em Proc. Natl. Acad. Sci. USA}, 110:13452--13456.

\bibitem[Luciani et~al., 2009]{luciani+sjft09}
Luciani, F., Sisson, S.~A., Jiang, H., Francis, A.~R., and Tanaka, M.~M.
  (2009).
\newblock The epidemiological fitness cost of drug resistance in
  {Mycobacterium} tuberculosis.
\newblock {\em Proceedings of the National Academy of the Sciences of the USA},
  106:14711--14715.

\bibitem[Marchand et~al., 2017]{marchand2017stochastic}
Marchand, P., Boenke, M., and Green, D.~M. (2017).
\newblock A stochastic movement model reproduces patterns of site fidelity and
  long-distance dispersal in a population of {F}owler's toads ({A}naxyrus
  fowleri).
\newblock {\em Ecological Modelling}, 360:63--69.

\bibitem[Nott et~al., 2018]{nott+ofs17}
Nott, D.~J., Ong, V. M.-H., Fan, Y., and Sisson, S.~A. (2018).
\newblock High-dimensional approximate {Bayesian} computation.
\newblock In Sisson, S.~A., Fan, Y., and Beaumont, M.~A., editors, {\em
  Handbook of Approximate Bayesian Computation}, pages 211--241.

\bibitem[Ong et~al., 2018a]{ong2018likelihood}
Ong, V.~M., Nott, D.~J., Tran, M.-N., Sisson, S.~A., and Drovandi, C.~C.
  (2018a).
\newblock Likelihood-free inference in high dimensions with synthetic
  likelihood.
\newblock {\em Computational Statistics \& Data Analysis}, 128:271--291.

\bibitem[Ong et~al., 2018b]{ong+ntsd18}
Ong, V. M.-H., Nott, D.~J., Tran, M.-N., Sisson, S.~A., and Drovandi, C.~C.
  (2018b).
\newblock Variational {Bayes} with synthetic likelihood.
\newblock {\em Statistics and Computing}, 28:971--988.

\bibitem[Prangle, 2018]{prangle2018summary}
Prangle, D. (2018).
\newblock Summary statistics.
\newblock In {\em Handbook of Approximate Bayesian Computation}, pages
  125--152. Chapman and Hall/CRC.

\bibitem[Price et~al., 2018]{price2018bayesian}
Price, L.~F., Drovandi, C.~C., Lee, A., and Nott, D.~J. (2018).
\newblock Bayesian synthetic likelihood.
\newblock {\em Journal of Computational and Graphical Statistics}, 27(1):1--11.

\bibitem[Shaw and Martin, 2009]{shaw2009wound}
Shaw, T.~J. and Martin, P. (2009).
\newblock Wound repair at a glance.
\newblock {\em Journal of {C}ell {S}cience}, 122(18):3209--3213.

\bibitem[Sisson et~al., 2018a]{sisson2018handbook}
Sisson, S.~A., Fan, Y., and Beaumont, M.~A. (2018a).
\newblock {\em Handbook of Approximate Bayesian Computation}.
\newblock Chapman and Hall.

\bibitem[Sisson et~al., 2018b]{sisson+f18}
Sisson, S.~A., Fan, Y., and Beaumont, M.~A. (2018b).
\newblock Overview of approximate {Bayesian} computation.
\newblock In Sisson, S.~A., Fan, Y., and Beaumont, M.~A., editors, {\em
  Handbook of Approximate Bayesian Computation}, pages 3--54. Chapman and
  Hall/CRC Press.

\bibitem[Vo et~al., 2015]{vo2015quantifying}
Vo, B.~N., Drovandi, C.~C., Pettitt, A.~N., and Simpson, M.~J. (2015).
\newblock Quantifying uncertainty in parameter estimates for stochastic models
  of collective cell spreading using approximate {B}ayesian computation.
\newblock {\em Mathematical {B}iosciences}, 263:133--142.

\bibitem[Warton, 2008]{warton2008penalized}
Warton, D.~I. (2008).
\newblock Penalized normal likelihood and ridge regularization of correlation
  and covariance matrices.
\newblock {\em Journal of the American Statistical Association},
  103(481):340--349.

\bibitem[Wood, 2010]{wood2010statistical}
Wood, S.~N. (2010).
\newblock Statistical inference for noisy nonlinear ecological dynamic systems.
\newblock {\em Nature}, 466(7310):1102.

\end{thebibliography}

\appendix
\section{Proof of Theorem 1}
For completeness, we first restate Assumption 1, Assumption 2 and Theorem 1. Let $\boldsymbol{s}_{\boldsymbol{y}}\in\mathbb{R}^{d}$ denote the observed summary statistic. Consider the standard BSL likelihood with  $\boldsymbol{\mu}_n(\boldsymbol{\theta})$ and $\boldsymbol{\Sigma}_n(\boldsymbol{\theta})$ denoting the sample mean and variance of the simulated summary statistics. To deduce the behavior of the variance we consider the following assumption. 
\begin{assumption}\label{ass:one}
For $i=1,\dots,n$, the simulated summaries $\boldsymbol{s}_i(\boldsymbol{\theta})$ are generated i.i.d.\ and satisfy	
$$\boldsymbol{s}_i(\boldsymbol{\theta})\sim_{}\mathcal{N}\left(\boldsymbol{\mu}(\boldsymbol{\theta}),\boldsymbol{\Sigma}(\boldsymbol{\theta})\right),$$ where $\sup_{\boldsymbol{\theta}\in\boldsymbol{\Theta}}\|\boldsymbol{\mu}(\boldsymbol{\theta})\|<\infty$ and $\boldsymbol{\Sigma}(\boldsymbol{\theta})$ is positive-definite for all $\boldsymbol{\theta}\in\boldsymbol{\Theta}$.	
\end{assumption}

In addition, consider a version of BSL that uses simulated summaries with a diagonal covariance structure, i.e., summaries that are uncorrelated across the $d$ dimensions. Denote these simulated summaries as $\boldsymbol{\zeta}_i(\boldsymbol{\theta})$. We maintain the following assumption about these simulated summaries. 
\begin{assumption}\label{ass:two}
For $i=1,\dots, n$, the simulated summaries $\boldsymbol{\zeta}_i(\boldsymbol{\theta})$ are generated i.i.d.\ and satisfy	
$$
\boldsymbol{\zeta}_i(\boldsymbol{\theta})\sim \mathcal{N}\left(\boldsymbol{\zeta}(\boldsymbol{\theta}),\boldsymbol{\Omega}(\boldsymbol{\theta})\right),\text{ where } \boldsymbol{\Omega}(\boldsymbol{\theta}):=\mathrm{diag}(\omega_{11}(\boldsymbol{\theta}),\dots,\omega_{dd}(\boldsymbol{\theta})),
$$where $$\sup_{\boldsymbol{\theta}\in\boldsymbol{\Theta}}\|\boldsymbol{\zeta}(\boldsymbol{\theta})\|<\infty,\mbox{ and }\max_{j\leq d}\sup_{\boldsymbol{\theta}\in\boldsymbol{\Theta}}\omega_{jj}(\boldsymbol{\theta})<\infty,\;\min_{j\leq d}\inf_{\boldsymbol{\theta}\in\boldsymbol{\Theta}}\omega_{jj}(\boldsymbol{\theta})>0.$$ 	
\end{assumption}
Let $\boldsymbol{\zeta}_n(\boldsymbol{\theta})$ denote the sample mean of these uncorrelated summary statistics and let $\boldsymbol{\Omega}_{n}(\boldsymbol{\theta})$ denote the sample variance. Throughout the remainder we explicitly consider that $\boldsymbol{\Omega}_{n}(\boldsymbol{\theta})$ is a diagonal matrix. 

Denote the BSL likelihood based on the uncorrelated summaries as $$p_{A,n,w}\left(\boldsymbol{s}_{\boldsymbol{y}}|\boldsymbol{\theta}\right)=\mathcal{N}(\boldsymbol{s}_{\boldsymbol{y}}|\boldsymbol{\zeta}_n(\boldsymbol{\theta}),\boldsymbol{\Omega}_{n}(\boldsymbol{\theta})),$$ and recall the standard BSL likelihood is given by:
$$p_{A,n}\left(\boldsymbol{s}_{\boldsymbol{y}}|\boldsymbol{\theta}\right)=\mathcal{N}(\boldsymbol{s}_{\boldsymbol{y}}|\boldsymbol{\mu}_n(\boldsymbol{\theta}),\boldsymbol{\Sigma}_{n}(\boldsymbol{\theta})).$$

For two real sequences $a_n,b_n$, we say that $b_n=\mathcal{O}(a_n)$ if the sequence $|b_n/a_n|$ is bounded.

\begin{theorem} For $d$ and $n$ large, but finite, with $n>d+4$, under Assumption \ref{ass:one},
\begin{flalign*}
\text{\em Var}\left[\log\left(p_{A,n}\left(\boldsymbol{s}_{\boldsymbol{y}}|\boldsymbol{\theta}\right) 
\right)\right]&=\mathcal{O}\left(\frac{d^2\cdot n^2}{(n-d)^3}\right),
\end{flalign*}however, under Assumption \ref{ass:two},
\begin{flalign*}
\text{\em Var}\left[\log\left(p_{A,n,w}\left(\boldsymbol{s}_{\boldsymbol{y}}|\boldsymbol{\theta}\right) 
\right)\right]&=\mathcal{O}\left(\frac{d\cdot n^2}{(n-d)^3}\right).
\end{flalign*}Therefore, for $d$ and $n$ large,
$$
\text{\em Var}\left[\log\left(p_{A,n,w}\left(\boldsymbol{s}_{\boldsymbol{y}}|\boldsymbol{\theta}\right) 
\right)\right] \leq \text{\em Var}\left[\log\left(p_{A,n}\left(\boldsymbol{s}_{\boldsymbol{y}}|\boldsymbol{\theta}\right) 
\right)\right].
$$
\end{theorem}

\begin{proof}
We deduce the order for standard BSL, $p_{A,n}$, and the diagonal counterpart, $p_{A,n,w}$ separately, starting with standard BSL.
\\

\noindent\textbf{Standard BSL}
\\

\noindent Decompose $-\log(p_{A,n}) $ as 
\begin{flalign*}
-\log\left(p_{A,n}\left(\boldsymbol{s}_{\boldsymbol{y}}|\boldsymbol{\theta}\right)
\right) &\propto\log\text{det}\left\{\boldsymbol{\Sigma}_n(\boldsymbol{\theta})\right\}+
\left(\boldsymbol{s}_{\boldsymbol{y}}-\boldsymbol{\mu}_n(\boldsymbol{\theta})\right)^\top\boldsymbol{\Sigma}_n^{-1}(\boldsymbol{\theta})\left(\boldsymbol{s}_{\boldsymbol{y}}-\boldsymbol{\mu}_n(\boldsymbol{\theta})\right)\\&:=X_{1n}+X_{2n},
\end{flalign*}
where the proportionality disregards terms that do not depend on $\boldsymbol{\theta}$. First, we deduce the variances of $X_{1n}$ and $X_{2n}$.  In what follows, we disregard the functions dependence on $\boldsymbol{\theta}$ to simplify notations.\footnote{We note that Assumptions \ref{ass:one}, \ref{ass:two} ensure that the ensuing derivations are valid uniformly in $\boldsymbol{\theta}$.}
\\

\noindent\textbf{(1) $X_{1n}$ term:} Consider the variance of $X_{1n}$. Let $\mathcal{W}(\boldsymbol{\Sigma},n)$ denote the Wishart distribution with $n$ degrees of freedom and scale matrix $\boldsymbol{\Sigma}$. Under Assumption \ref{ass:one}, for $W\sim\mathcal{W}(\boldsymbol{\Sigma},n-1)$,
\begin{flalign*}
\log\text{det}\left(\boldsymbol{\Sigma}_n\right)&=\log\text{det}\left\{\frac{W}{(n-1)}\right\}=-d\log(n-1)+\log\text{det}\left\{W\right\}.
\end{flalign*}
From \cite{bishop2006pattern}, for $W\sim \mathcal{W}(\boldsymbol{\Sigma},n),$
\begin{flalign}
\text{Var}\left[\log\text{det}\{W\}\right]&=\sum_{i=1}^{d}\psi_1\left(\frac{n-i}{2}\right),\label{eq:wishart_var}
\end{flalign}	where $\psi_1$ denotes the trigamma function. Applying equation \eqref{eq:wishart_var}, we have that 
\begin{equation}
\text{Var}\left[\log\text{det}\left\{\boldsymbol{\Sigma}_n\right\}\right]= \sum_{i=1}^{d}\psi_1\left(\frac{n-i}{2}\right)= \mathcal{O}(1),\label{eq:det_order}
\end{equation}where the latter follows from the asymptotic properties of the trigamma function. 
\\

\noindent\textbf{(2) $X_{2n}$ term:} Recalling that $X_{2n}:=\left(\boldsymbol{s}_{\boldsymbol{y}}-\boldsymbol{\mu}_n\right)^\top\boldsymbol{\Sigma}_n^{-1}\left(\boldsymbol{s}_{\boldsymbol{y}}-\boldsymbol{\mu}_n\right)$, we bound this term via 
$$
X_{2n}\leq 2\left(\boldsymbol{\mu}_n-\boldsymbol{\mu}\right)^\top\boldsymbol{\Sigma}_n^{-1}\left(\boldsymbol{\mu}_n-\boldsymbol{\mu}\right)+2(\boldsymbol{s}_{\boldsymbol{y}}-\boldsymbol{\mu})^\top\boldsymbol{\Sigma}_n^{-1}(\boldsymbol{s}_{\boldsymbol{y}}-\boldsymbol{\mu}).
$$Apply the Cauchy-Schwarz inequality to obtain
\begin{flalign*}
\text{Var}\left[X_{2n}\right]&\leq 4\text{Var}\left[\left(\boldsymbol{\mu}_n-\boldsymbol{\mu}\right)^\top\boldsymbol{\Sigma}_n^{-1}\left(\boldsymbol{\mu}_n-\boldsymbol{\mu}\right)\right]+4\text{Var}\left[(\boldsymbol{s}_{\boldsymbol{y}}-\boldsymbol{\mu})^\top\boldsymbol{\Sigma}_n^{-1}(\boldsymbol{s}_{\boldsymbol{y}}-\boldsymbol{\mu})\right]\\&+2\sqrt{\text{Var}\left[\left(\boldsymbol{\mu}_n-\boldsymbol{\mu}\right)^\top\boldsymbol{\Sigma}_n^{-1}\left(\boldsymbol{\mu}_n-\boldsymbol{\mu}\right)\right]}\times\sqrt{\text{Var}\left[(\boldsymbol{s}_{\boldsymbol{y}}-\boldsymbol{\mu})^\top\boldsymbol{\Sigma}_n^{-1}(\boldsymbol{s}_{\boldsymbol{y}}-\boldsymbol{\mu})\right]}.
\end{flalign*}

Rewrite the first term, $\left(\boldsymbol{\mu}_n-\boldsymbol{\mu}\right)^\top\boldsymbol{\Sigma}_n^{-1}\left(\boldsymbol{\mu}_n-\boldsymbol{\mu}\right)$, as 
$$
t^2:=\frac{1}{n}\left[\left(\boldsymbol{\mu}_n-\boldsymbol{\mu}\right)^\top\left[\boldsymbol{\Sigma}_n/n\right]^{-1}\left(\boldsymbol{\mu}_n-\boldsymbol{\mu}\right)\right].
$$
Under Assumption \ref{ass:one}, $$n\cdot t^2\sim \text{T}^2_{d,n-1},$$where $\text{T}^2_{d,n-1}$ denotes Hotelling's T-squared distribution with $d$ and $n-1$ degrees of freedom, and where we note that
$$
\text{T}^2_{d,n-1}\sim \frac{d(n-1)}{(n-d)}\mathcal{F}_{d,n-d},
$$
with $\mathcal{F}_{d,n-1}$ denoting the $\mathcal{F}$-distribution with $d$ numerator and $n-1$ denominator degrees of freedom, respectively. The variance of $t^2$ can now be obtained using the moments of the $\mathcal{F}$-distribution. To this end, for $n>d+4$,
\begin{flalign}
\text{Var}[t^2]&=\frac{1}{n^2}\left(\frac{d(n-1)}{n-d}\right)^2\text{Var}\left[\mathcal{F}_{d,n-d}\right]\nonumber\\&=\frac{1}{n^2}\left(\frac{d(n-1)}{n-d}\right)^2\frac{2(n-d)^2(n-2)}{d(n-d-2)^2(n-d-4)}\nonumber\\&=\frac{1}{n^2}\frac{2d(n - 1)^2(n - 2)}{(n-d - 2)^2(n-d - 4)}\nonumber\\&=\mathcal{O}\left( \frac{d\cdot n}{(n-d)^3}\right)\label{eq:t2_order}.
\end{flalign}

To deduce the variance of the second term, $(\boldsymbol{s}_{\boldsymbol{y}}-\boldsymbol{\mu})^\top\boldsymbol{\Sigma}_n^{-1}(\boldsymbol{s}_{\boldsymbol{y}}-\boldsymbol{\mu})$, we can use the following result for the inverse-Wishart distribution (see \citealp{fujikoshi2011multivariate}, Ch. 2): if $W\sim \mathcal{W}^{-1}(\boldsymbol{\Sigma}^{-1},n)$, it follows that, for a constant vector $\boldsymbol{z}\in\mathbb{R}^{d}$, with $\|\boldsymbol{z}\|^2<M$, and for $\|\boldsymbol{\Sigma}^{-1}\|<M$, 
\begin{flalign}
%\mathbb{E}\left[z'Wz\right]&=\frac{z'\Sigma^{-1}z}{n-d-1}\\
\text{Var}\left[\boldsymbol{z}^\top W \boldsymbol{z}\right]&=\sum_{i=1}^{d}\sum_{j=1}^{d}z_iz_j\frac{(n-d+1)[\boldsymbol{\Sigma}^{-1}]_{ij}^2+(n-d+1)[\boldsymbol{\Sigma}^{-1}]_{ii}[\boldsymbol{\Sigma}^{-1}]_{jj}}{(n-d)(n-d-1)^2(n-d-3)}=\mathcal{O}\left( \frac{d^2}{(n-d)^3}\right)\label{eq:inv_wishart} .
\end{flalign}
Now, rewrite the term  $(\boldsymbol{s}_{\boldsymbol{y}}-\boldsymbol{\mu})^\top\boldsymbol{\Sigma}_n^{-1}(\boldsymbol{s}_{\boldsymbol{y}}-\boldsymbol{\mu})$ as 
$$
(n-1)(\boldsymbol{s}_{\boldsymbol{y}}-\boldsymbol{\mu})^\top\left[(n-1)\boldsymbol{\Sigma}_n\right]^{-1}(\boldsymbol{s}_{\boldsymbol{y}}-\boldsymbol{\mu}).
$$ Use the fact that $[(n-1)\boldsymbol{\Sigma}_n]^{-1}\sim \mathcal{W}^{-1}(\boldsymbol{\Sigma}^{-1},n-1)$ and equation \eqref{eq:inv_wishart} to deduce that, for $\boldsymbol{z}:=(\boldsymbol{s}_{\boldsymbol{y}}-\boldsymbol{\mu})$,
\begin{flalign}
\text{Var}\left[(n-1)(\boldsymbol{s}_{\boldsymbol{y}}-\boldsymbol{\mu})^\top\left[(n-1)\boldsymbol{\Sigma}_n\right]^{-1}(\boldsymbol{s}_{\boldsymbol{y}}-\boldsymbol{\mu})\right]&=(n-1)^2\text{Var}\left[\boldsymbol{z}^\top W \boldsymbol{z}\right]\nonumber\\&=\mathcal{O}\left(   \frac{d^2 n^2}{(n-d)^3}\right).\label{eq:invwishart_qf_order}
\end{flalign} 
\\
\noindent\textbf{$\text{Var}[X_{1n}+X_{2n}]$:} To obtain the order of $\text{Var}[X_{1n}+X_{2n}]$, again apply the Cauchy-Schwarz inequality to obtain:
\begin{flalign*}
\text{Var}\left[\log\left(p_{A,n}\left(\boldsymbol{s}_{\boldsymbol{y}}|\boldsymbol{\theta}\right)
\right)\right]&\leq\text{Var}[X_{1n}]+\text{Var}[X_{2n}]+\sqrt{\text{Var}[X_{1n}]}\times \sqrt{\text{Var}[X_{2n}]}. 
\end{flalign*}The dominant order is determined by the variance terms, and we can then apply the results in \eqref{eq:det_order}, \eqref{eq:t2_order}, \eqref{eq:invwishart_qf_order} to obtain 
\begin{flalign}
\text{Var}\left[\log\left(p_{A,n}\left(\boldsymbol{s}_{\boldsymbol{y}}|\boldsymbol{\theta}\right)
\right)\right]=\mathcal{O}\left( \frac{d^2 n^2}{(n-d)^3}\right)\label{eq:bsl_order}
\end{flalign}

\noindent\textbf{Case: Diagonal/Whitened BSL}
\\

\noindent Decompose $-\log(p_{A,n,w}) $ as 
\begin{flalign*}
-\log\left(p_{A,n,w}\left(\boldsymbol{s}_{\boldsymbol{y}}|\boldsymbol{\theta}\right) 
\right) &\propto\log\text{det}\{\boldsymbol{\Omega}_n(\boldsymbol{\theta})\}+
\left(\boldsymbol{s}_{\boldsymbol{y}}-\boldsymbol{\zeta}_n(\boldsymbol{\theta})\right)^\top \boldsymbol{\Omega}_n^{-1}(\boldsymbol{\theta})\left(\boldsymbol{s}_{\boldsymbol{y}}-\boldsymbol{\zeta}_n(\boldsymbol{\theta})\right)\\&:=Y_{1n}+Y_{2n},
\end{flalign*}
where the proportionality disregards terms that do not depend on $\boldsymbol{\theta}$. Similar to the case of standard BSL, we disregard the functions dependence on $\boldsymbol{\theta}$ to simplify notations. The result in question is obtained by specializing the results obtained for standard BSL to this variant. 
\\

\noindent\textbf{(1) $Y_{1n}$ term:}
Under Assumption \ref{ass:two}, $(n-1)\boldsymbol{\Omega}_n\sim \mathcal{W}(\boldsymbol{\Omega},n-1)$, where  $\boldsymbol{\Omega}:=\text{diag}(\omega_{11},\dots,\omega_{dd})$. Therefore, for $W\sim\mathcal{W}(\boldsymbol{\Omega},n-1)$, by arguments similar to those used to obtain equation \eqref{eq:det_order},
\begin{equation}
\text{Var}\left[\log\text{det}\left\{\boldsymbol{\Omega}_n\right\}\right]= \sum_{i=1}^{d}\psi_1\left(\frac{n-i}{2}\right)= \mathcal{O}(1)\label{eq:Wdet_order}.
\end{equation}

\noindent\textbf{(2) $Y_{2n}$ term:} Again, we specialize the results for the standard BSL. For 
$$Y_{2n}:=\left(\boldsymbol{s}_{\boldsymbol{y}}-\boldsymbol{\zeta}_n(\boldsymbol{\theta})\right)^\top \boldsymbol{\Omega}_n^{-1}(\boldsymbol{\theta})\left(\boldsymbol{s}_{\boldsymbol{y}}-\boldsymbol{\zeta}_n(\boldsymbol{\theta})\right),$$
similar arguments to those used in the standard BSL case yield 
\begin{flalign*}
\text{Var}\left[Y_{2n}\right]&\leq 4\text{Var}\left[\left(\boldsymbol{\zeta}_n-\boldsymbol{\zeta}\right)^\top \boldsymbol{\Omega}_n^{-1}\left(\boldsymbol{\zeta}_n-\boldsymbol{\zeta}\right)\right]+4\text{Var}\left[\left(\boldsymbol{s}_{\boldsymbol{y}}-\boldsymbol{\zeta}\right)^\top\boldsymbol{\Omega}_n^{-1}\left(\boldsymbol{s}_{\boldsymbol{y}}-\boldsymbol{\zeta}\right)\right]\\&+2\sqrt{\text{Var}\left[\left(\boldsymbol{\zeta}_n-\boldsymbol{\zeta}\right)^\top \boldsymbol{\Omega}_n^{-1}\left(\boldsymbol{\zeta}_n-\boldsymbol{\zeta}\right)\right]}\times\sqrt{\text{Var}\left[\left(\boldsymbol{s}_{\boldsymbol{y}}-\boldsymbol{\zeta}\right)^\top\boldsymbol{\Omega}_n^{-1}\left(\boldsymbol{s}_{\boldsymbol{y}}-\boldsymbol{\zeta}\right)\right]}.
\end{flalign*}

For the first term, $\left(\boldsymbol{\zeta}_n-\boldsymbol{\zeta}\right)^\top \boldsymbol{\Omega}_n^{-1}\left(\boldsymbol{\zeta}_n-\boldsymbol{\zeta}\right)$, note that
\begin{flalign*}
t^2_\text{diag}&:=\frac{1}{n}\left[\left(\boldsymbol{\zeta}_n-\boldsymbol{\zeta}\right)^\top \left[\boldsymbol{\Omega}_n/n\right]^{-1}\left(\boldsymbol{\zeta}_n-\boldsymbol{\zeta}\right)\right]=\sum_{i=1}^{d}\left(\frac{\sqrt{n}\{\boldsymbol{\zeta}_{i,n}-\boldsymbol{\zeta}_i\}}{\sqrt{s_i^2}}\right)^2,
\end{flalign*}
where $s_i^2:=[\boldsymbol{\Omega}_n]_{ii}$. Under Assumption \ref{ass:two}, 
$  
{\sqrt{n}\{\boldsymbol{\zeta}_{i,n}-\boldsymbol{\zeta}_i\}}/{\sqrt{s_i^2}}\sim t_{n},$ where $t_n$ denotes a student-t random variable with $n$ degrees of freedom, and where$${\sqrt{n}\{\boldsymbol{\zeta}_{i,n}-\boldsymbol{\zeta}_i\}}/{{\sqrt{s_i^2}}}\perp {\sqrt{n}\{\boldsymbol{\zeta}_{j,n}-\boldsymbol{\zeta}_j\}}/{{\sqrt{s_j^2}}},\text{ for all }i\neq j.$$ 

Now, use the fact that $t_n^2\equiv \mathcal{F}_{1,n}$ to obtain 
\begin{flalign*}
t^2_\text{Diag}&\sim\frac{1}{n}\sum_{i=1}^{d}\mathcal{F}_{1,n}^{i},
\end{flalign*}where $\mathcal{F}_{1,n}^{i}$ denotes an i.i.d.\ copy of an $\mathcal{F}_{1,n}$ random variable. Using the moments of the $\mathcal{F}$-distribution and independence across the $i=1,\dots, d$ components, we obtain 
\begin{flalign}
\text{Var}[t^2_{\text{Diag}}]&=\frac{1}{n^2}\sum_{i=1}^{d}\text{Var}[\mathcal{F}_{1,n}^i]=\frac{d}{n^2}\frac{2(n-1)n^2}{(n-2)^2(n-4)}=\mathcal{O}\left(  \frac{d}{n^2}\right)\label{eq:Wt2_order}.
\end{flalign}

To deduce the variance of the second term, $\left(\boldsymbol{s}_{\boldsymbol{y}}-\boldsymbol{\zeta}\right)^\top\boldsymbol{\Omega}_n^{-1}\left(\boldsymbol{s}_{\boldsymbol{y}}-\boldsymbol{\zeta}\right)$, we make use of the following fact (\citealp{fujikoshi2011multivariate}, Ch.2 ): if $W\sim \mathcal{W}^{-1}(\boldsymbol{\Omega}^{-1},n)$ with $$\boldsymbol{\Omega}^{-1}=\text{diag}\{1/\omega_{11},\dots,1/\omega_{dd}\},$$ then for $\nu^2_{ii}$ denoting the diagonal elements of $W$,
$$
\nu_{ii}^{2}\sim\text{Inv-}\chi^2\left(n-d+1,\frac{1/\omega^{}_{ii}}{n-d+1}\right),
$$ with $\text{Inv-}\chi^2\left(n-d+1,\frac{1/\omega^{}_{ii}}{n-d+1}\right)$ denoting the inverse scaled $\chi^2$ distribution with $(n-d+1)$ degrees of freedom and scale parameter $({1/\omega^{}_{ii}}({n-d+1}))$.

Now, rewrite the term $(\boldsymbol{s}_{\boldsymbol{y}}-\boldsymbol{\zeta})^\top\boldsymbol{\Omega}_n^{-1}(\boldsymbol{s}_{\boldsymbol{y}}-\boldsymbol{\zeta})$ as 
$$
(n-1)(\boldsymbol{s}_{\boldsymbol{y}}-\boldsymbol{\zeta})^\top\left[(n-1)\boldsymbol{\Omega}_n\right]^{-1}(\boldsymbol{s}_{\boldsymbol{y}}-\boldsymbol{\zeta}).
$$ By Assumption \ref{ass:two}, $$\left[(n-1)\boldsymbol{\Omega}_n\right]^{-1}\sim \mathcal{W}^{-1}(\boldsymbol{\Omega}^{-1},n-1),$$ and, for $\boldsymbol{z}:= (\boldsymbol{s}_{\boldsymbol{y}}-\boldsymbol{\zeta})$,
\begin{flalign*}
\text{Var}\left[(n-1)\boldsymbol{z}^\top\left[(n-1)\boldsymbol{\Omega}_n\right]^{-1}\boldsymbol{z}\right]&=(n-1)^2\text{Var}\left[\boldsymbol{z}^\top\text{diag}\{\nu^2_{11},\dots\nu^2_{dd}\}\boldsymbol{z}\right],\\&=(n-1)^2\sum_{i=1}^{d}z_{i}^2\text{Var}\left(\nu^2_{ii}\right).
\end{flalign*}From the properties of the inverse chi-squared distribution, 
$$
\text{Var}\left(\nu^2_{ii}\right)=\frac{2\nu_{ii}^{-2}}{(n-d-2)^2(n-d-4)}.
$$
Conclude that 
\begin{flalign}
\text{Var}\left[(n-1)(\boldsymbol{s}_{\boldsymbol{y}}-\boldsymbol{\zeta})^\top\left[(n-1)\boldsymbol{\Omega}_n\right]^{-1}(\boldsymbol{s}_{\boldsymbol{y}}-\boldsymbol{\zeta})\right]=  \mathcal{O}\left( \frac{d\cdot n^2}{(n-d)^3}\right).\label{eq:Winvwishart_qf_order}
\end{flalign} 
\\
\noindent\textbf{$\text{Var}[X_{1n}+X_{2n}]$:} To deduce the order of $\text{Var}[X_{1n}+X_{2n}]$, again apply the Cauchy-Schwarz inequality to obtain:
\begin{flalign*}
\log\left(p_{A,n,w}\left(\boldsymbol{s}_{\boldsymbol{y}}|\boldsymbol{\theta}\right) 
\right)&\leq\text{Var}[X_{1n}]+\text{Var}[X_{2n}]+\sqrt{\text{Var}[X_{1n}]}\times \sqrt{\text{Var}[X_{2n}]}. 
\end{flalign*}Now, apply the results in \eqref{eq:Wdet_order}, \eqref{eq:Wt2_order}, \eqref{eq:Winvwishart_qf_order} to obtain 
\begin{flalign}\label{eq:w_order}
\text{Var}\left[\log\left(p_{A,n,w}\left(\boldsymbol{s}_{\boldsymbol{y}}|\boldsymbol{\theta}\right) 
\right)\right]=\mathcal{O}\left( \frac{d\cdot n^2}{(n-d)^3}\right). 
\end{flalign}

Comparing the orders of the variance in equations \eqref{eq:bsl_order} and \eqref{eq:w_order}, the result follows.

\end{proof}

%%%%%%%%%%%%%%%%%%%%%%%%%%%%%%%%%%%%%%%%%%%
%%%%%%%%%%%%%%%%%%%%%%%%%%%%%%%%%%%%%%%%%%%
\section{Results for toy models}
%%%%%%%%%%%%%%%%%%%%%%%%%%%%%%%%%%%%%%%%%%%
%%%%%%%%%%%%%%%%%%%%%%%%%%%%%%%%%%%%%%%%%%%
\subsection{An AR(1) model}
%Another common time series model is the autoregressive model of order $p$, AR$(p)$. The $t^{\text{th}}$ value $z_t$ in the series is assumed to be a function of the past $p$ values $z_{t-1},...,z_{t-p}$ according to:
%\begin{align*}
%    z_t = \phi_1 z_{t-1} + \phi_2 z_{t-2} +\cdots+ \phi_p z_{t-p} + w_t,
%\end{align*}
%where $\phi_1,...,\phi_p$ are constants and $w_t$ is often assumed to be $\mathcal{N}(0,\sigma^2)$. To ensure stationarity, it is a requirement that the roots of the polynomial $1 - \phi_1 z - \cdots - \phi_p z^p$ lie outside of the unit circle. Like the moving average model, the AR$(p)$ model has a tractable likelihood function. \\

The autocorrelation function of an autoregressive model typically decays more slowly than that of a moving average model. As a result, taking the full observed dataset as summary statistics for an AR model will produce summary statistics with stronger dependences, thereby presenting a greater inferential challenge.
% dependencies.
%. Provided we again take the full data set to be the set of summary statistics, the covariances between summary statistics will be much larger in magnitude. This presents an even more challenging example for wBSL.\\
%
We consider an AR(1) model 
%with each observation taking 
of the form
\begin{align*}
    z_t = \phi z_{t-1} + w_t, 
\end{align*}
where $w_t\sim \mathcal{N}(0,\sigma^2)$ for $t = 1,...,T_0$ and $z_0 = 0$. The likelihood is again multivariate normal with zero mean vector and  covariance matrix constructed from $\gamma(h) = \text{cov}(z_{t+h},z_t) = \phi^h/(1-\phi^2)$ for $h\geq0$, subject to the constraint $\left\vert\phi\right\vert<1$. We generate 200 observations from the AR$(1)$ process with $\phi_{\text{true}} = 0.9$ and fixed $\sigma^2=1$. As before, we take the full dataset as summary statistics, use  $n_{\text{cov}} = 20\,000$ model simulations at $\phi^0=\phi_{\text{true}}$ to estimate $\boldsymbol{W}$ and implement the wBSL MCMC sampler for $T = 200\,000$ iterations. The prior is specified as $\phi\sim\mathcal{U}(-1,1)$.
%We use the prior $\phi\sim\mathcal{U}(-1,1)$, $T = 200000$ iterations of MCMC and $n_{\text{cov}} = 20000$\\

\begin{figure}[h!]
\centering
%\centering\includegraphics[width = 15cm]{AR1+tv.png}
\begin{subfigure}
\centering\includegraphics[width = 16cm]{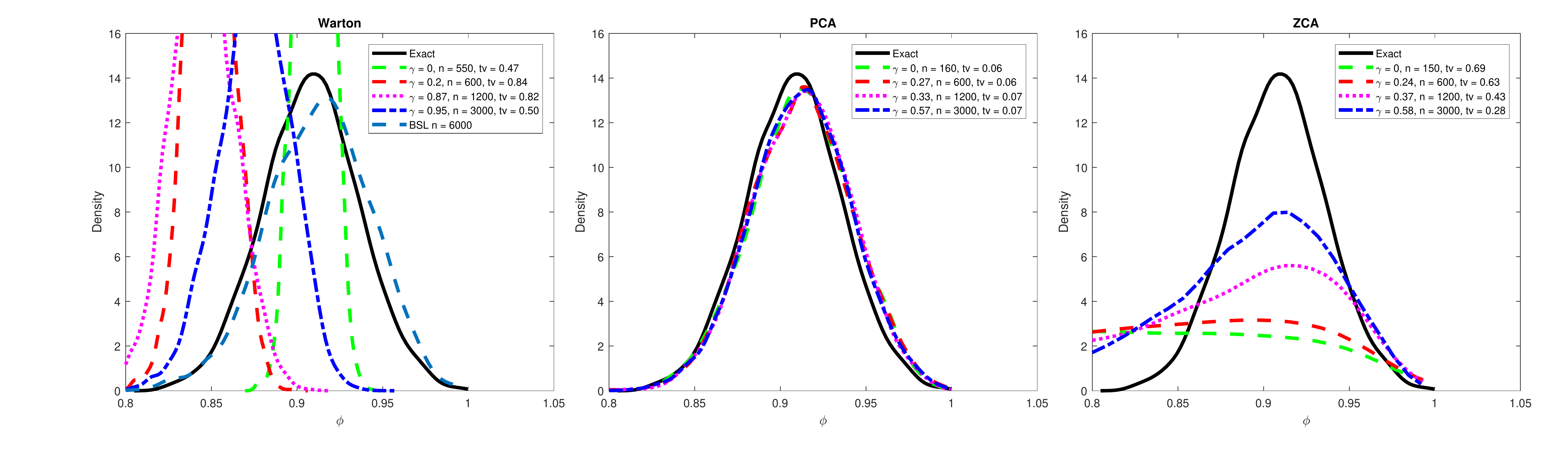}
\end{subfigure}
\begin{subfigure}
\centering\includegraphics[width = 16cm]{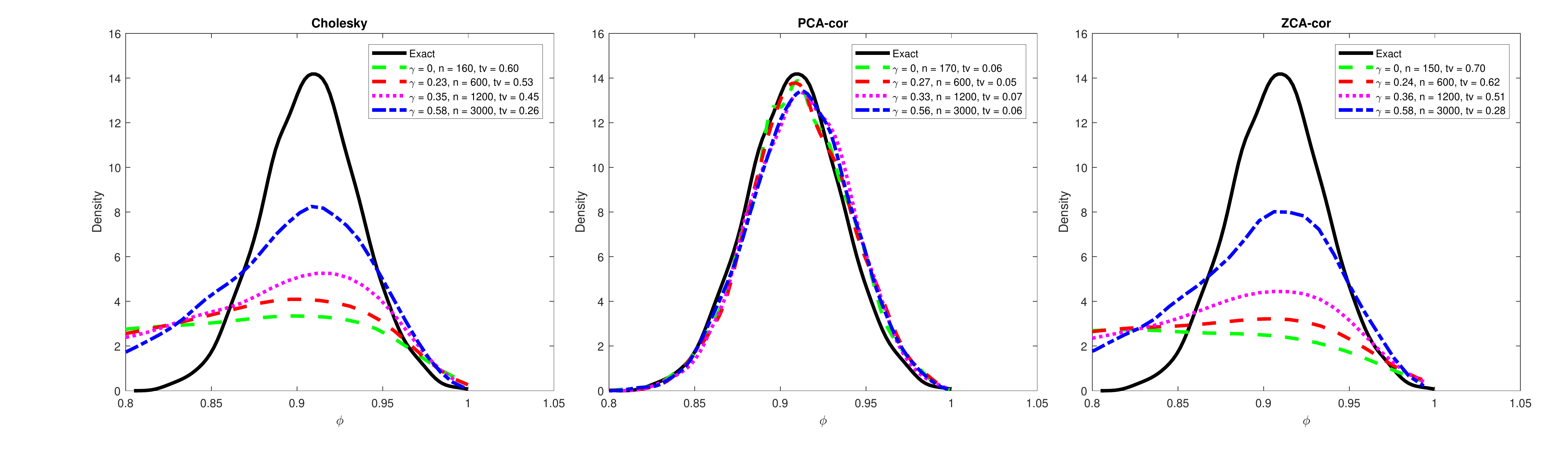}
\end{subfigure}
\caption{\small
Posterior density approximations for the AR(1) model. 
Panels correspond to the method used: either Warton shrinkage by itself (top left), or one of the five whitening transformations. 
The true posterior (solid line) estimated using MCMC is overlaid with the approximate posterior obtained for each considered level of shrinkage ($\gamma$) and number of model simulations ($n$) combination. The approximate standard BSL posterior with $n=6\,000$ is shown in the top left panel (blue dashed lines).
 tv denotes total variation distance between approximate and true  distributions.
}
\label{fig:ar1}
\end{figure}

%Posterior distributions for each of the methods are shown in 
The resulting estimated posterior approximations and the tv distance between these and the true posterior (solid lines) are illustrated in Figure \ref{fig:ar1}.
%, and the total variation distance between these and the true posterior (solid lines) is given in Table \ref{tab:ar1}. 
%
%The figure indicates that a
All whitening methods produce more accurate approximations to the true posterior than Warton shrinkage by itself. 
PCA and PCA-cor achieve the best posterior approximations for all levels of shrinkage. Remarkably, this means that PCA and PCA-cor whitening allows the number of model simulations to be reduced from $n=6\,000$ for standard BSL to just $n=160$ and $n=170$ (for $\gamma=0$) respectively, with no detrimental effect on the inferred posterior approximation. While outperforming Warton shrinkage alone (for $\gamma>0$), the remaining three whitening methods, all have poorly-estimated means and variances, and generally perform worse than for the MA(2) model.
%overall they seem to be performing worse in this example than in the MA$(2)$ example. 

%Contour plots of the wBSL posterior approximations for the MA(2) model. 
%Columns denote (left to right) Warton shrinkage alone, and the whitening methods PCA, PCA-cor, Cholesky, ZCA and ZCA-cor. Rows correspond to complete shrinkage ($\gamma=0$; top row) and 
% $90\%$, $80\%$ and $50\%$ reductions in the number of model simulations (rows 2--4).}

%\begin{table}[h!]
%\caption{\small 
%Total variation distances between the true posterior and the BSL posterior approximation for the AR(1) model under different whitening and shrinkage scenarios.
%\textcolor{red}{[Typically these figures would be mean values (with associated standard errors ($s/\sqrt{n_{rep}}$ in parentheses) obtained from a moderate number (100? 50? 20?) of replicate simulations. Suggest we consider doing that here, as these numbers are currently unreliable (from a Reviewer perspective).]}
% of Warton and whitening methods to the MCMC posterior for the AR$(1)$ example.
%}
%\centering
% \begin{tabular}{ p{2.5cm}|p{2.5cm}p{2cm}p{2cm}p{2cm}  }
% %\hline
%   & $\gamma = 0$ & $n=600$ & $n=1200$ & $n=3000$\\
%% \hline
% \hline
% Warton & 0.47 & 0.84 & 0.82 & 0.50\\
% Cholesky & 0.60 & 0.53 & 0.45 & 0.26\\
% PCA & 0.06 & 0.06 & 0.07 & 0.07 \\
% PCA-cor & 0.06 & 0.05 & 0.07 & 0.06\\
% ZCA & 0.69 & 0.63 & 0.43 & 0.28 \\
% ZCA-cor  & 0.70 & 0.62 & 0.51 & 0.28\\
%% \hline
%\end{tabular}
%\label{tab:ar1}
%\end{table}

%%%%%%%%%%%%%%%%%%%%%%%%%%%%%%%%%%%%%%%%%%%
%%%%%%%%%%%%%%%%%%%%%%%%%%%%%%%%%%%%%%%%%%%
\subsection{Normal model}
%%%%%%%%%%%%%%%%%%%%%%%%%%%%%%%%%%%%%%%%%%%
%%%%%%%%%%%%%%%%%%%%%%%%%%%%%%%%%%%%%%%%%%%

The final simulated example examines how $\boldsymbol{\Sigma}$'s dependence on model parameters affects the wBSL posterior. We consider data drawn from a $k=200$ dimensional multivariate normal distribution $\mathcal{N}_k(\boldsymbol{y}|\boldsymbol{\mu},\boldsymbol{\Sigma})$ where the mean vector $\boldsymbol{\mu}$ has all elements equal to $\theta_1$ and the covariance matrix is $\boldsymbol{\Sigma} = \boldsymbol{\Psi} + \theta_2\boldsymbol{I}_k$ with $\theta_2>0$. The $(i,j)^{\text{th}}$ element of $\boldsymbol{\Psi}$ is given by $\Psi_{i,j} = 0.5^{\left|i-j\right|}$, for $i,j = 1,...,k$. Note that the covariance depends on $\theta_2$ but not $\theta_1$. 
We generate 200 observations from this normal model with $\boldsymbol{\theta}_{\text{true}} = (\theta_1,\theta_2)^\top = (0.5,0.1)^\top$, and use the full dataset as summary statistics. As previously we use $n_{cov}=20\,000$ model simulations drawn at $\boldsymbol{\theta}^0=\boldsymbol{\theta}_{\text{true}}$ to estimate $\boldsymbol{W}$, and implement standard and wBSL MCMC samplers for $T=200\,000$ iterations.
The joint prior is specified $p(\theta_1,\theta_2)\propto 1$ over the parameter support.

%We use the parameter value $\boldsymbol{\theta} = (\theta_1,\theta_2)^\top = (0.5,0.1)^\top$ to generate the data set. The summary statistics are taken to be the full data set and again, the availability of the likelihood function allows for the comparison of each method to the true posterior obtained by an MCMC run. We use flat improper priors $p(\theta_1)\propto 1$ and $p(\theta_2)\propto 1$ with $\theta_2>0$, $T=200000$ iterations of MCMC and $n_{\text{cov}}=20000$ to accurately estimate $\boldsymbol{W}$.\\

\begin{figure}[h!]
\centering
\begin{subfigure}
\centering\includegraphics[width = 15cm]{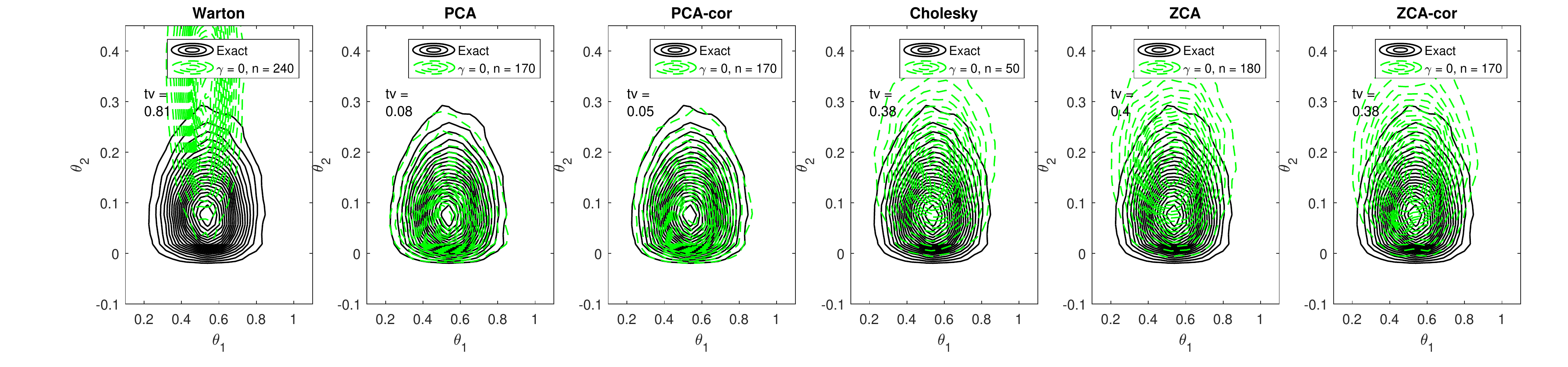}
\end{subfigure}
\begin{subfigure}
\centering\includegraphics[width = 15cm]{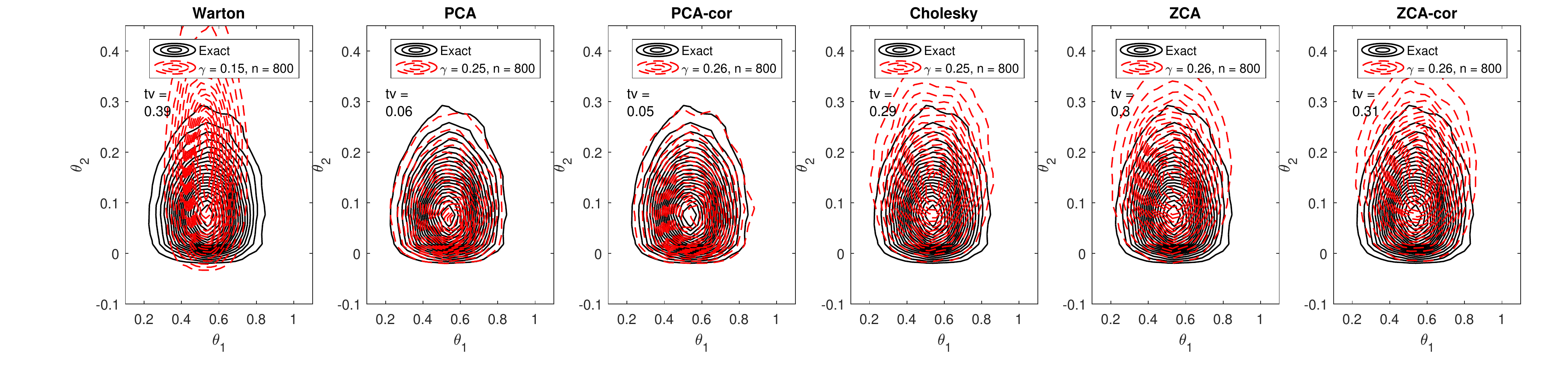}

\end{subfigure}
\begin{subfigure}
\centering\includegraphics[width = 15cm]{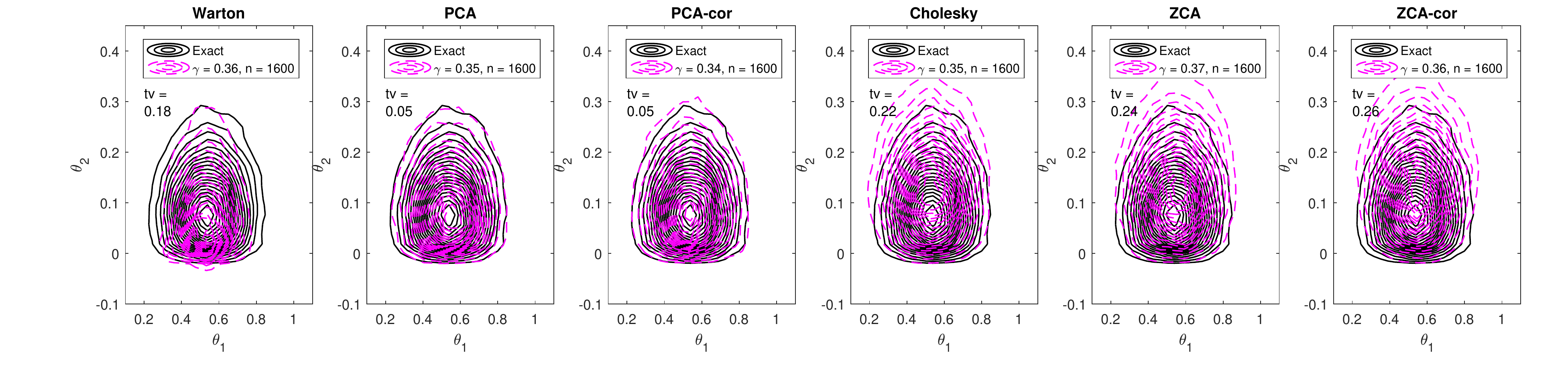}
\end{subfigure}

\begin{subfigure}
\centering\includegraphics[width = 15cm]{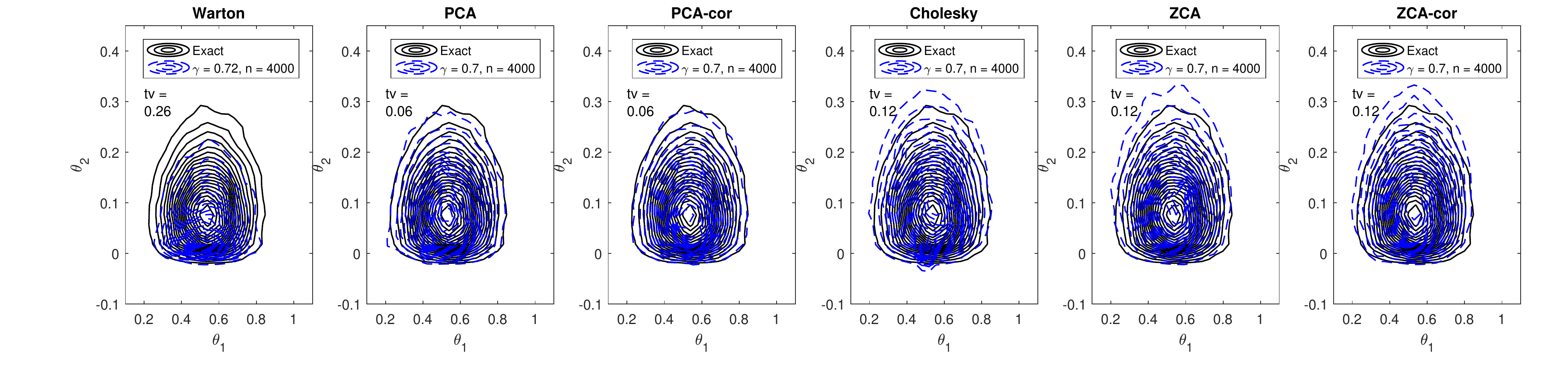}
\end{subfigure}

\caption{\small
Contour plots of the wBSL posterior approximations for the normal model. 
Columns denote (left to right) Warton shrinkage alone, and the whitening methods PCA, PCA-cor, Cholesky, ZCA and ZCA-cor. Rows correspond to complete shrinkage ($\gamma=0$; top row) and 
 $90\%$, $80\%$ and $50\%$ reductions in the number of model simulations (rows 2--4).
 tv denotes total variation distance between approximate and true  distributions.
 %
% Normal example bivariate contour plots of the posterior approximations. Rows correspond to the $\gamma$/$n$ combinations, with the rows from top to bottom being $\gamma = 0$, $n = 800$, $n = 1600$ and $n = 4000$. The columns correspond to the method/whitening transformation, with left to right being Warton shrinkage by itself, PCA, PCA-cor, Cholesky, ZCA and ZCA-cor.
}
\label{fig:toy1}
\end{figure}

The resulting estimated posterior approximations are illustrated in Figure \ref{fig:toy1}.
It is evident that the marginal distribution for $\theta_1$ is estimated accurately for each of the whitening transforms. 
This is not the case for Warton shrinkage alone, for which the variance is clearly under-estimated.
As expected, this demonstrates that when the covariance of the summary statistics $\boldsymbol{s}_i$ depends weakly (or in this case, not at all) on a parameter, then any whitening transformation will perform well. 
For $\theta_2$, which has an effect on the covariance of $\boldsymbol{s}_i$, all whitening transforms perform better than Warton shrinkage alone, with PCA and PCA-cor outperforming all other whitening methods. ZCA, ZCA-cor and Cholesky whitening clearly find it more challenging to estimate parameters that have an influence on the covariance. In terms of posterior dependence structure, the parameters are largely independent of each other, and this is reflected for all results.

We find that $n=8\,000$ model simulations are required for standard BSL. Using wBSL with either PCA or PCA-cor whitening  the number of model simulations can be reduced to just $n=170$ and produce  an essentially identical posterior approximation.
\newpage

%%%%%%%%%%%%%%%%%%%%%%%%%%%%%%%%%%%%%%%%%%%
%%%%%%%%%%%%%%%%%%%%%%%%%%%%%%%%%%%%%%%%%%%
\newpage
\section{Additional figures}
\subsection{Movement models for Fowler's toads}
\begin{figure}[h!]
\centering
\begin{subfigure}
\centering\includegraphics[width = 15cm]{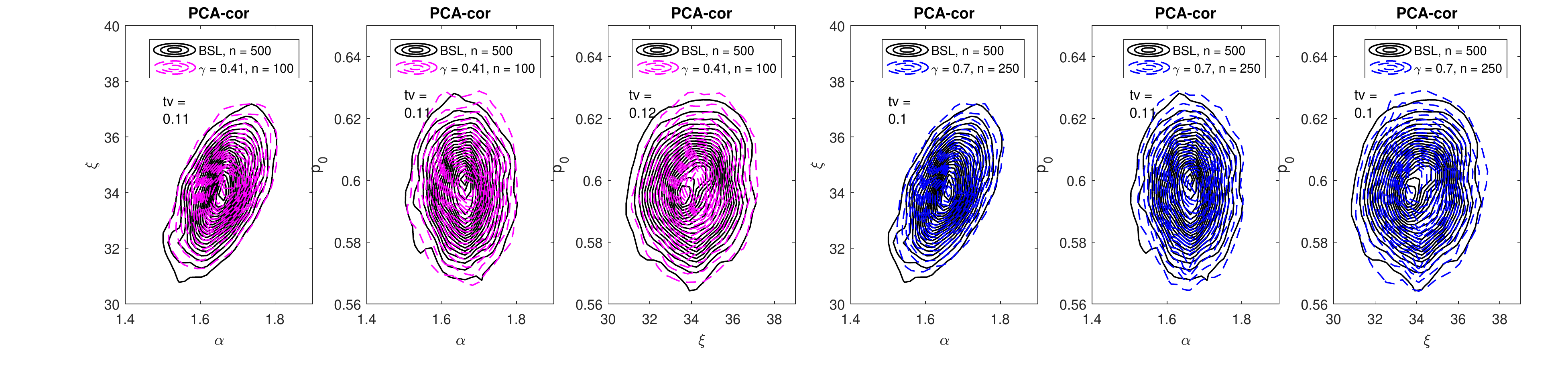}
\end{subfigure}
\begin{subfigure}
\centering\includegraphics[width = 15cm]{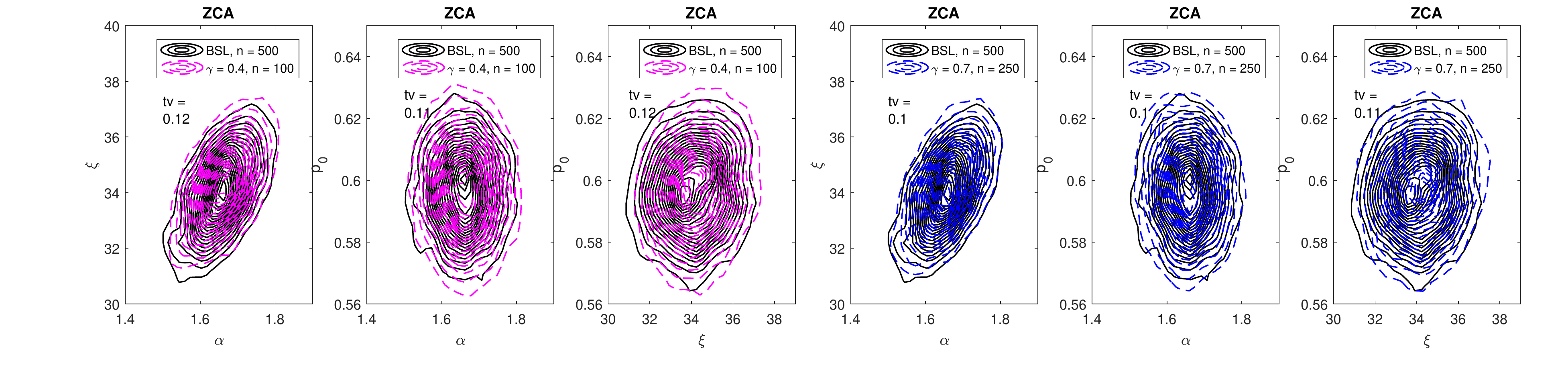}
\end{subfigure}
\begin{subfigure}
\centering\includegraphics[width = 15cm]{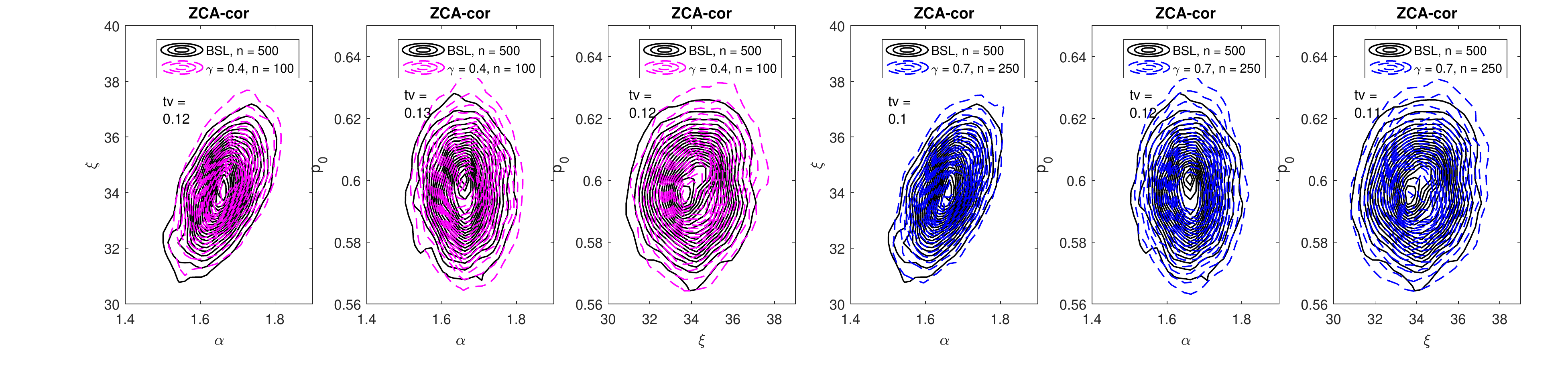}
\end{subfigure}
\begin{subfigure}
\centering\includegraphics[width = 15cm]{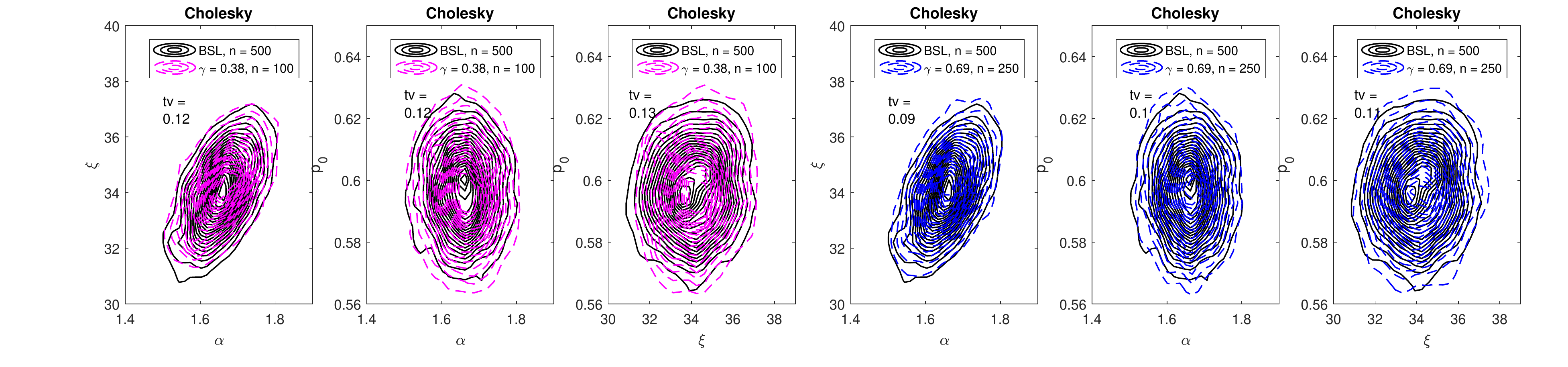}
\end{subfigure}
\caption{\small As Figure 2 in the main text, but for 50\% (right columns) and 80\% (left columns) reductions in the number of model simulations for PCA-cor, ZCA, ZCA-cor and Cholesky whitening.
%
%Toad example approximate posterior distributions. Left three columns are for $n=100$ and right three columns are for $n=250$. Rows correspond to the method used -- PCA-cor, ZCA, ZCA-cor and Choleksy, from top to bottom.
}
\label{fig:Toad3}
\end{figure}

\begin{figure}[h!]
\centering
\begin{subfigure}
\centering\includegraphics[width = 15cm]{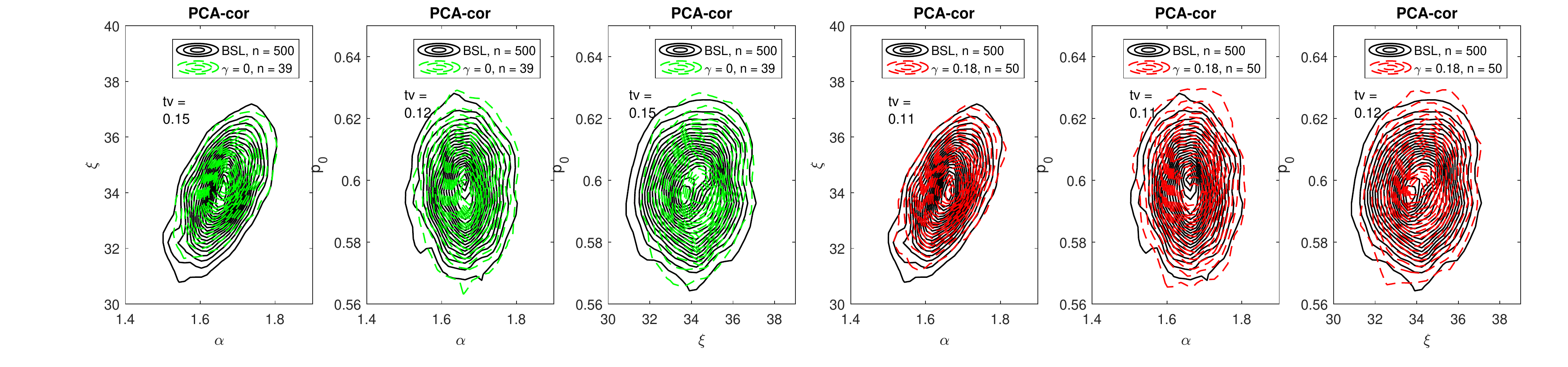}
\end{subfigure}
\begin{subfigure}
\centering\includegraphics[width = 15cm]{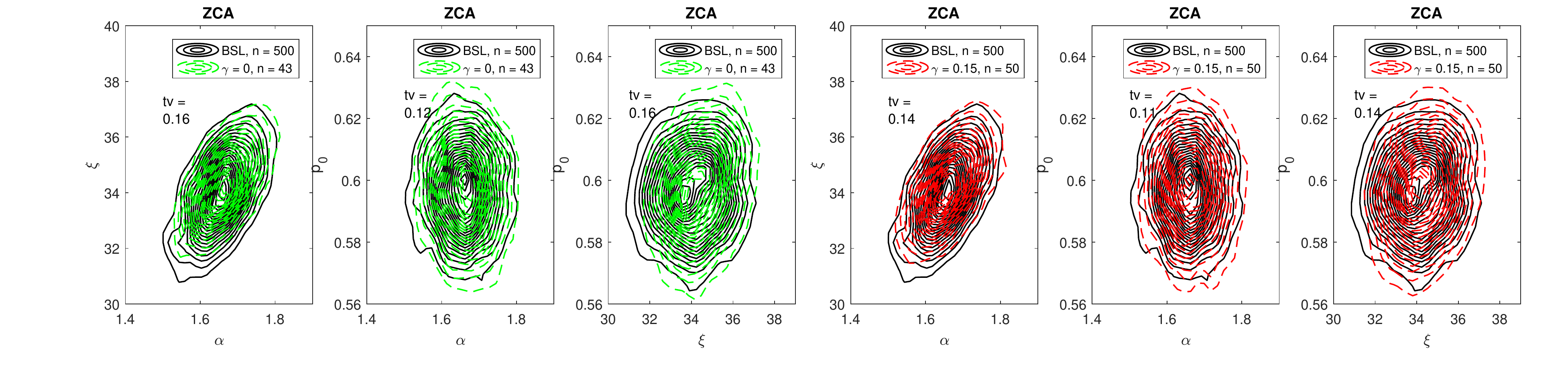}
\end{subfigure}
\begin{subfigure}
\centering\includegraphics[width = 15cm]{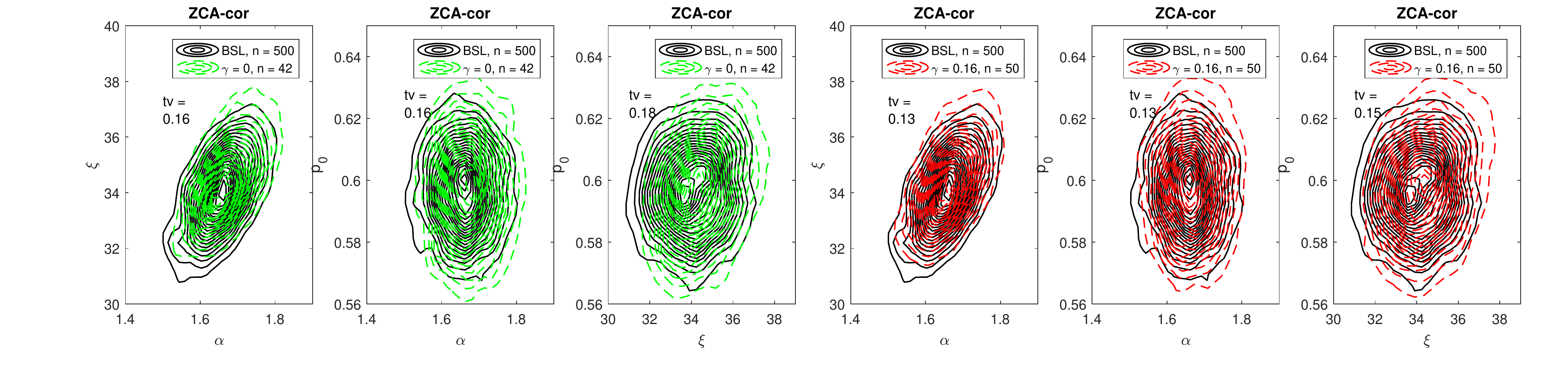}
\end{subfigure}
\begin{subfigure}
\centering\includegraphics[width = 15cm]{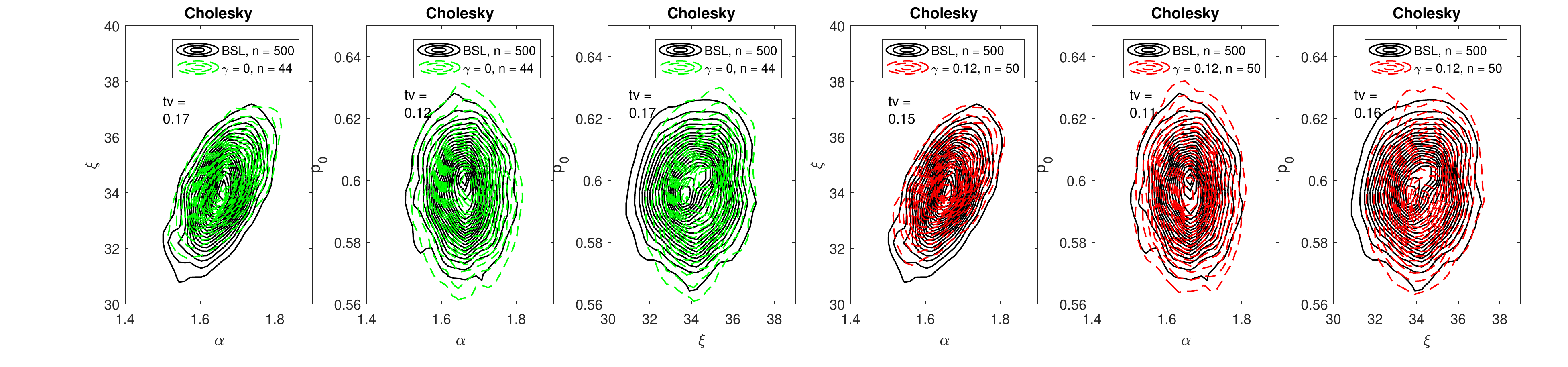}
\end{subfigure}
\caption{\small As Figure 2 in the main text, but for 90\% (right columns) reductions in the number of model simulations, and complete shrinkage ($\gamma=0$; left columns) for PCA-cor, ZCA, ZCA-cor and Cholesky whitening.
%Toad example approximate posterior distributions. Left three columns are for complete shrinkage $\gamma = 0$ and right three columns are for $n=50$ model simulations. Again, rows correspond to the method for wBSL.
}
\label{fig:Toad4}
\end{figure}

\clearpage
%\section*{Appendix B}
\subsection{Choice of Whitening Method}
\subsubsection{An MA(2) model}
\begin{figure}[h!]
\centering
\begin{subfigure}
\centering\includegraphics[width = 5cm]{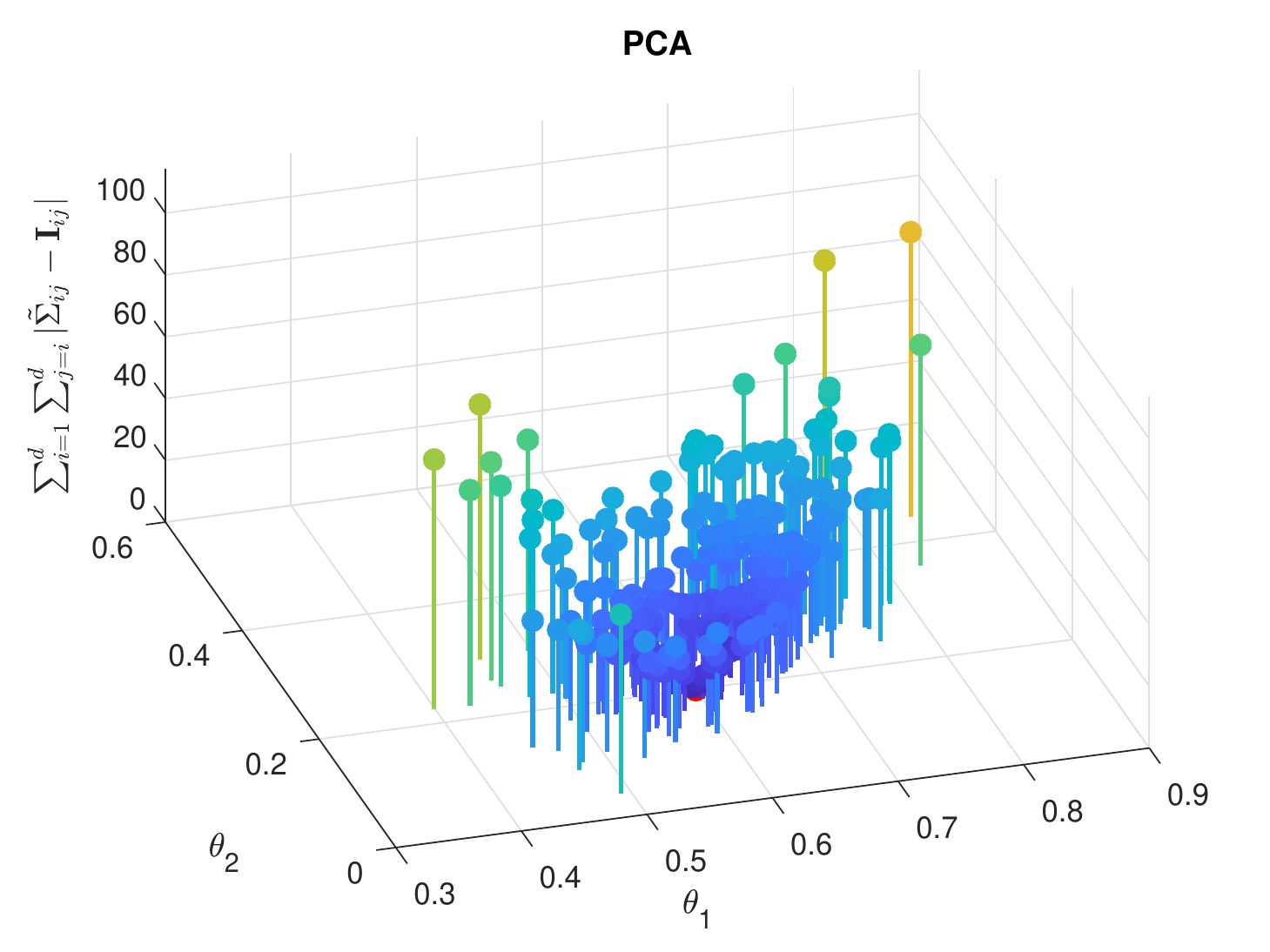}
\end{subfigure}
\begin{subfigure}
\centering\includegraphics[width = 5cm]{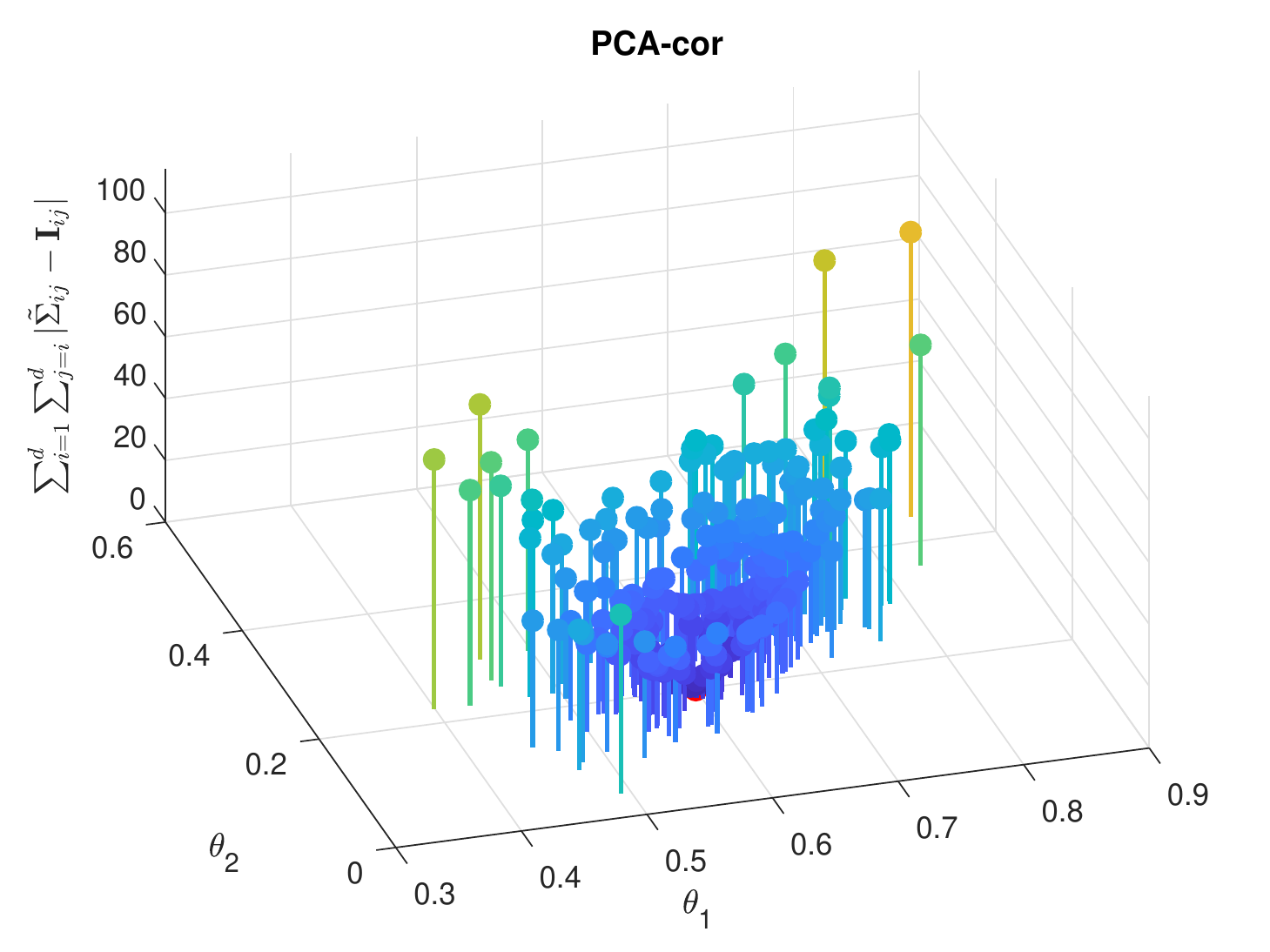}
\end{subfigure}
\begin{subfigure}
\centering\includegraphics[width = 5cm]{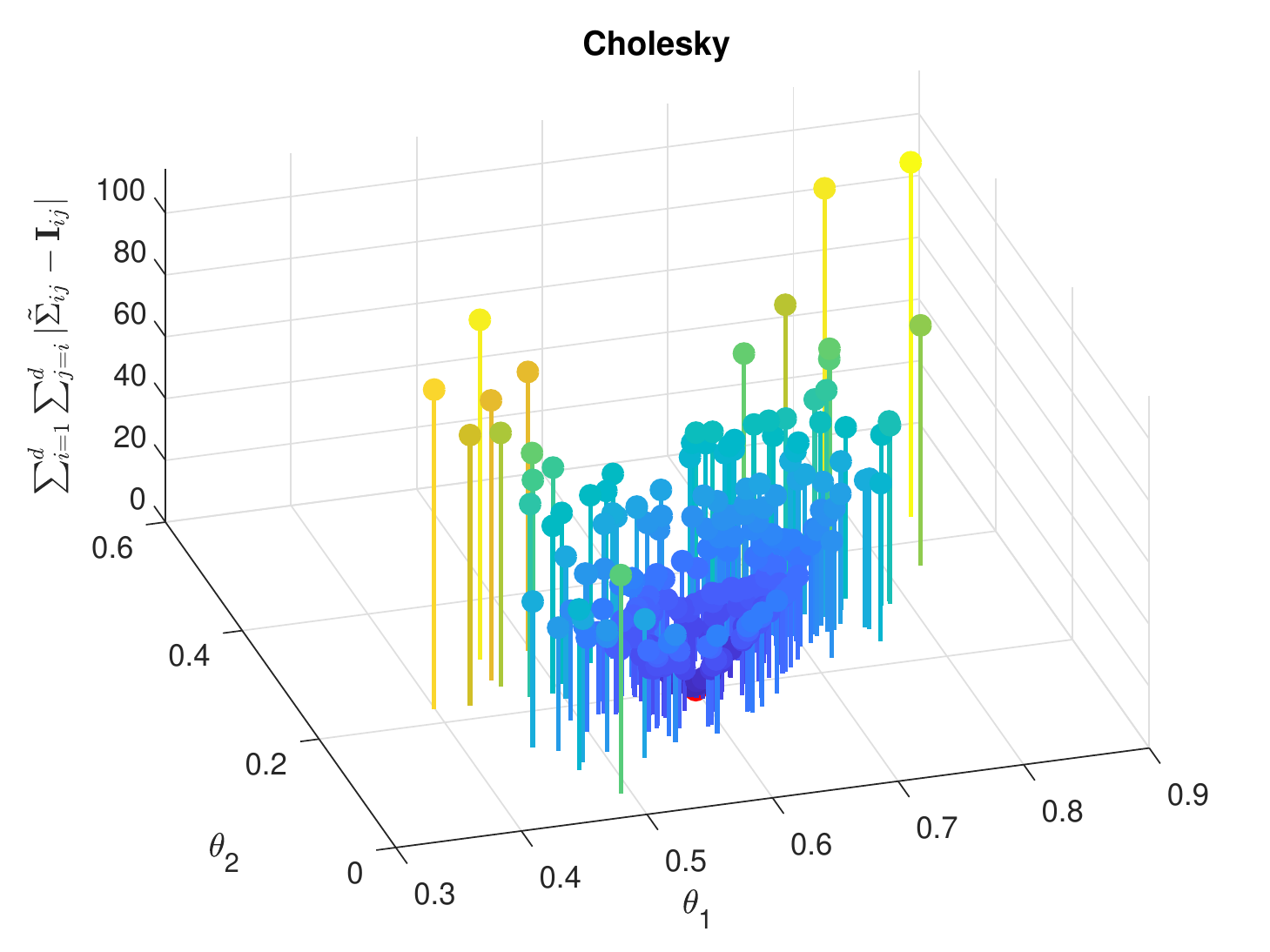}
\end{subfigure}
\begin{subfigure}
\centering\includegraphics[width = 5cm]{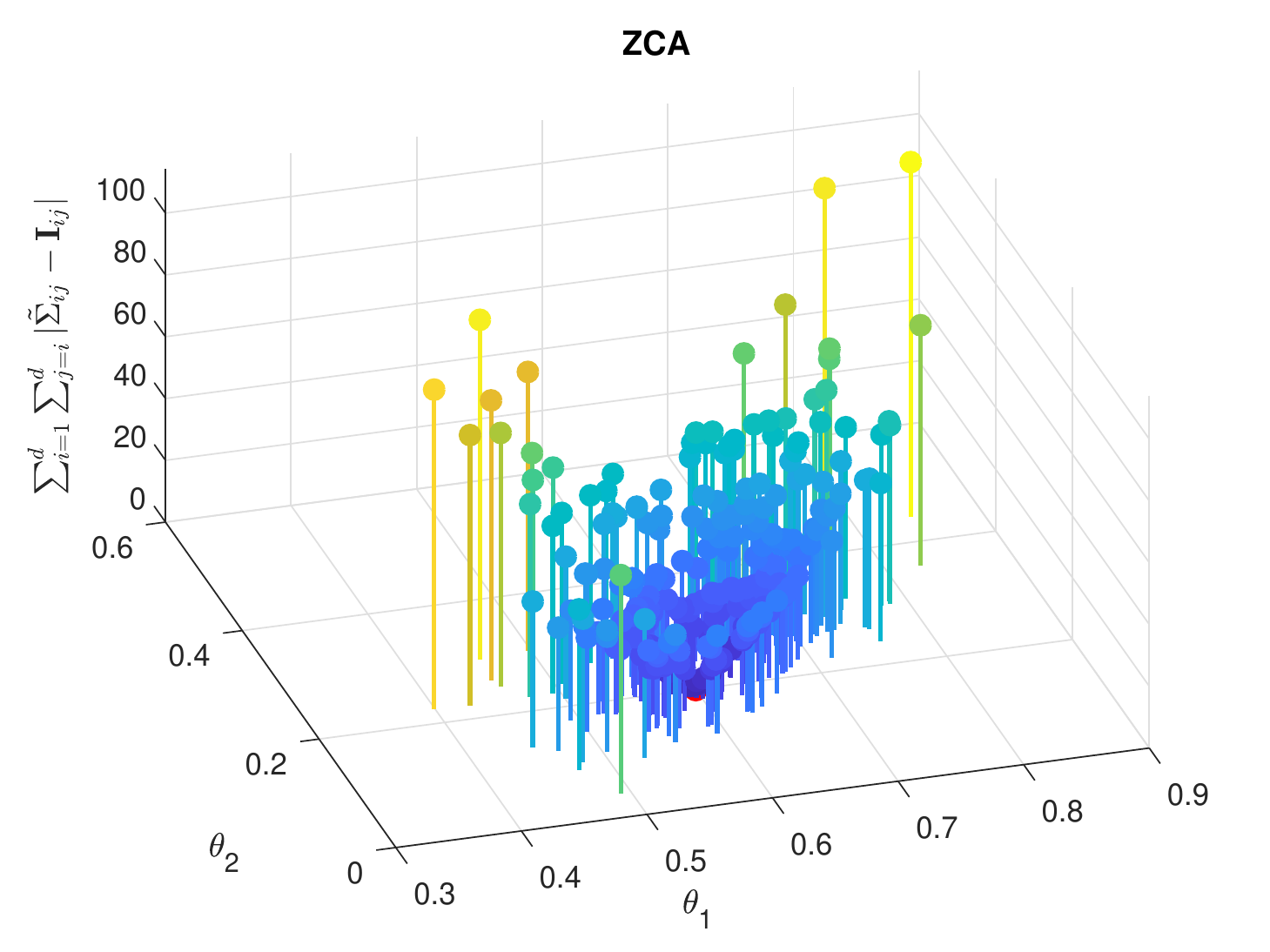}
\end{subfigure}
\begin{subfigure}
\centering\includegraphics[width = 5cm]{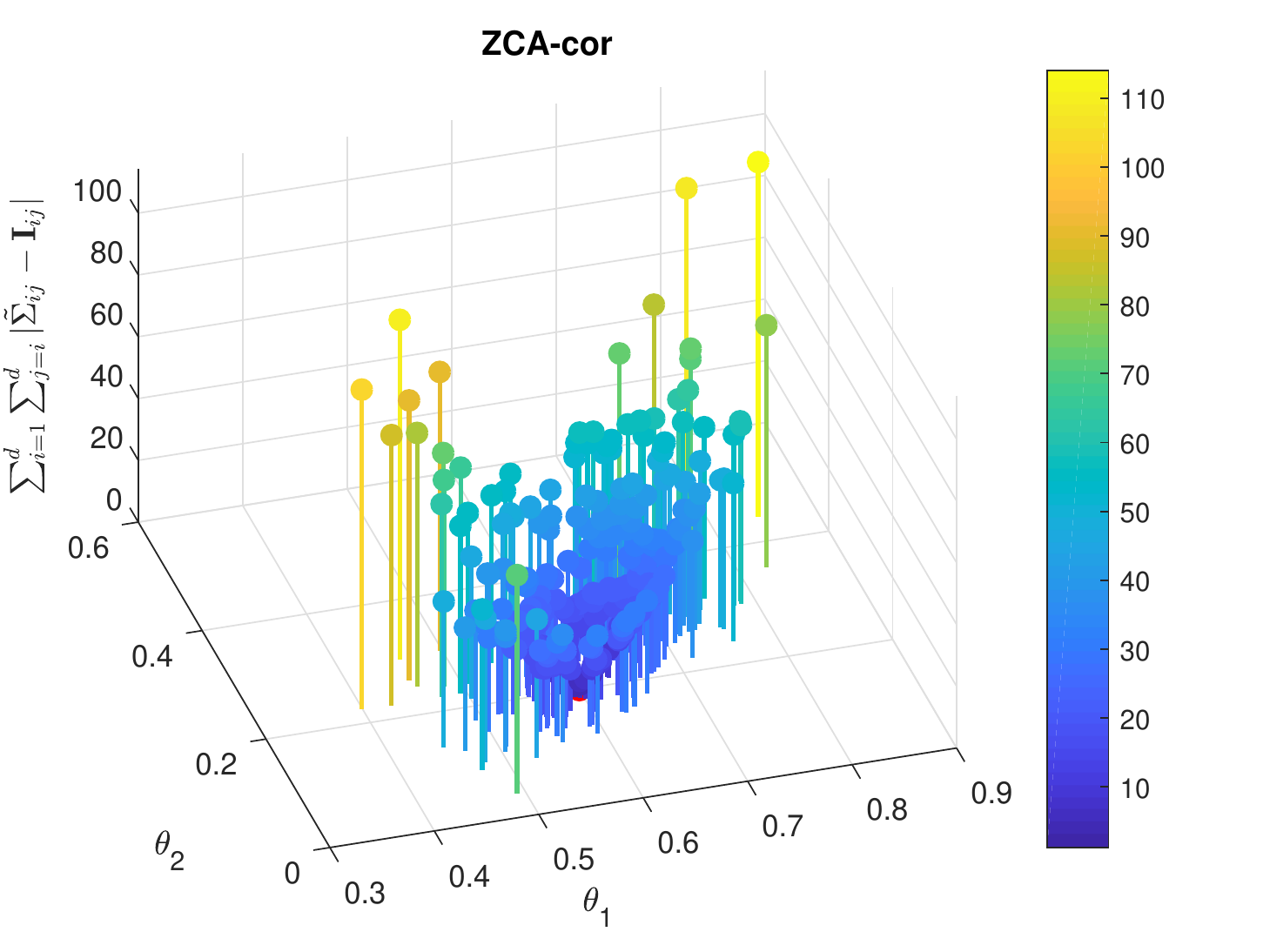}
\end{subfigure}
\caption{\small As Figure 4 in the main text, but including variance terms in the $L_1$ norm deviation.
%Covariance deviation (including variances) from the identity matrix for transformed summary statistics over parameter values sampled from the MCMC posterior for the MA$(2)$ example. Deviation is measured using the $L_1$ matrix norm. Bar colour represents error, and the `true' parameter value $\boldsymbol{\theta} = (0.6,0.2)^\top$ (where $\boldsymbol{W}$ is estimated) is shown in red.
}
\label{fig:choiceofwhiteningsampleswithdiagsMA}
\end{figure}

\begin{figure}[h!]
\centering
\begin{subfigure}
\centering\includegraphics[width = 5cm]{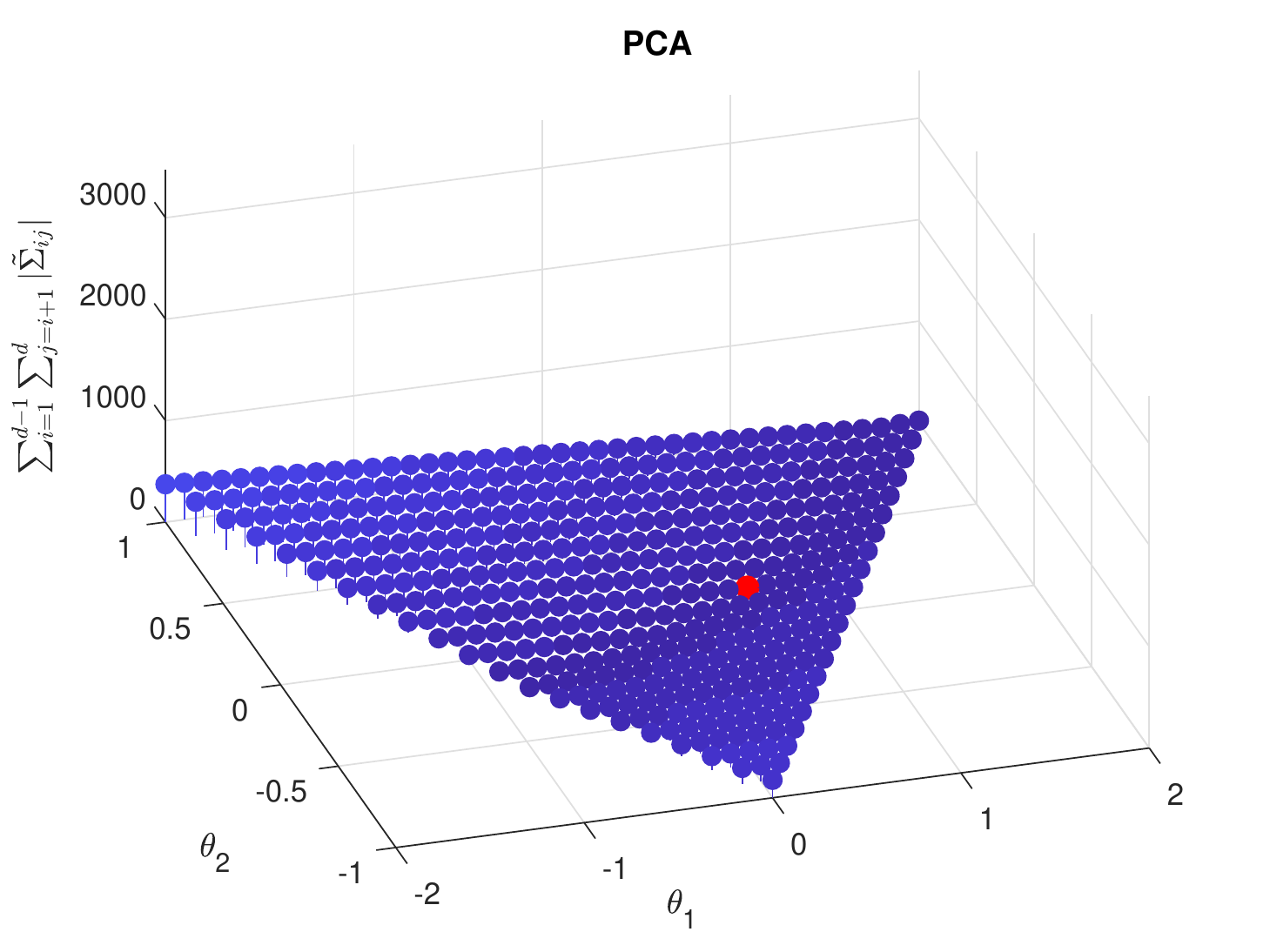}
\end{subfigure}
\begin{subfigure}
\centering\includegraphics[width = 5cm]{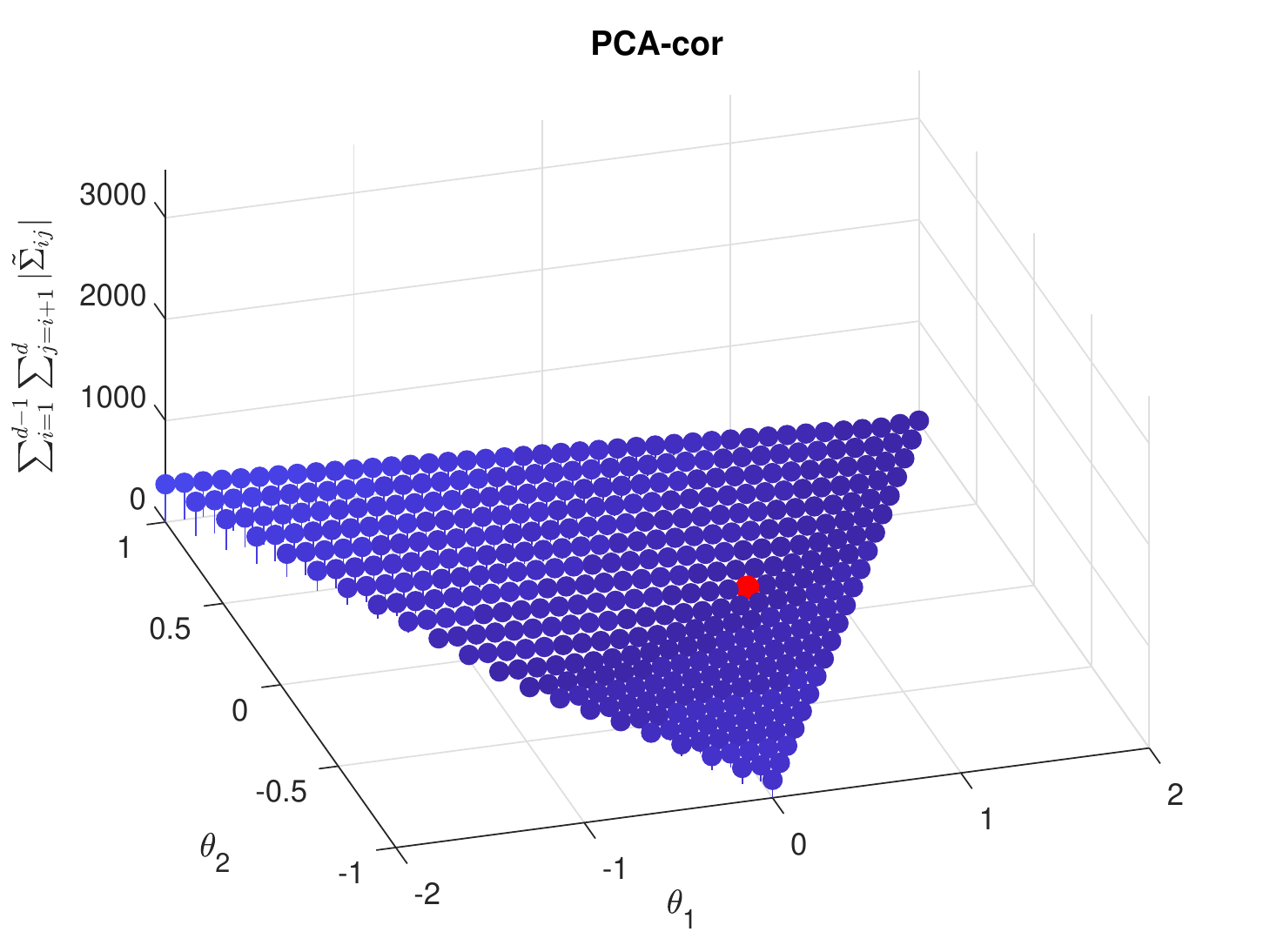}
\end{subfigure}
\begin{subfigure}
\centering\includegraphics[width = 5cm]{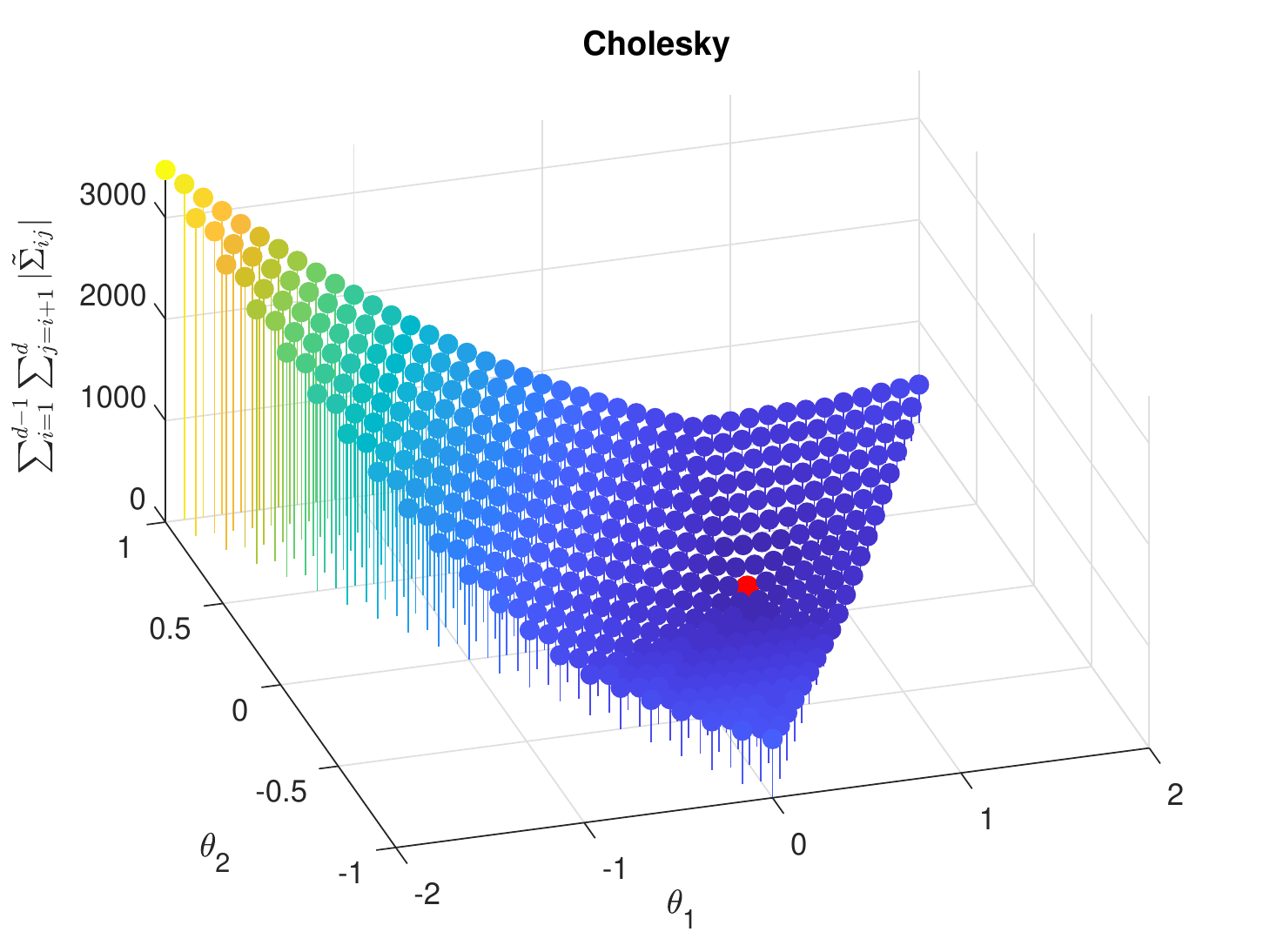}
\end{subfigure}
\begin{subfigure}
\centering\includegraphics[width = 5cm]{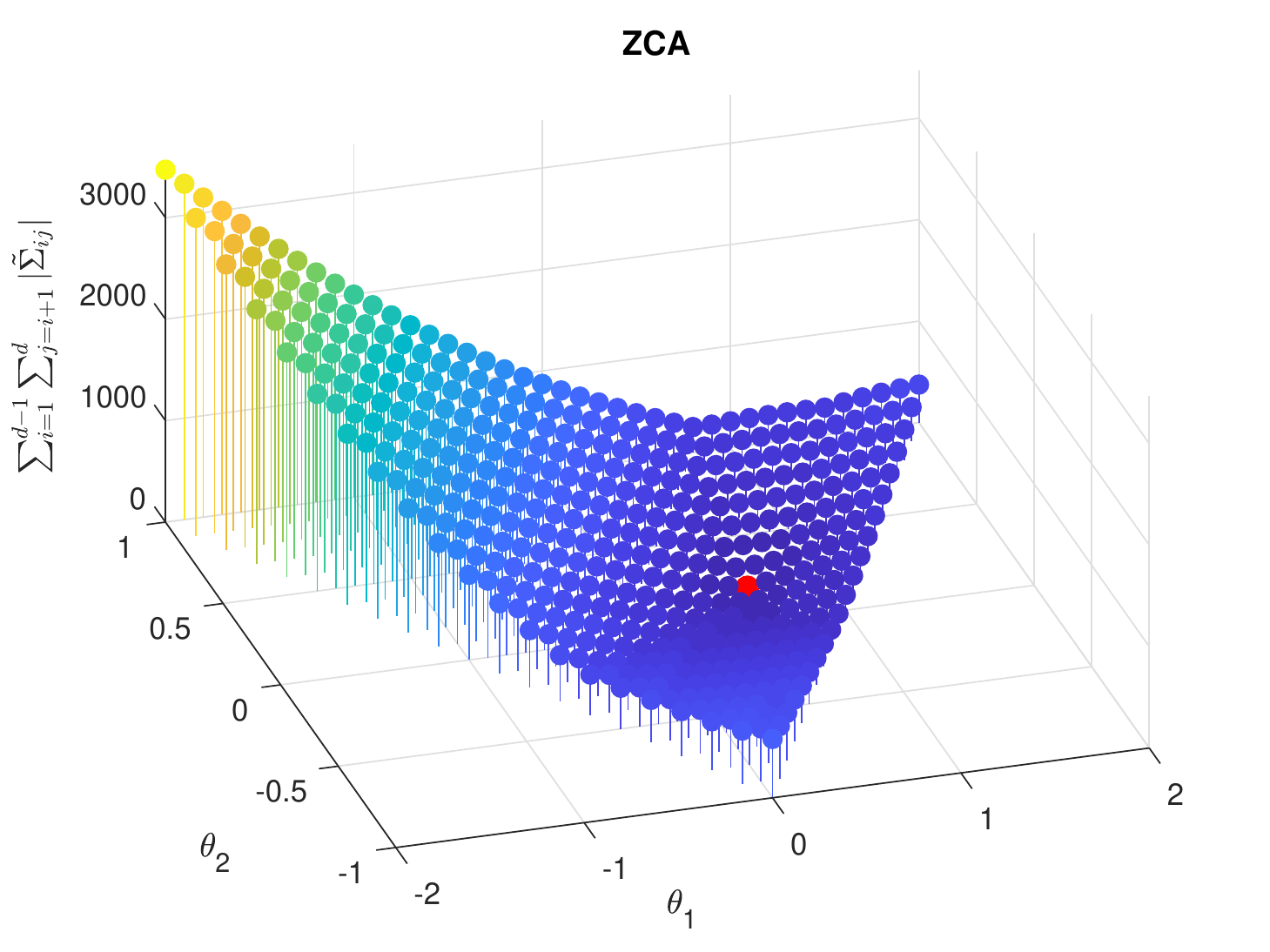}
\end{subfigure}
\begin{subfigure}
\centering\includegraphics[width = 5cm]{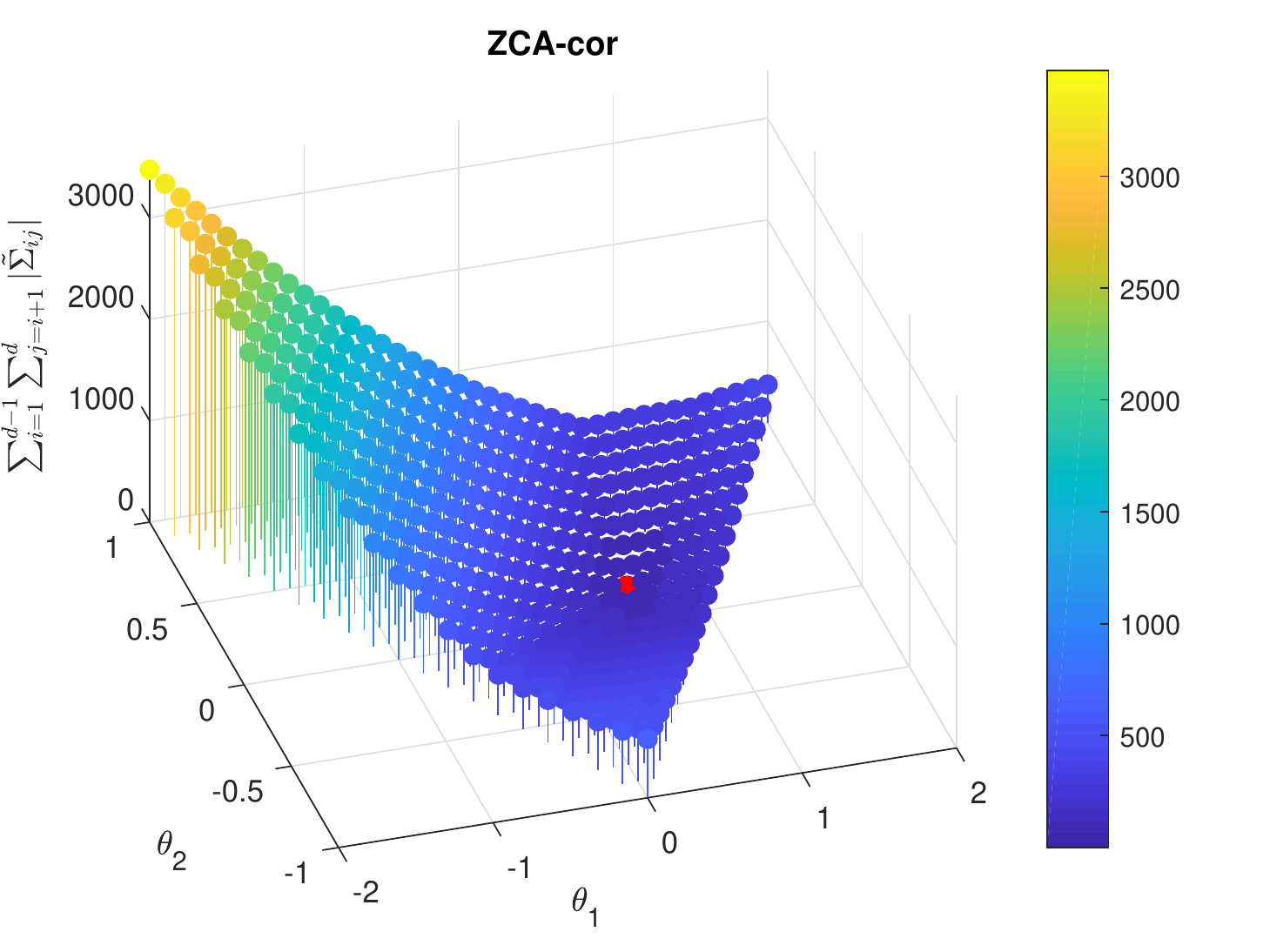}
\end{subfigure}
\caption{\small As Figure 4 in the main text, but covering the full parameter space.%Covariance deviation (without variances) from the zero matrix for the whitened summary statistics over the entire parameter space for the MA$(2)$ example. Deviation is measured using the $L_1$ matrix norm. Bar colour represents the error and the `true' parameter value $\boldsymbol{\theta} = (0.6,0.2)^\top$ is shown in red.
}
\label{fig:choiceofwhiteningparamspaceMA}
\end{figure}

\subsubsection{An AR(1) model}

\begin{figure}[h!]
\centering
\begin{subfigure}
\centering\includegraphics[width = 5cm]{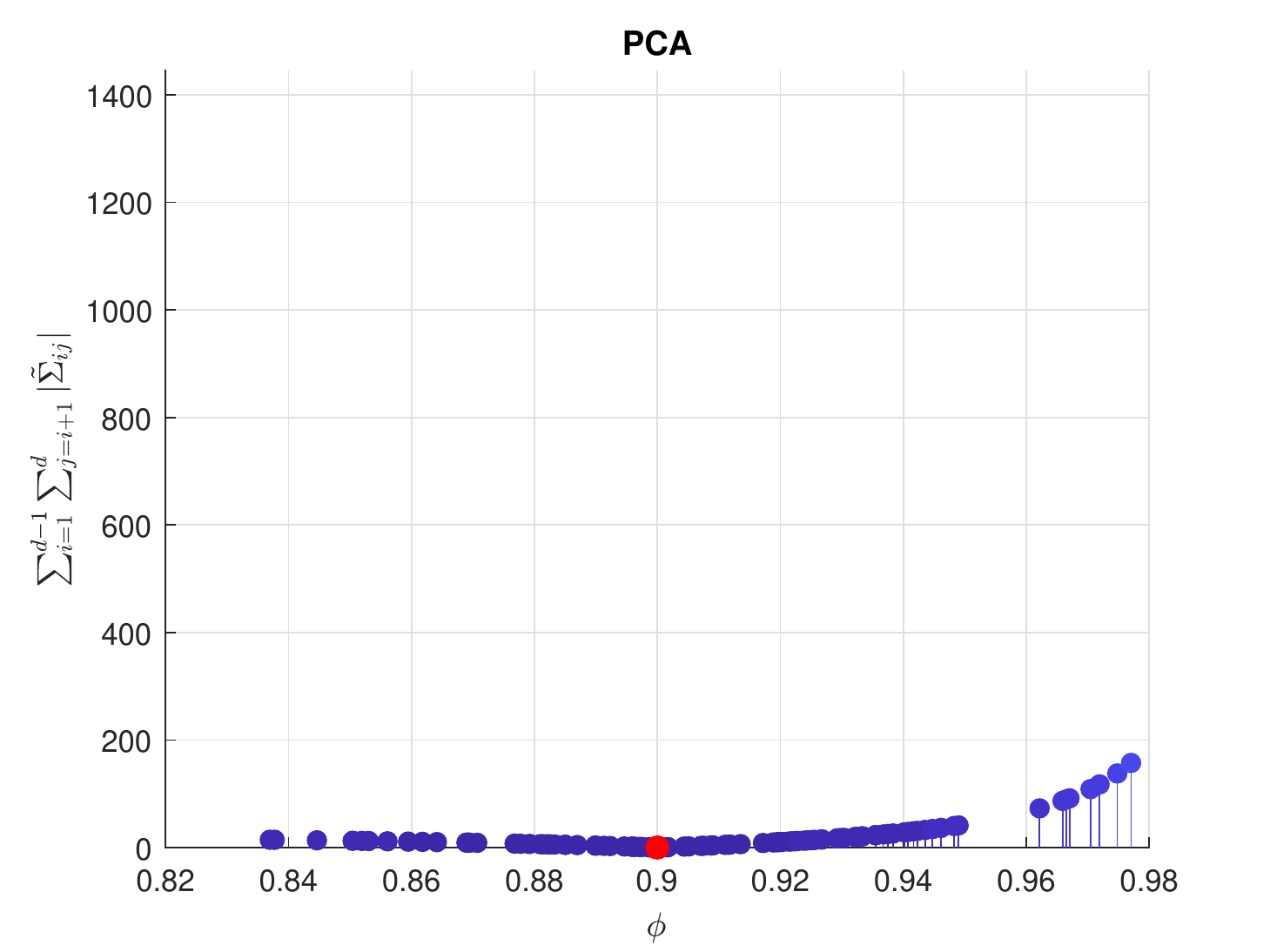}
\end{subfigure}
\begin{subfigure}
\centering\includegraphics[width = 5cm]{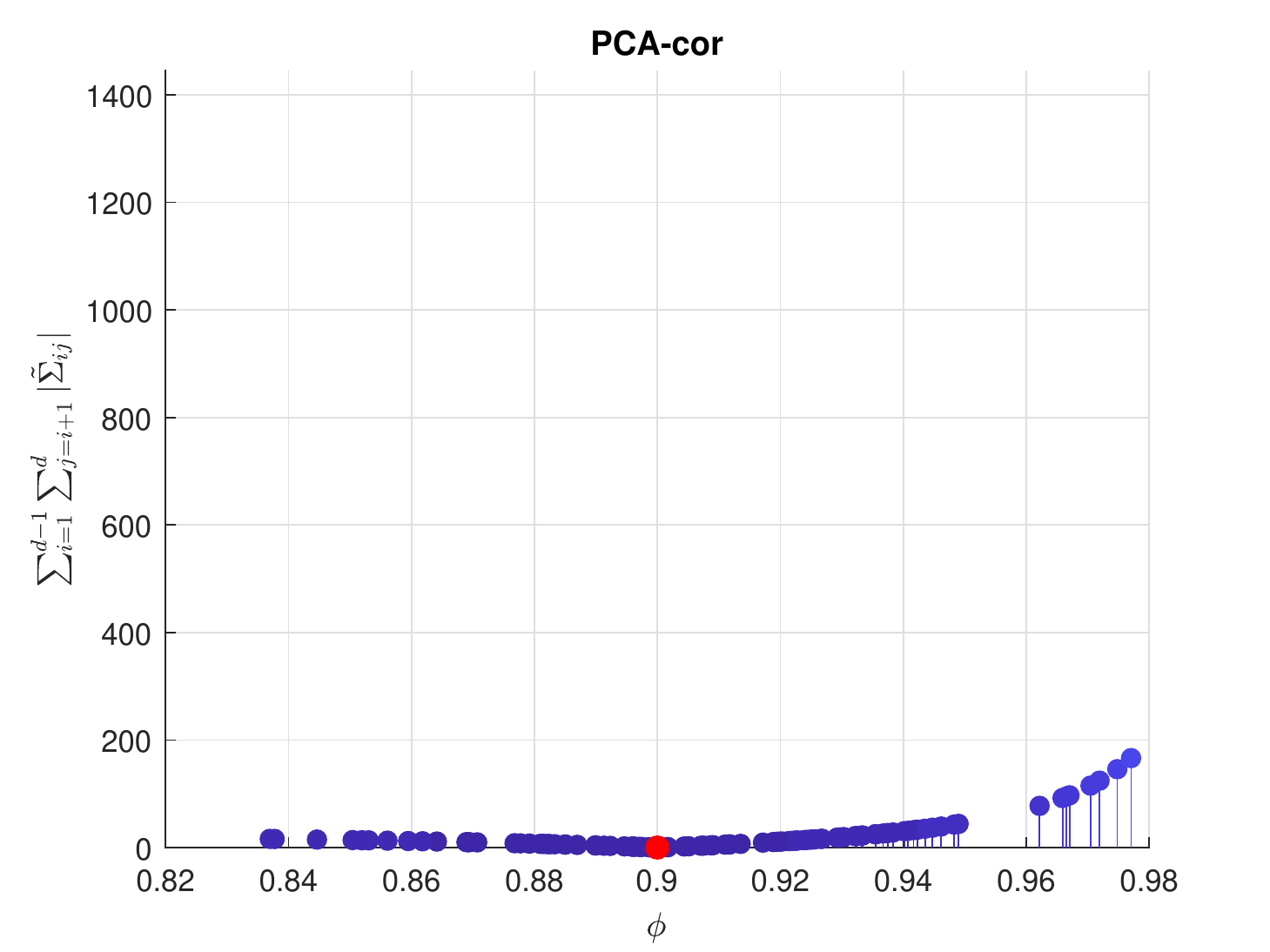}
\end{subfigure}
\begin{subfigure}
\centering\includegraphics[width = 5cm]{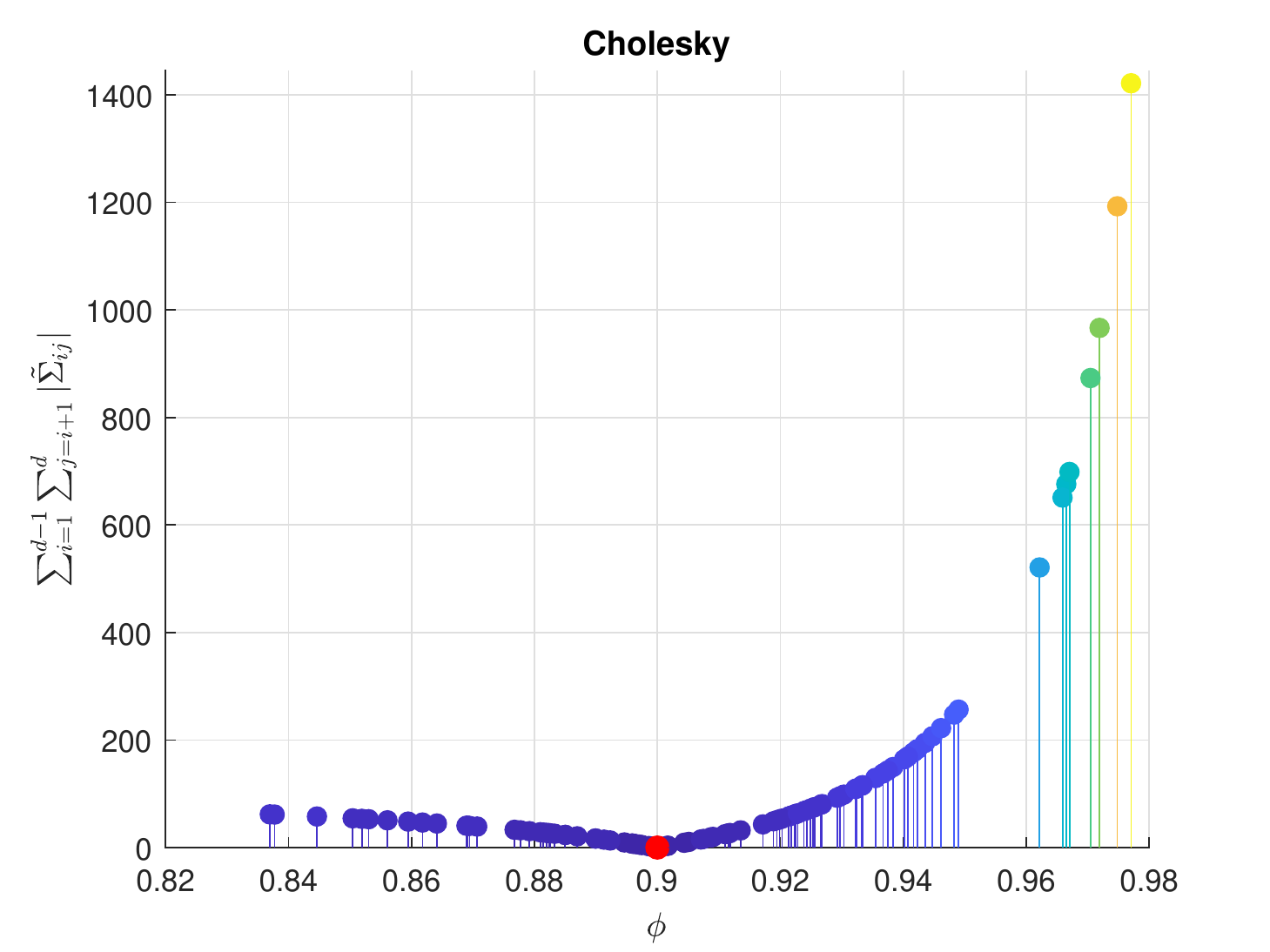}
\end{subfigure}
\begin{subfigure}
\centering\includegraphics[width = 5cm]{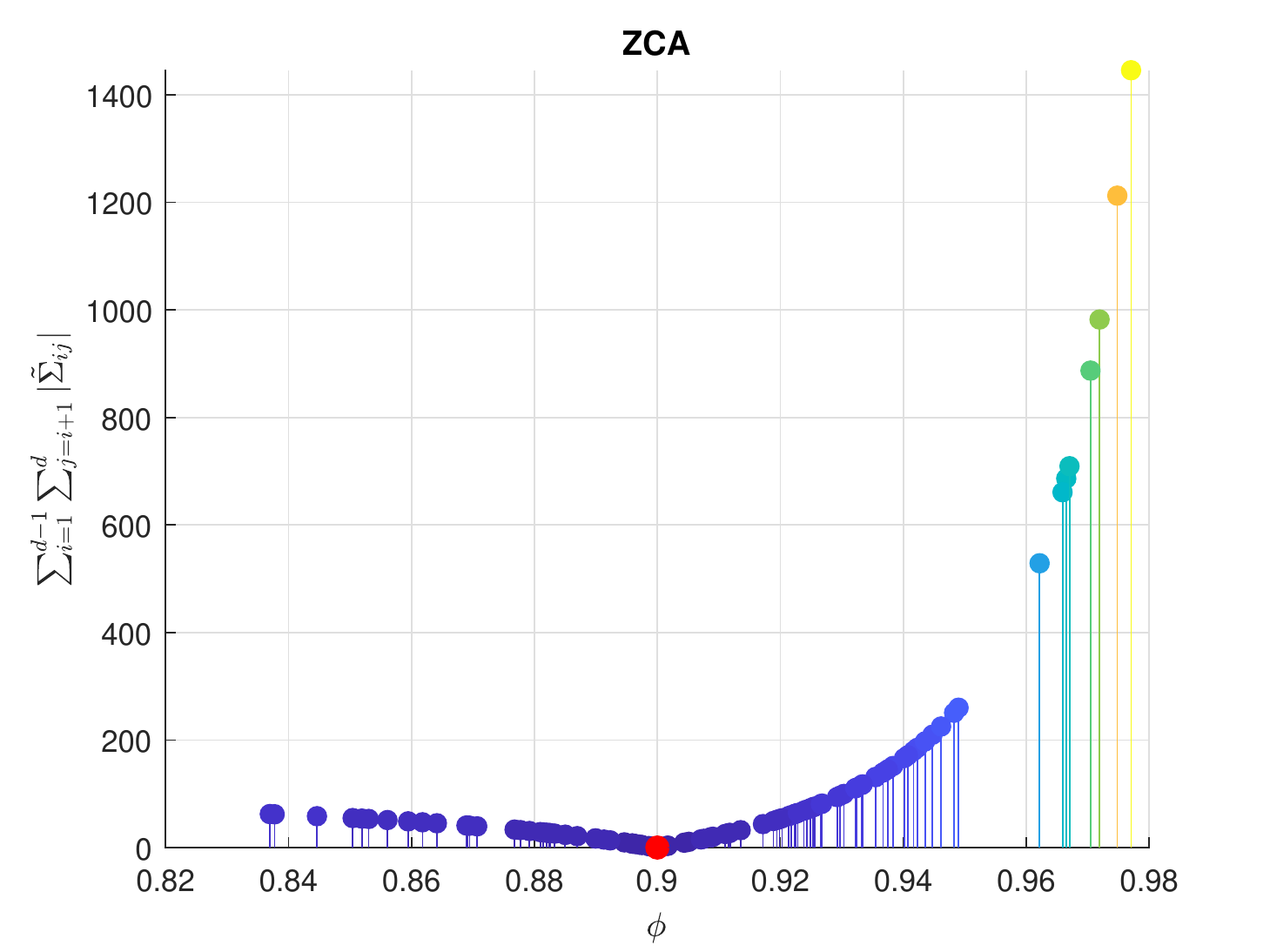}
\end{subfigure}
\begin{subfigure}
\centering\includegraphics[width = 5cm]{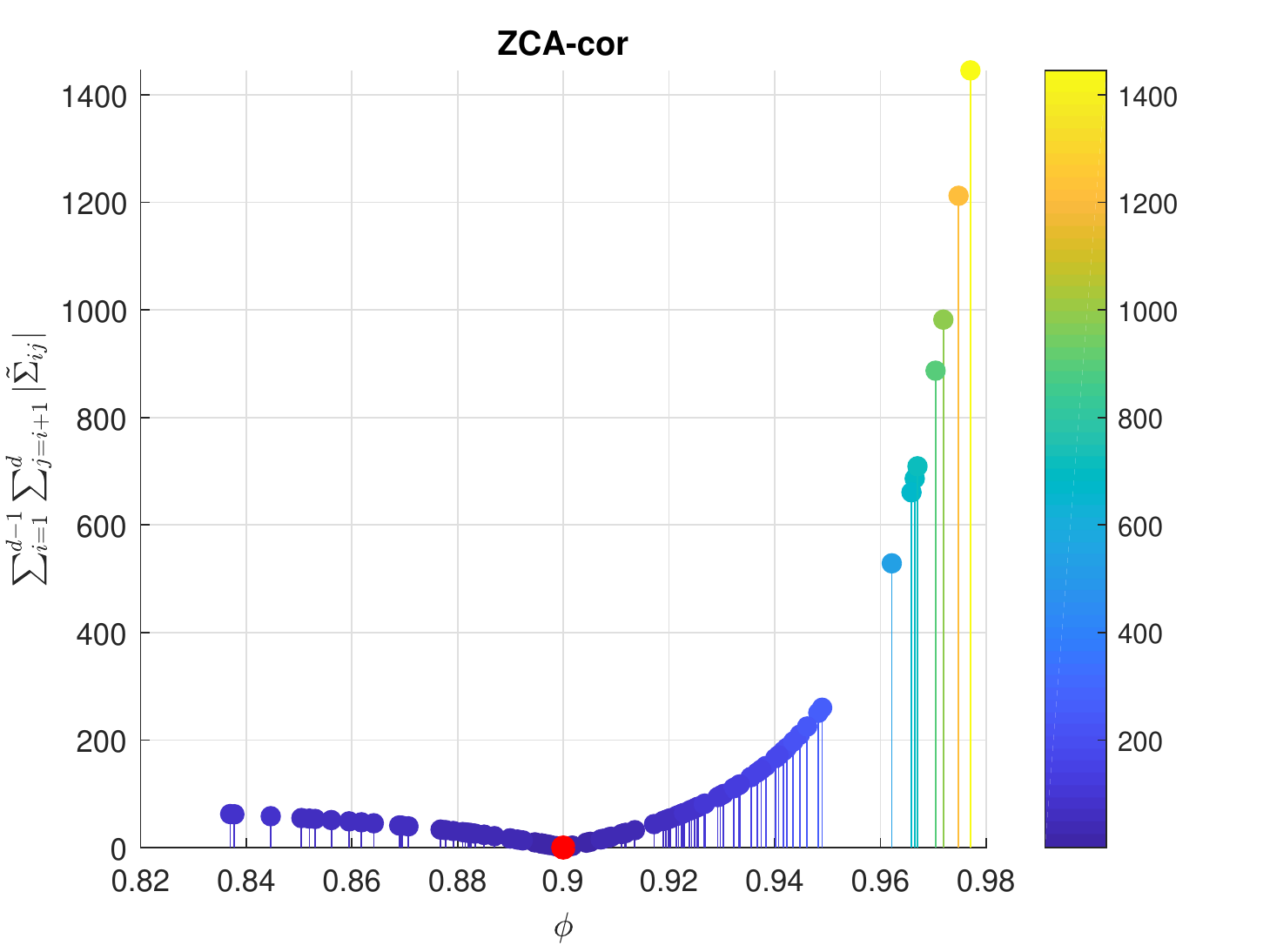}
\end{subfigure}
\caption{
\small 
$L_1$ matrix norm deviation of the upper-triangular (excluding diagonals) elements of $\tilde{\boldsymbol{\Sigma}}(\phi)$ from the zero matrix $\boldsymbol{0}_{d\times d}$, for each of the whitening methods, under the AR(1) model.
Values of $\phi$ are drawn from the true posterior distribution.
Bar height and colour indicate the magnitude of the deviation. The true parameter value $\phi_{\text{true}} = 0.9$ is shown by the red dot.
}
\label{fig:choiceofwhiteningsamplesAR}
\end{figure}

\begin{figure}[h!]
\centering
\begin{subfigure}
\centering\includegraphics[width = 5cm]{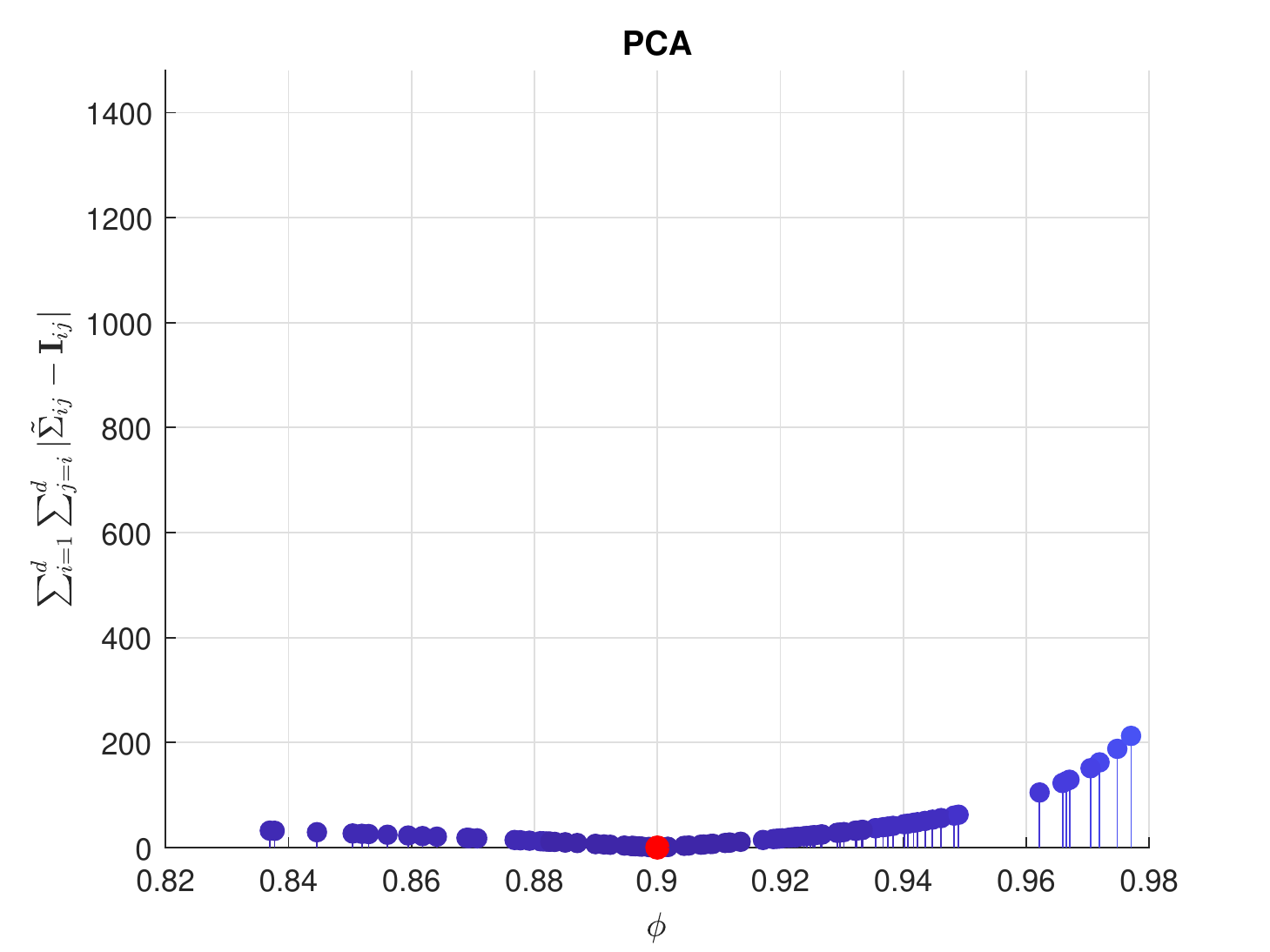}
\end{subfigure}
\begin{subfigure}
\centering\includegraphics[width = 5cm]{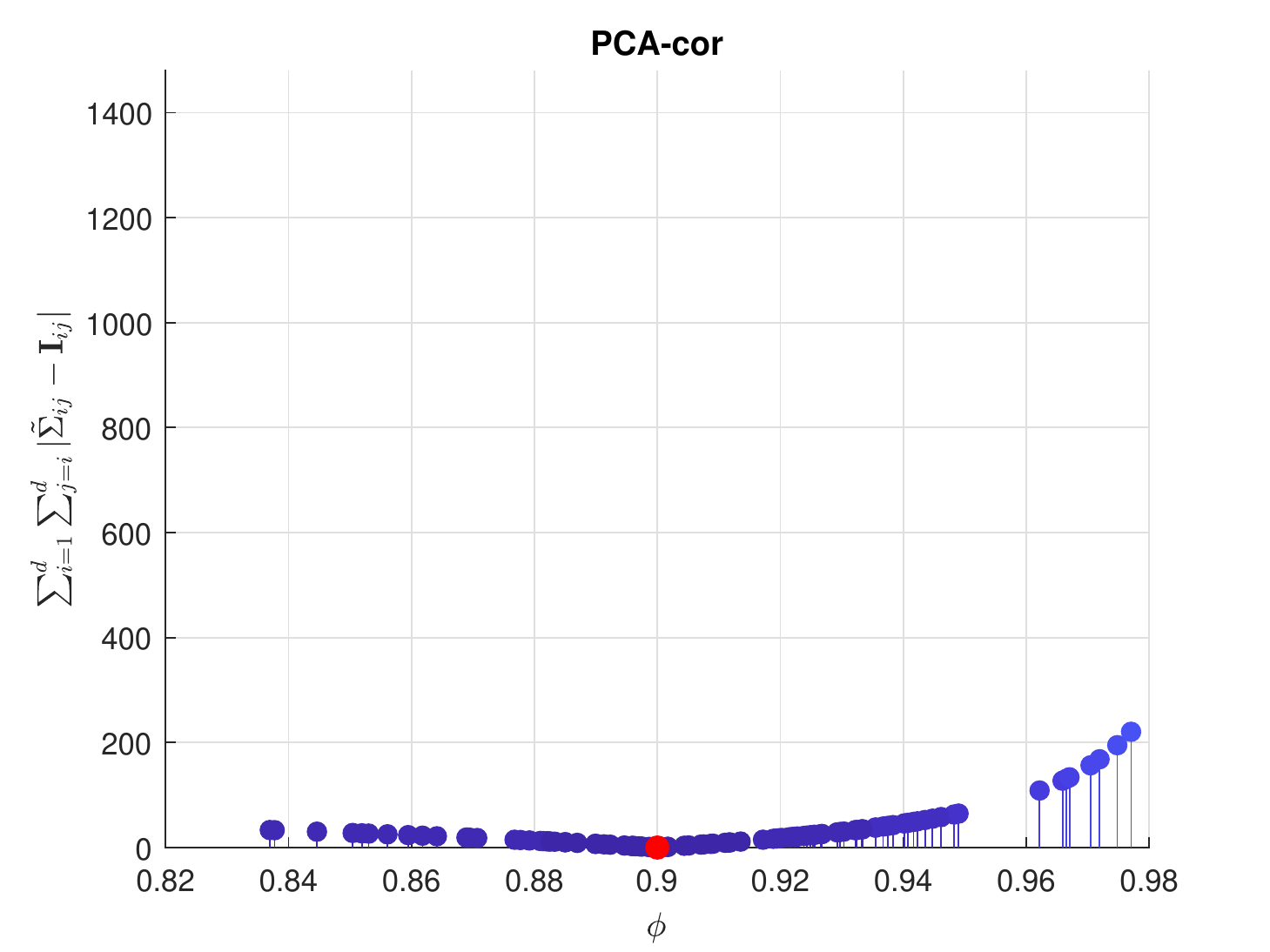}
\end{subfigure}
\begin{subfigure}
\centering\includegraphics[width = 5cm]{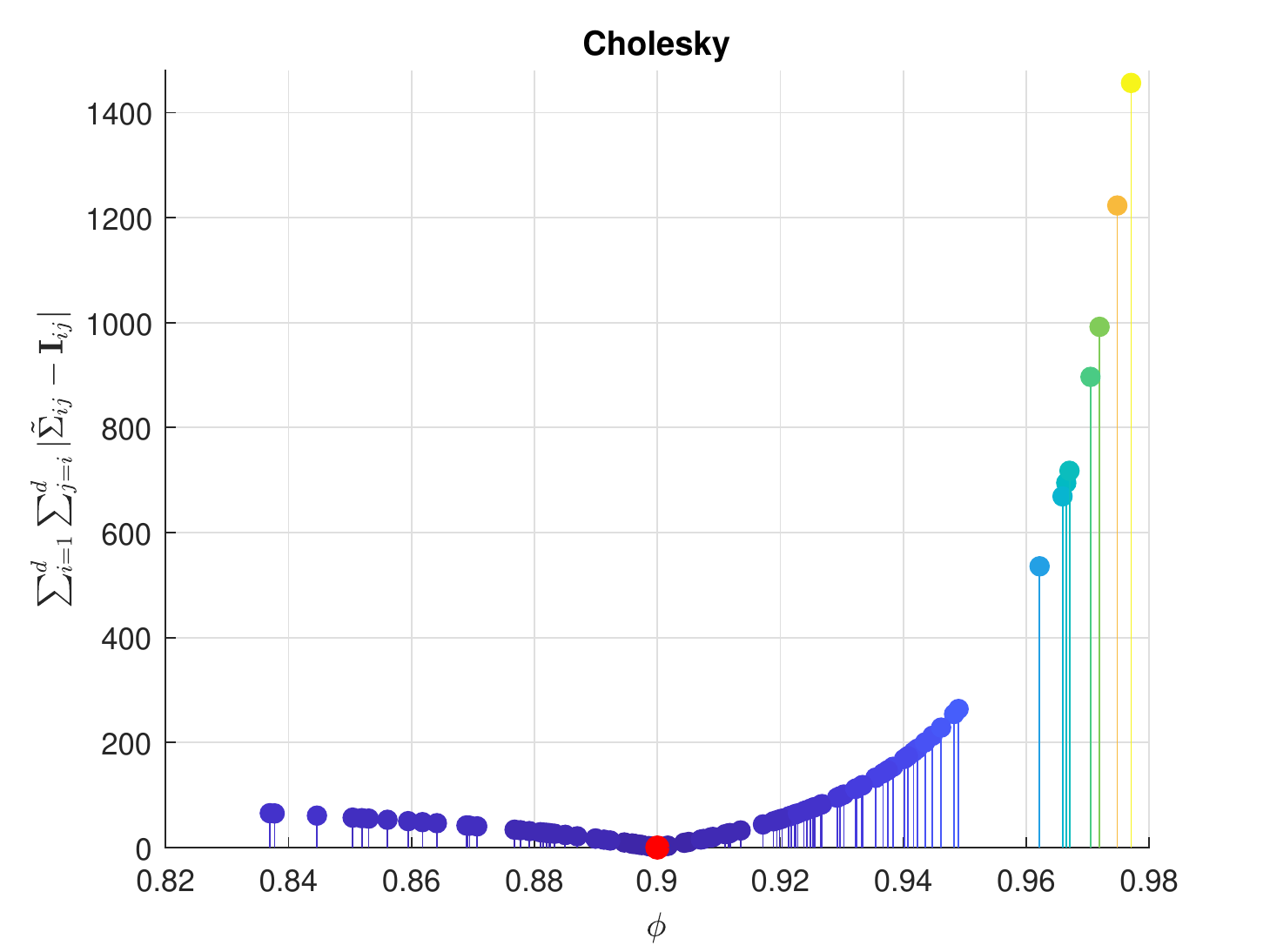}
\end{subfigure}
\begin{subfigure}
\centering\includegraphics[width = 5cm]{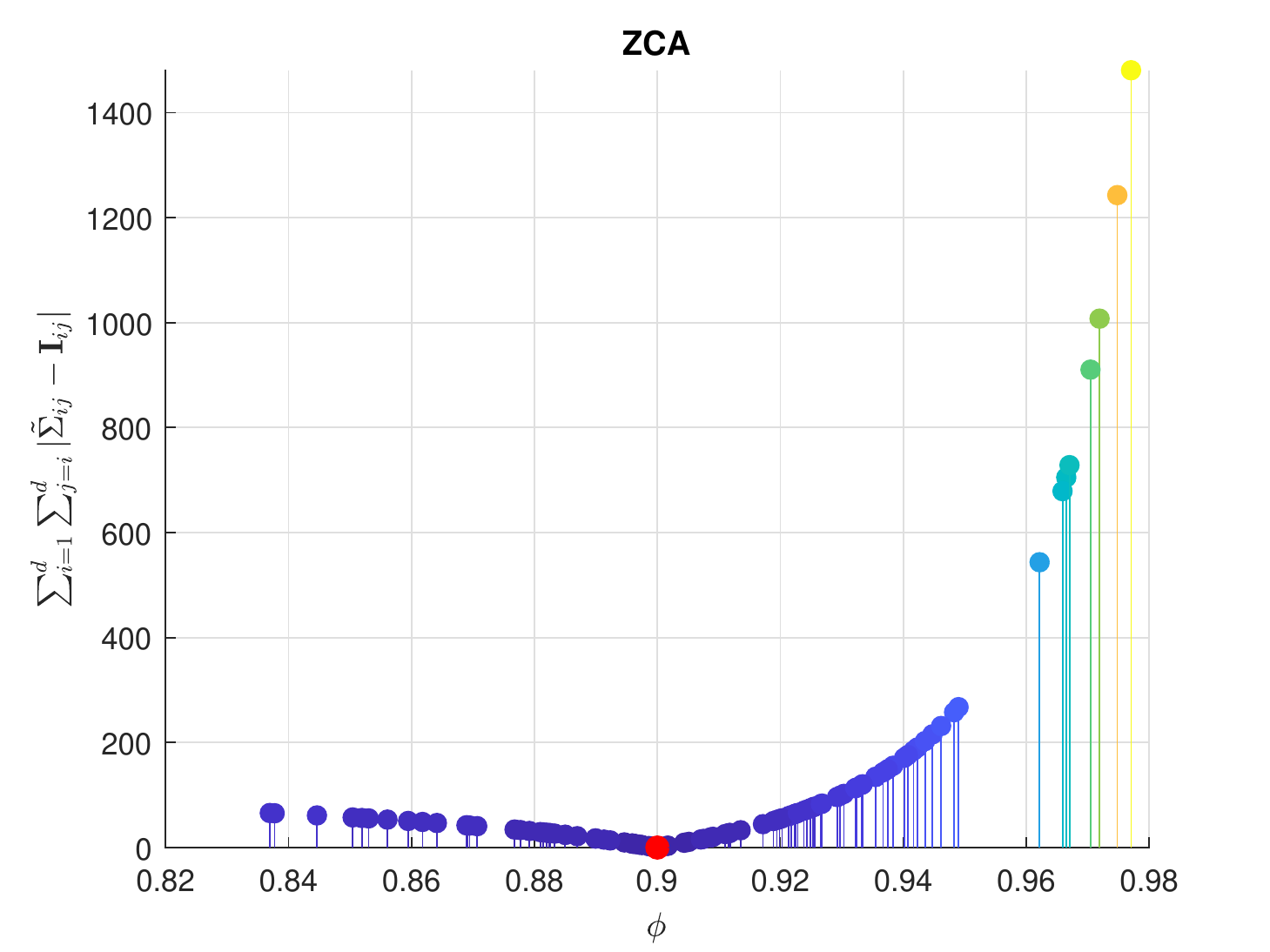}
\end{subfigure}
\begin{subfigure}
\centering\includegraphics[width = 5cm]{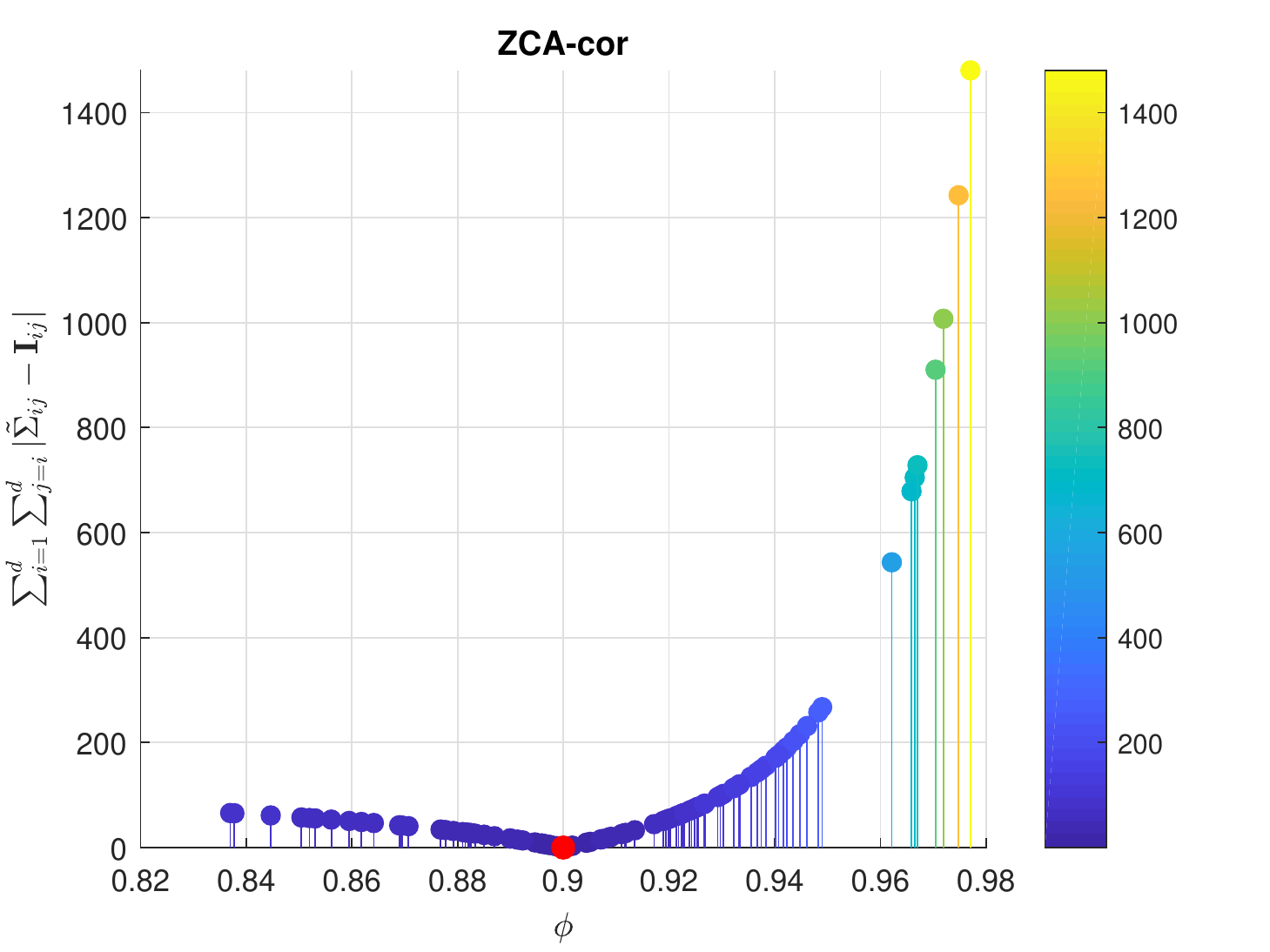}
\end{subfigure}
\caption{\small As Figure \ref{fig:choiceofwhiteningsamplesAR}, but including variance terms in the $L_1$ norm deviation.
%Covariance deviation (including variances) from the identity matrix for transformed summary statistics over parameter values sampled from the MCMC posterior for the AR$(1)$ example. Deviation is measured using the $L_1$ matrix norm. Bar colour represents the error, and the 'true' parameter value $\phi = 0.9$ (where $\boldsymbol{W}$ is estimated) is shown in red.
}
\label{fig:choiceofwhiteningsampleswithdiagsAR}
\end{figure}

\begin{figure}[h!]
\centering
\begin{subfigure}
\centering\includegraphics[width = 5cm]{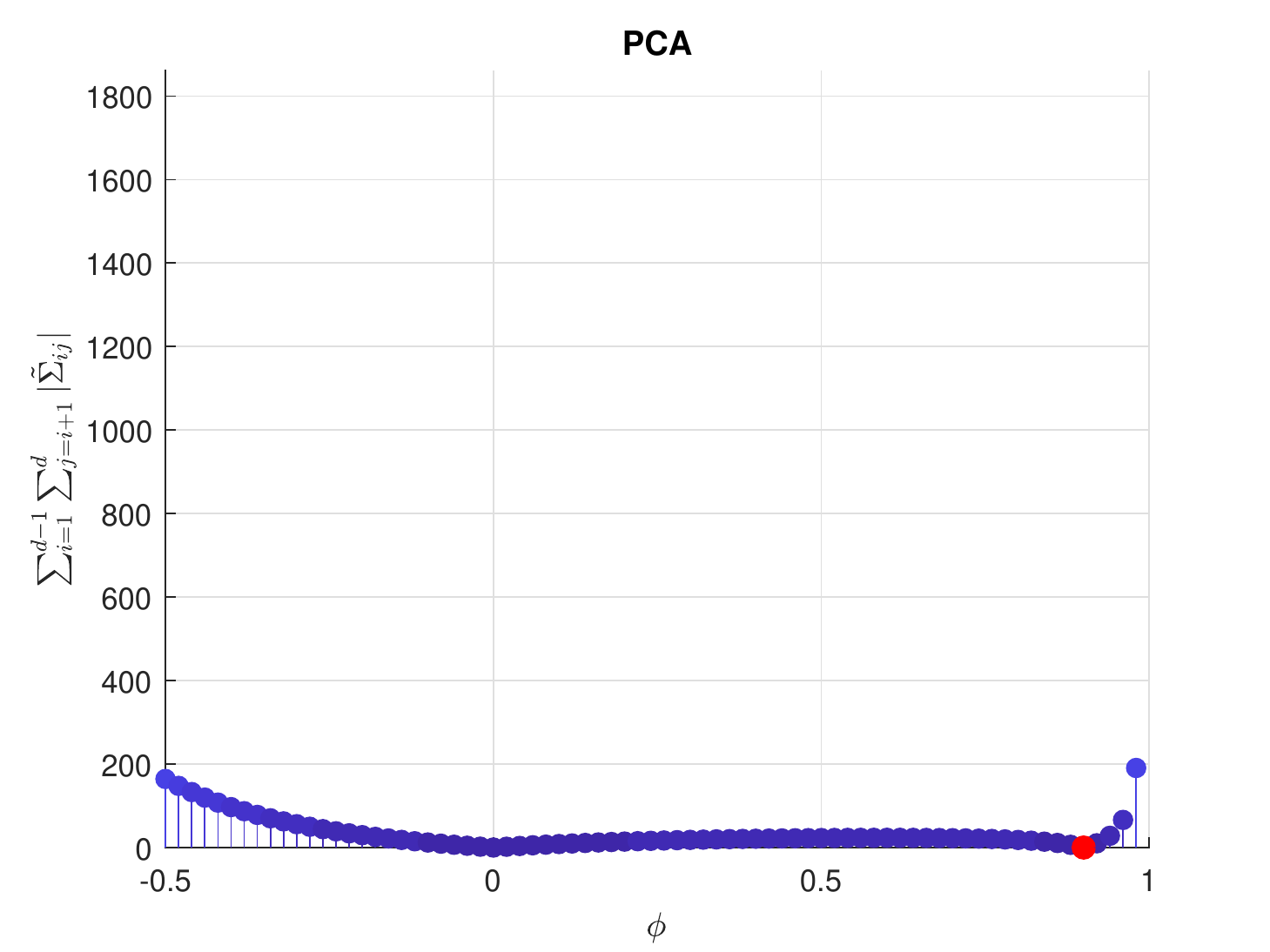}
\end{subfigure}
\begin{subfigure}
\centering\includegraphics[width = 5cm]{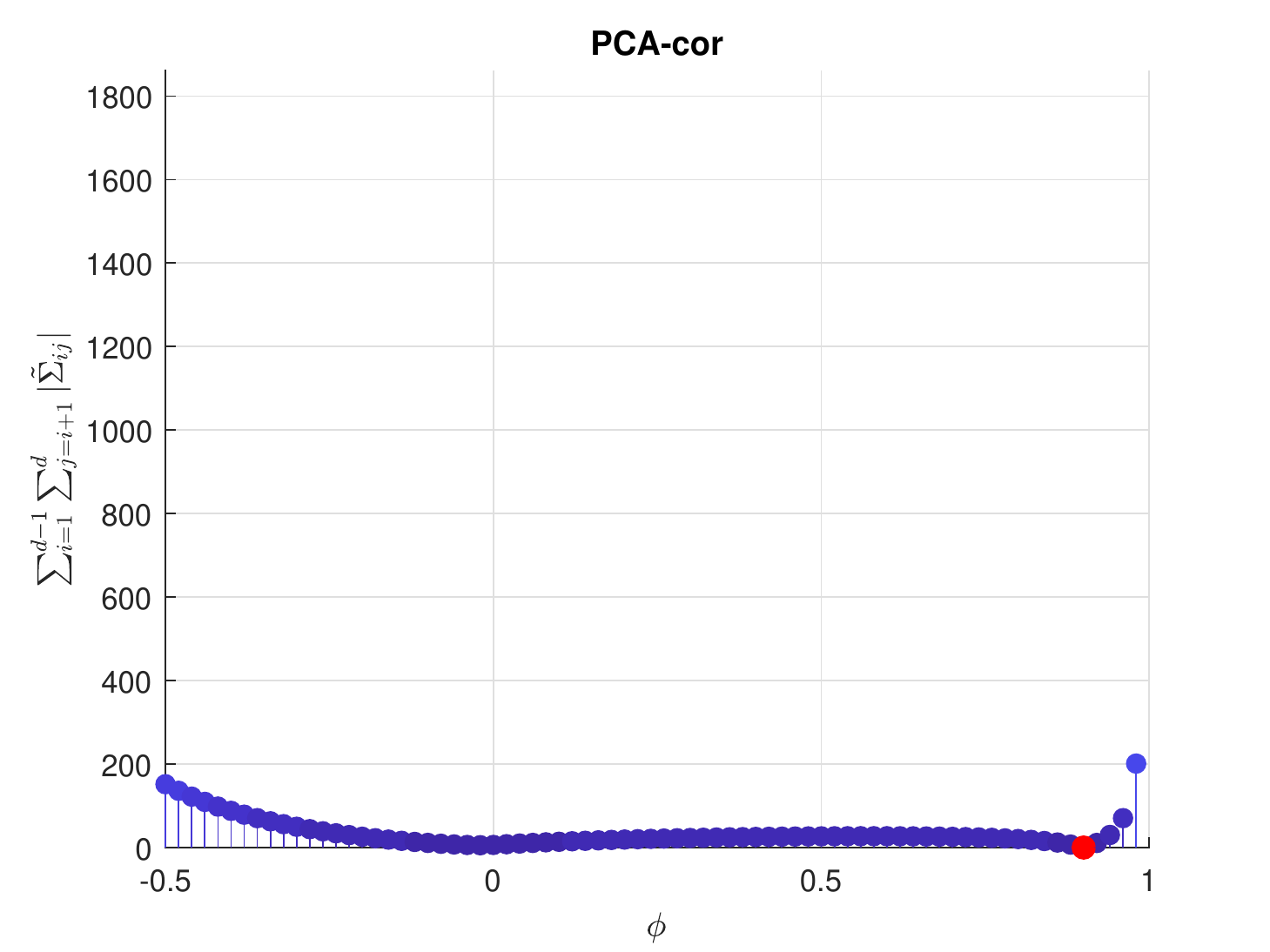}
\end{subfigure}
\begin{subfigure}
\centering\includegraphics[width = 5cm]{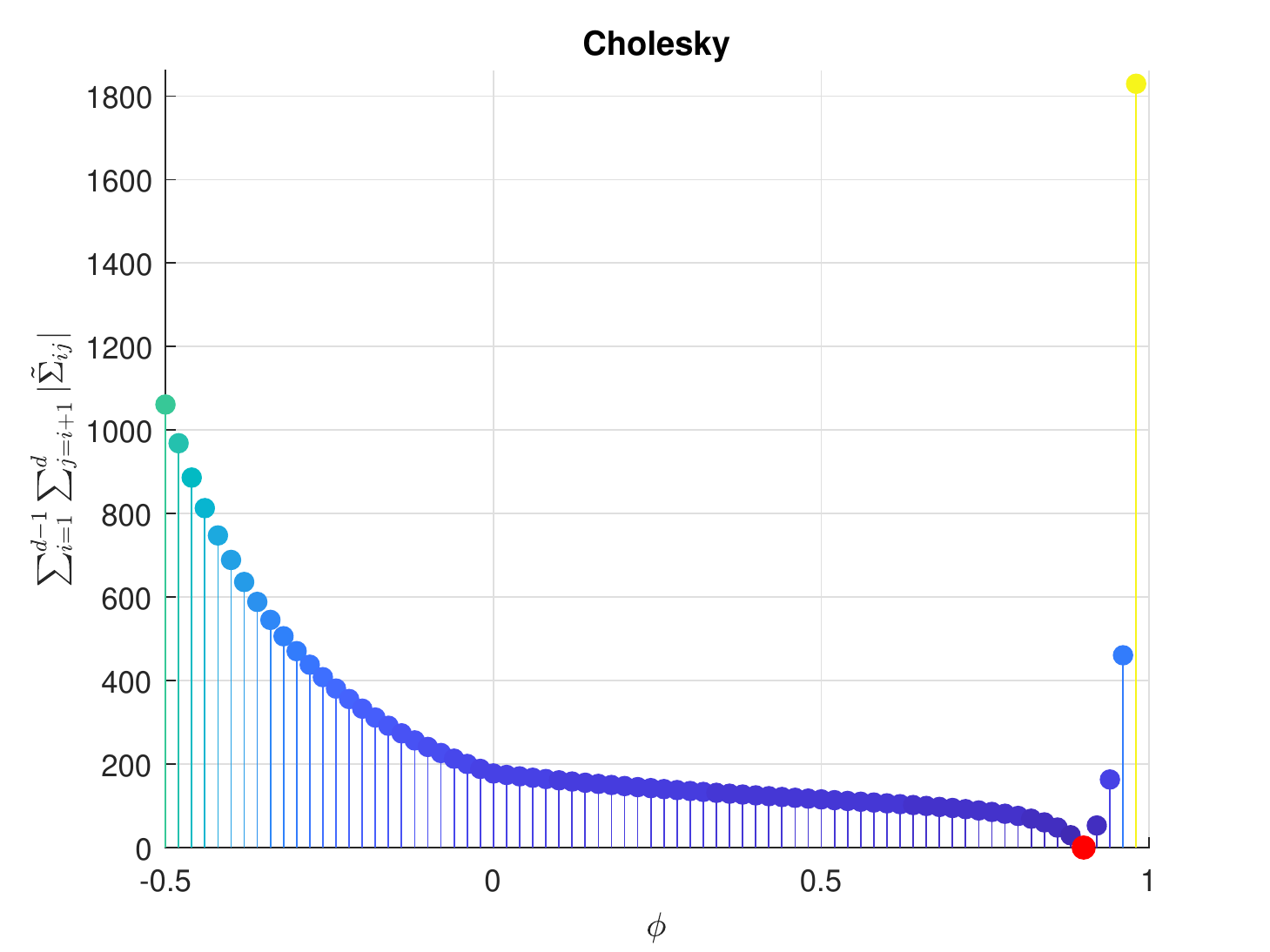}
\end{subfigure}
\begin{subfigure}
\centering\includegraphics[width = 5cm]{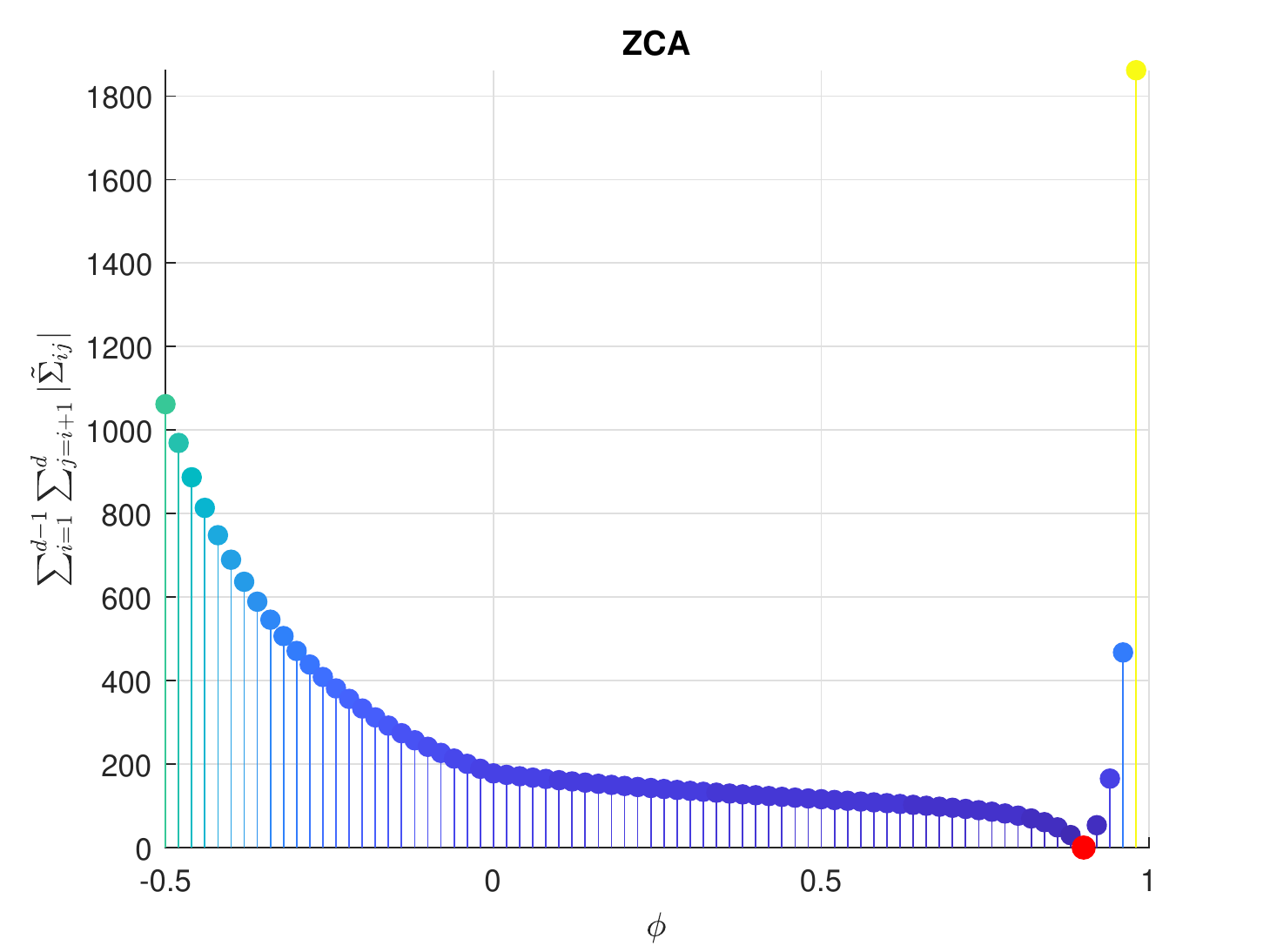}
\end{subfigure}
\begin{subfigure}
\centering\includegraphics[width = 5cm]{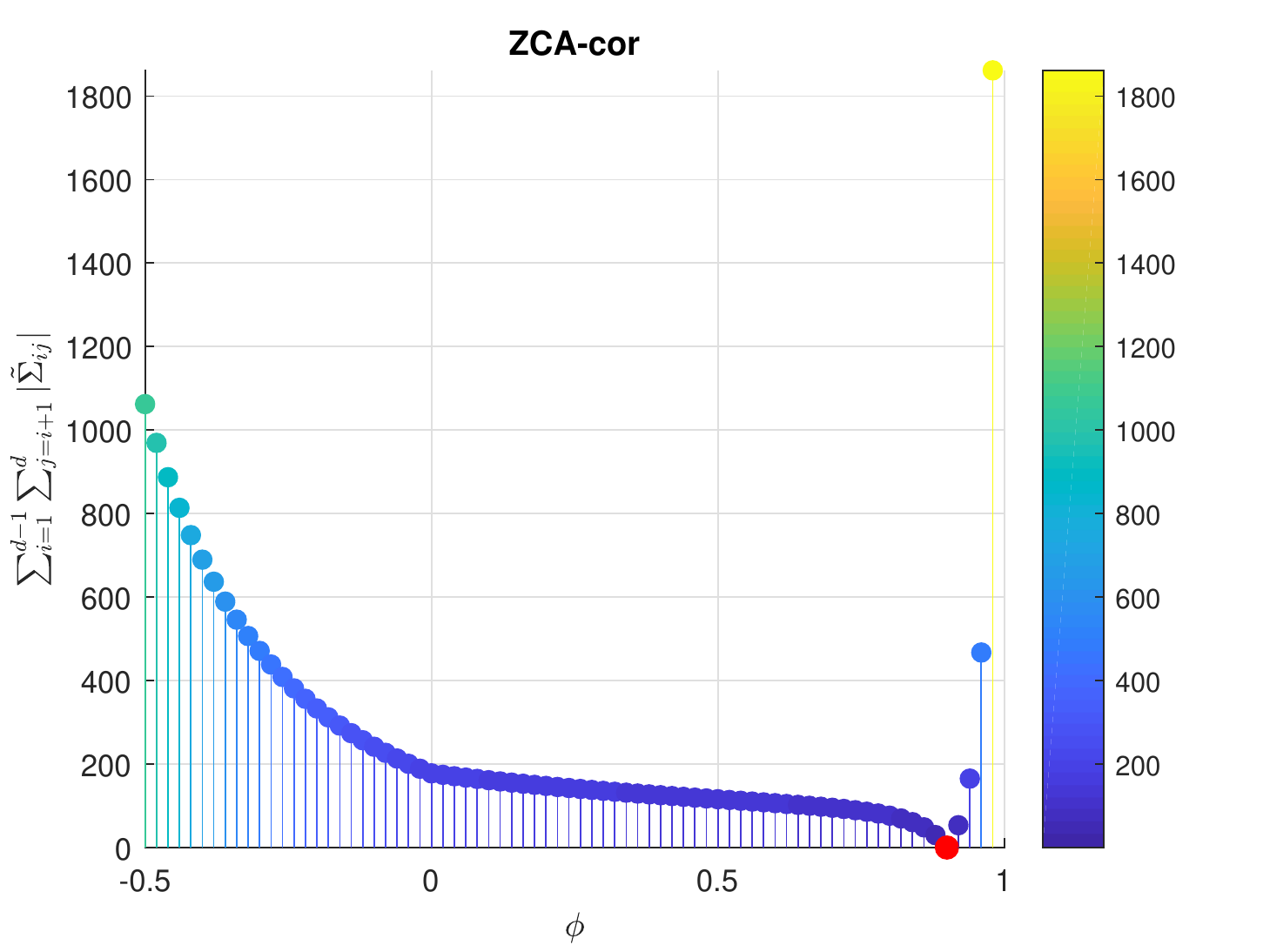}
\end{subfigure}
\caption{As Figure \ref{fig:choiceofwhiteningsamplesAR}, but covering the full parameter space.
%Covariance deviation (without variances) from the zero matrix for the whitened summary statistics over the parameter space for the AR$(1)$ example. The results for $-1<\phi<-0.5$ are not shown, but the error continues to increase as $\phi$ decreases. Deviation is measured using the $L_1$ matrix norm. Bar colour represents the error and the `true' parameter value $\phi = 0.9$ is shown in red.
}
\label{fig:choiceofwhiteningparamspaceAR}
\end{figure}

\clearpage
\subsection{Sensitivity to the value of $\boldsymbol{\theta}^0$}
\subsubsection{An MA(2) model}
%\section*{Appendix C}
\begin{figure}[h!]
\centering
\begin{subfigure}
\centering\includegraphics[width = 11cm]{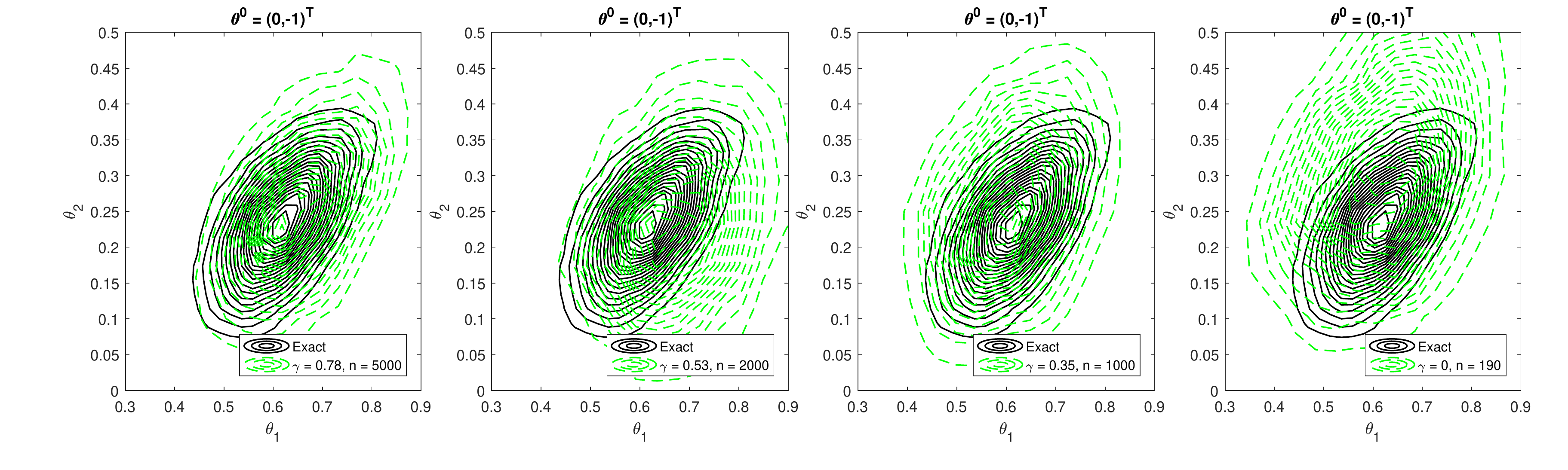}
\end{subfigure}

\begin{subfigure}
\centering\includegraphics[width = 11cm]{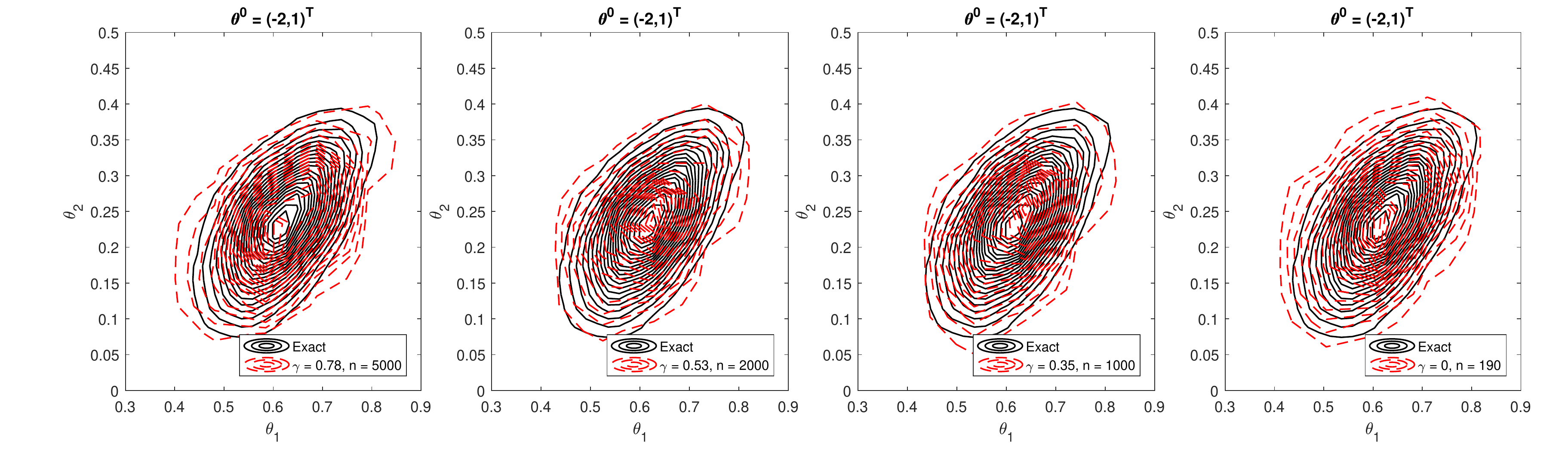}
\end{subfigure}

\begin{subfigure}
\centering\includegraphics[width = 11cm]{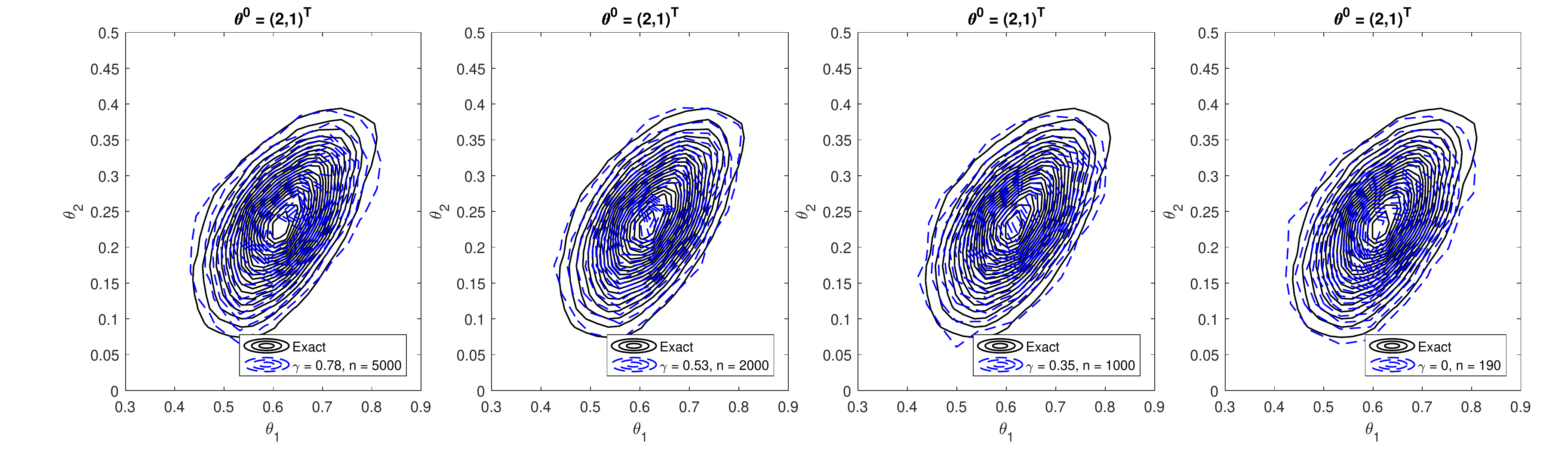}
\end{subfigure}

\begin{subfigure}
\centering\includegraphics[width = 11cm]{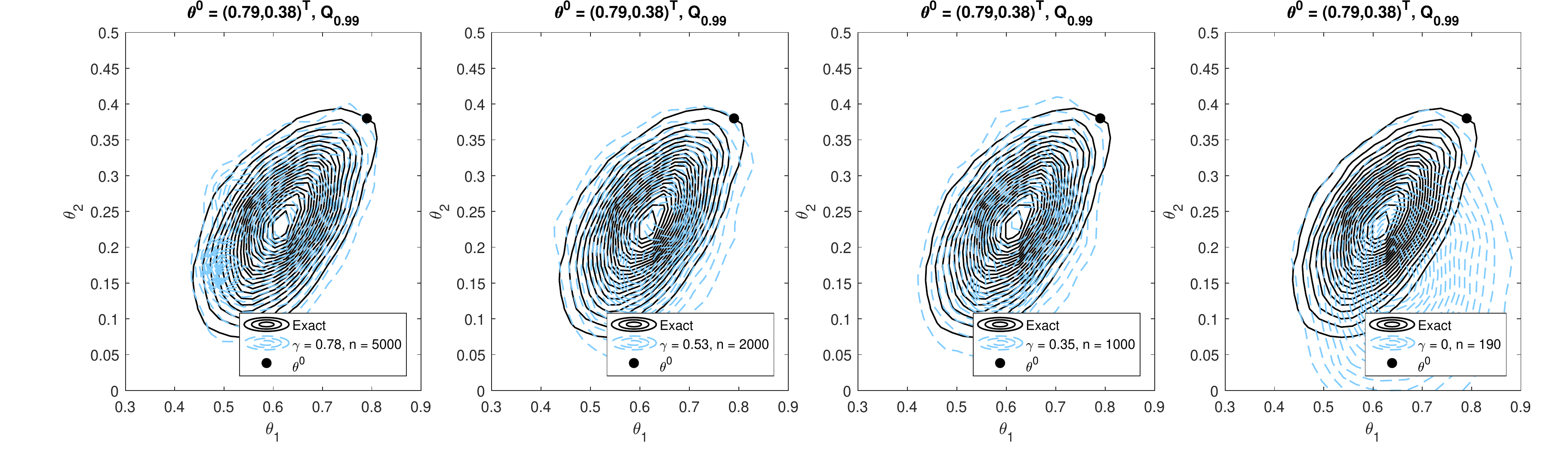}
\end{subfigure}

\begin{subfigure}
\centering\includegraphics[width = 11cm]{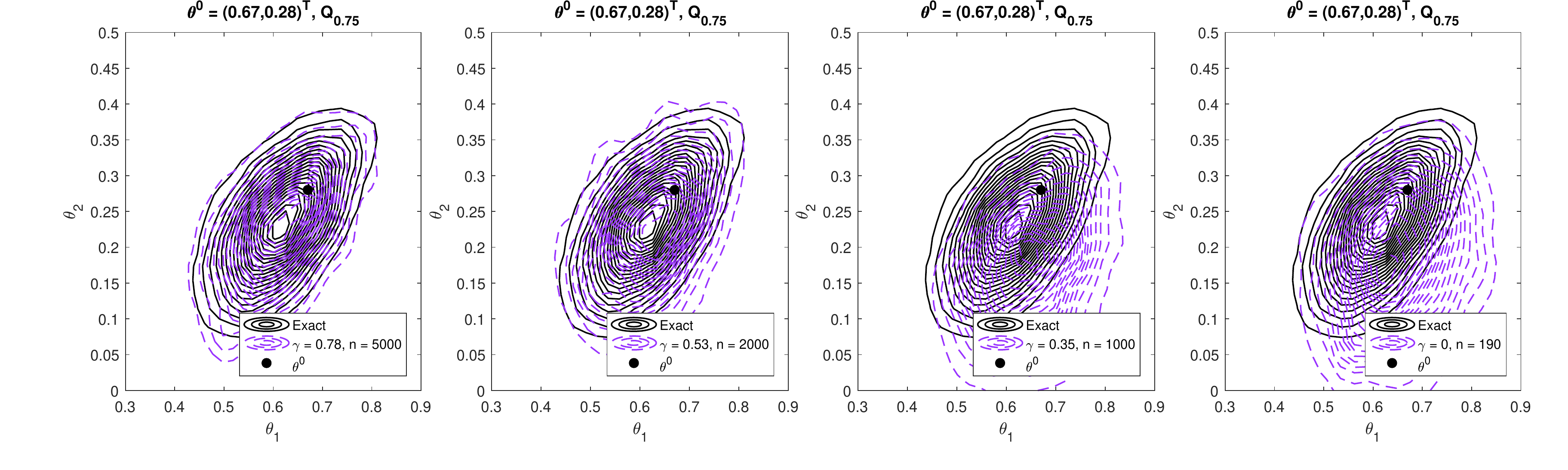}
\end{subfigure}

\caption{\small As Figure 5 in the main text, but for PCA-cor whitening.
%Sensitivity of wBSL posterior with PCA-cor whitening to point estimate for MA(2) example.
}
\label{fig:sensMA2pcacor2}
\end{figure}

\begin{figure}[h!]
\centering
\begin{subfigure}
\centering\includegraphics[width = 11cm]{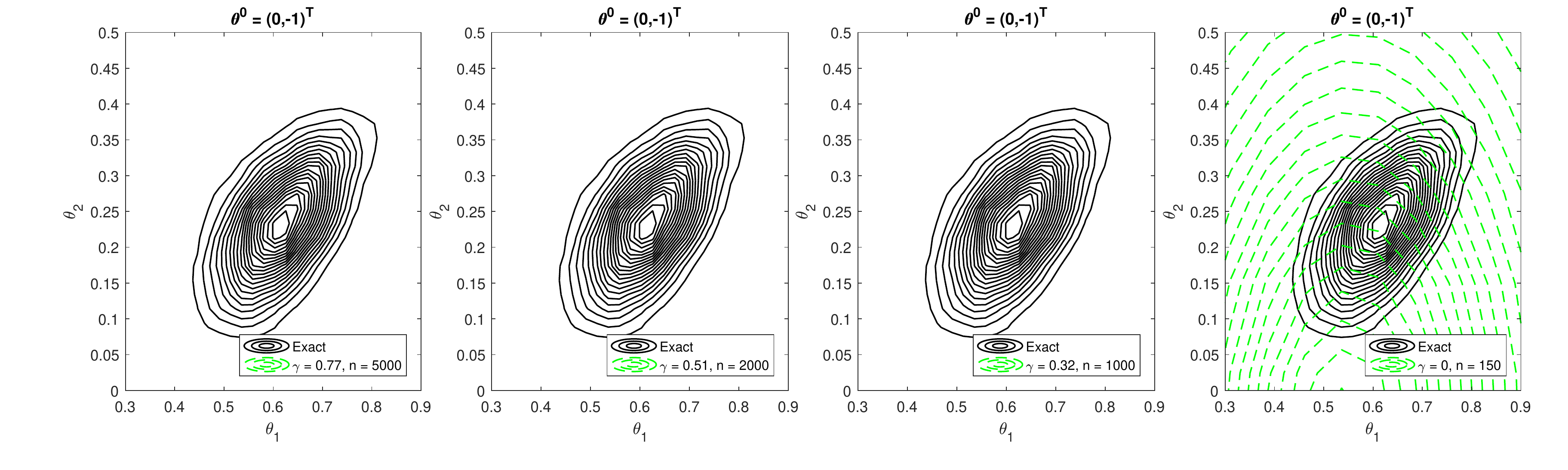}
\end{subfigure}

\begin{subfigure}
\centering\includegraphics[width = 11cm]{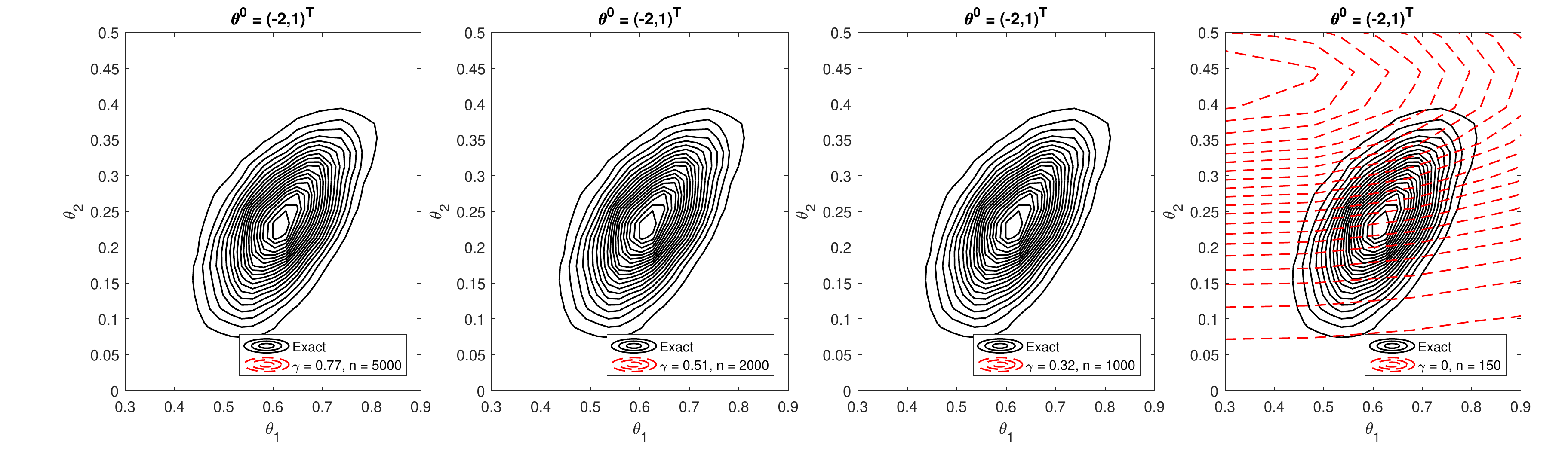}
\end{subfigure}

\begin{subfigure}
\centering\includegraphics[width = 11cm]{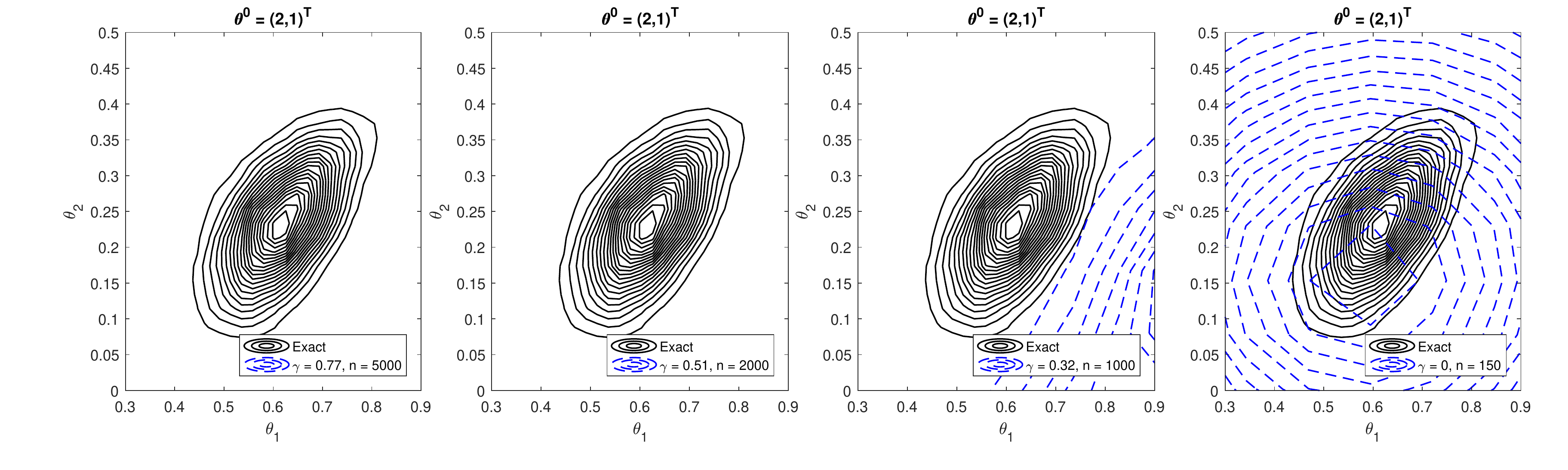}
\end{subfigure}

\begin{subfigure}
\centering\includegraphics[width = 11cm]{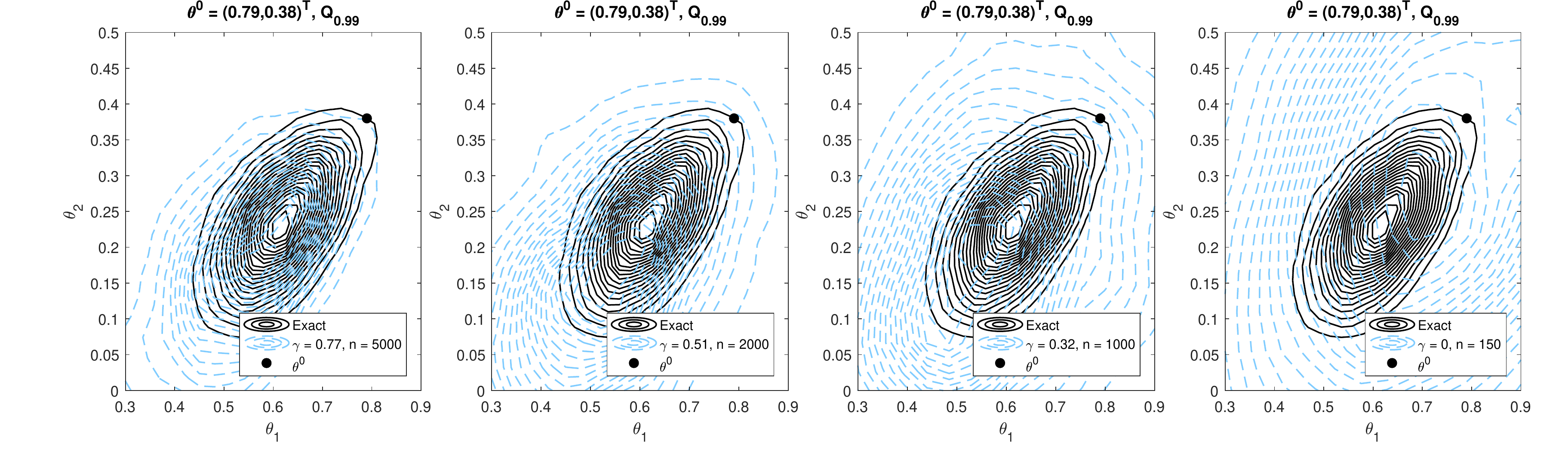}
\end{subfigure}

\begin{subfigure}
\centering\includegraphics[width = 11cm]{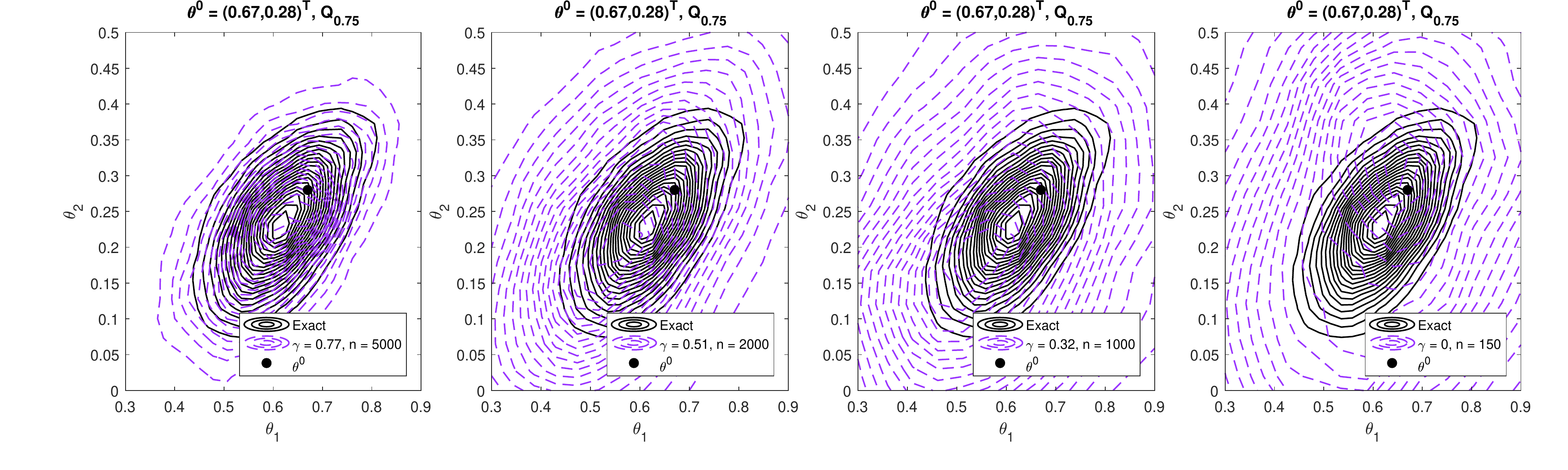}
\end{subfigure}

\caption{%Sensitivity of wBSL posterior with ZCA whitening to point estimate for MA(2) example. 
\small As Figure 5 in the main text, but for ZCA whitening. Note some panels do not show any posterior support within the plotted bounds.
}
\label{fig:sensMA2pcacor3}
\end{figure}

\begin{figure}[h!]
\centering
\begin{subfigure}
\centering\includegraphics[width = 11cm]{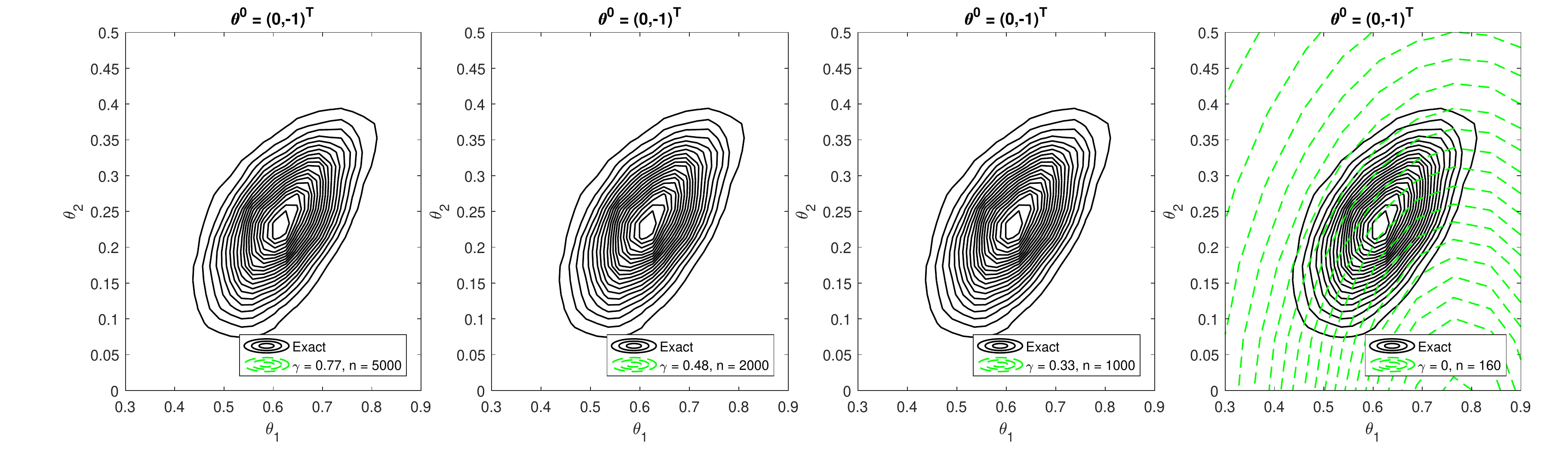}
\end{subfigure}

\begin{subfigure}
\centering\includegraphics[width = 11cm]{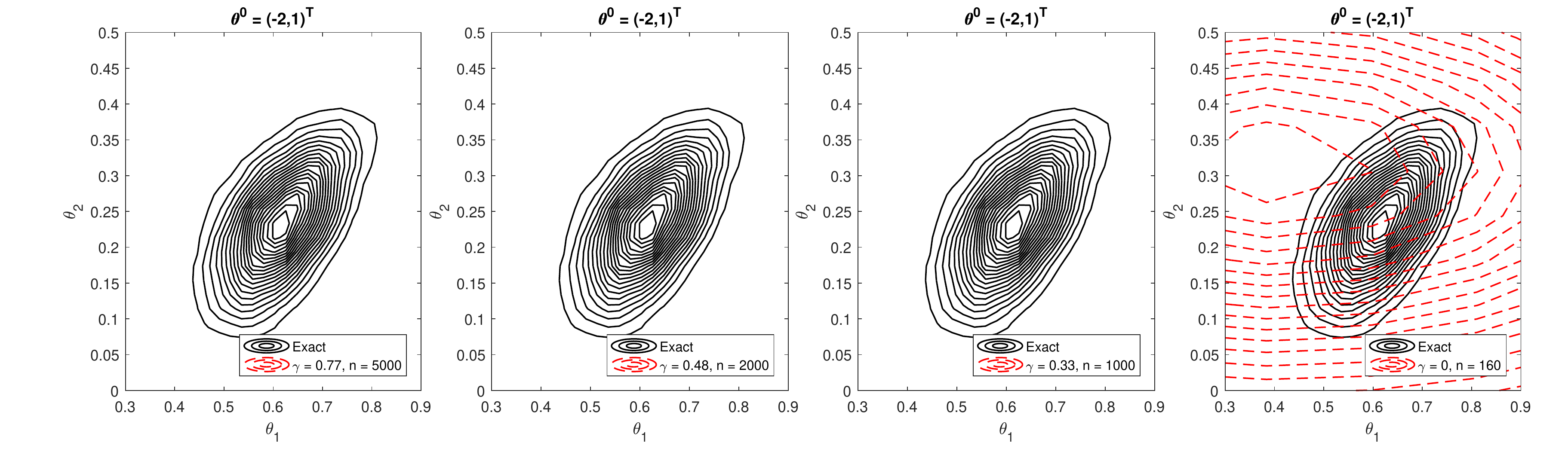}
\end{subfigure}

\begin{subfigure}
\centering\includegraphics[width = 11cm]{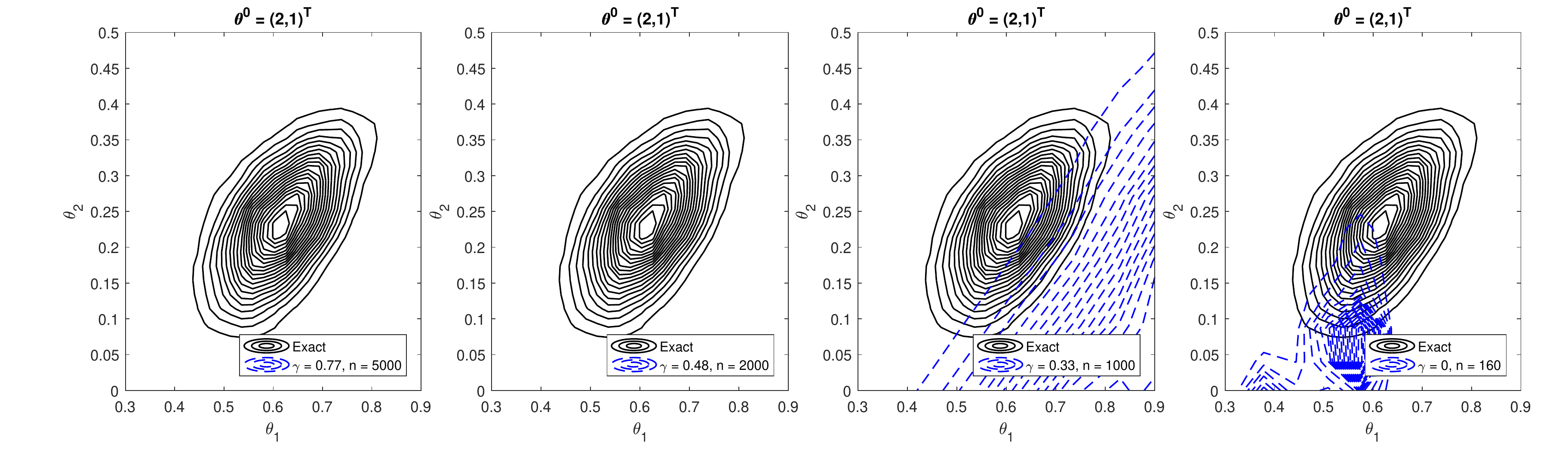}
\end{subfigure}

\begin{subfigure}
\centering\includegraphics[width = 11cm]{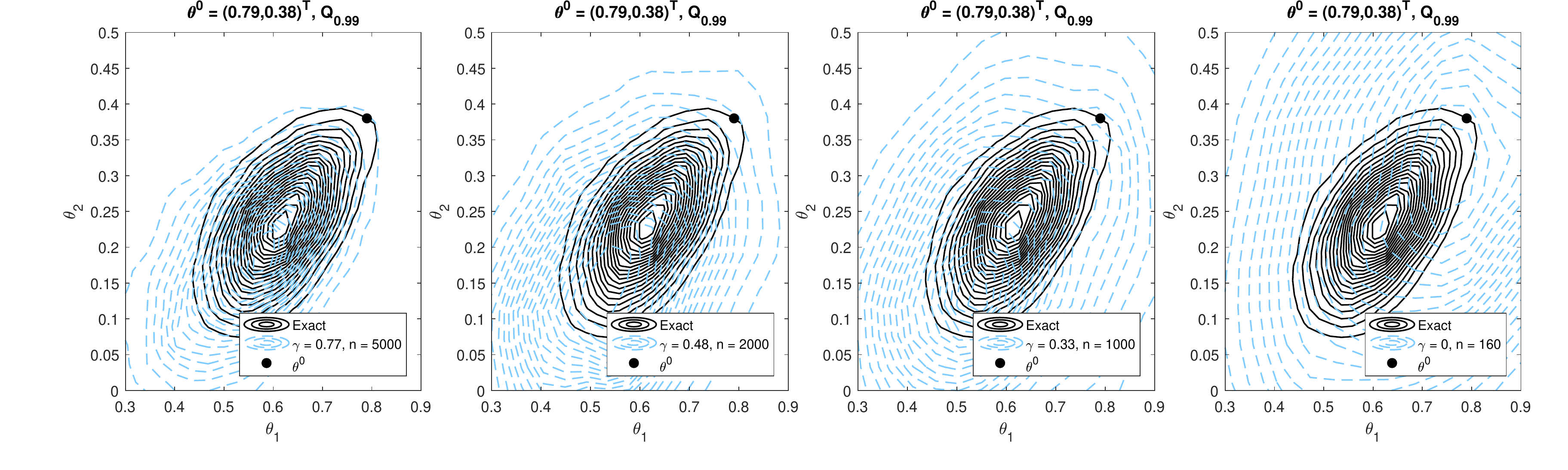}
\end{subfigure}

\begin{subfigure}
\centering\includegraphics[width = 11cm]{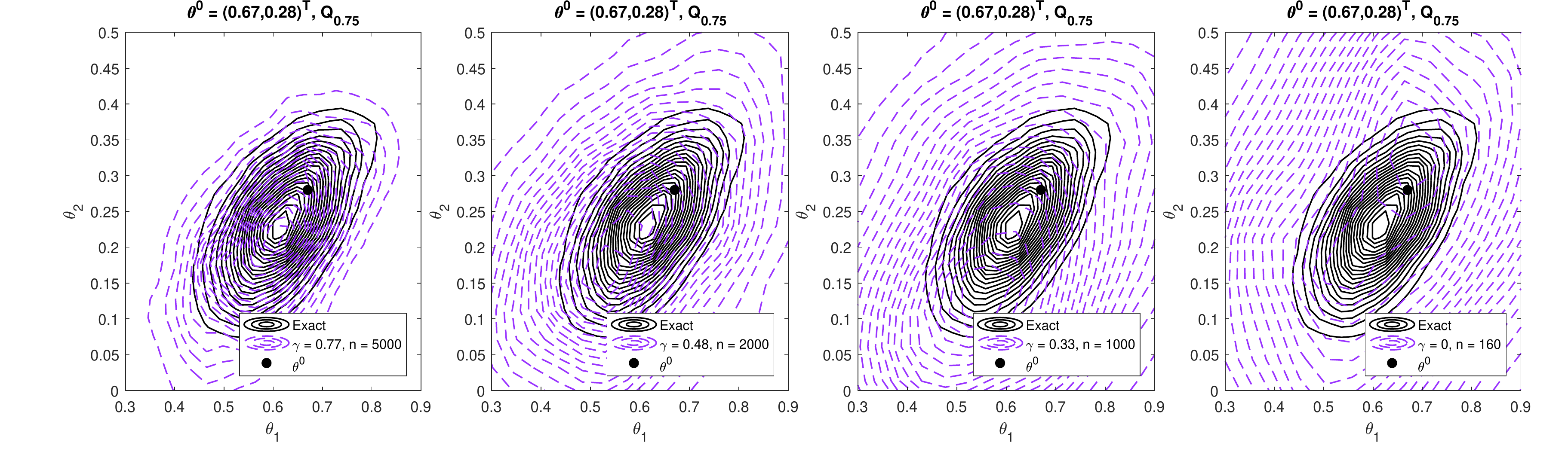}
\end{subfigure}

\caption{\small As Figure 5 in the main text, but for ZCA-cor whitening. Note some panels do not show any posterior support within the plotted bounds.
%Sensitivity of wBSL posterior with ZCA-cor whitening to point estimate for MA(2) example. Note some results do not show any posterior support within the plotted bounds.
}
\label{fig:sensMA2pcacor4}
\end{figure}

\begin{figure}[h!]
\centering
\begin{subfigure}
\centering\includegraphics[width = 11cm]{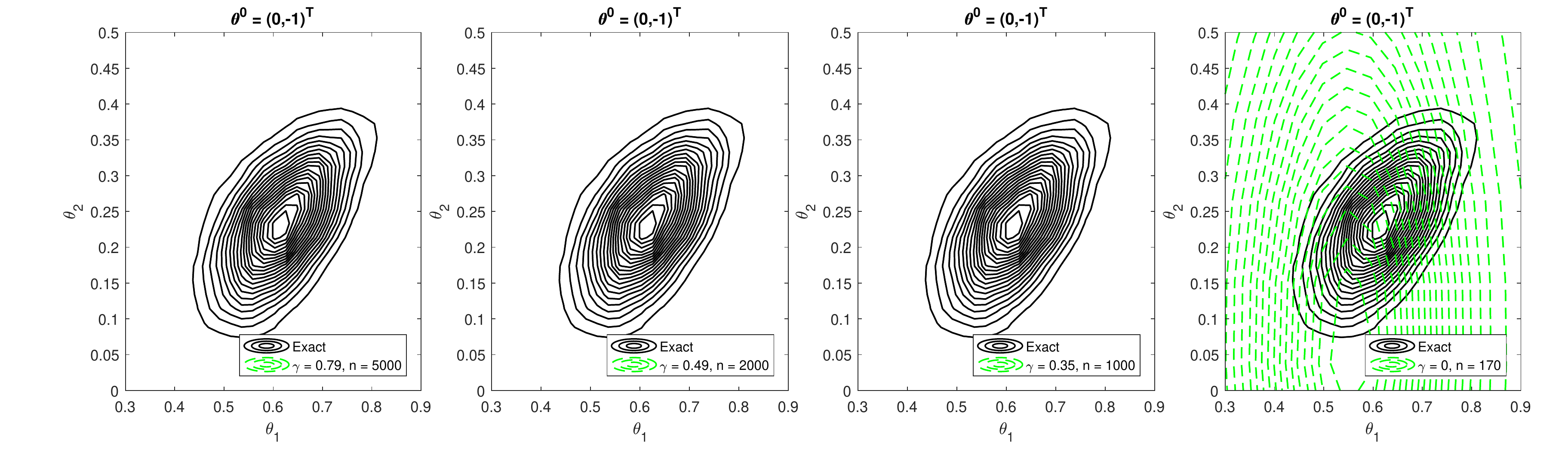}
\end{subfigure}

\begin{subfigure}
\centering\includegraphics[width = 11cm]{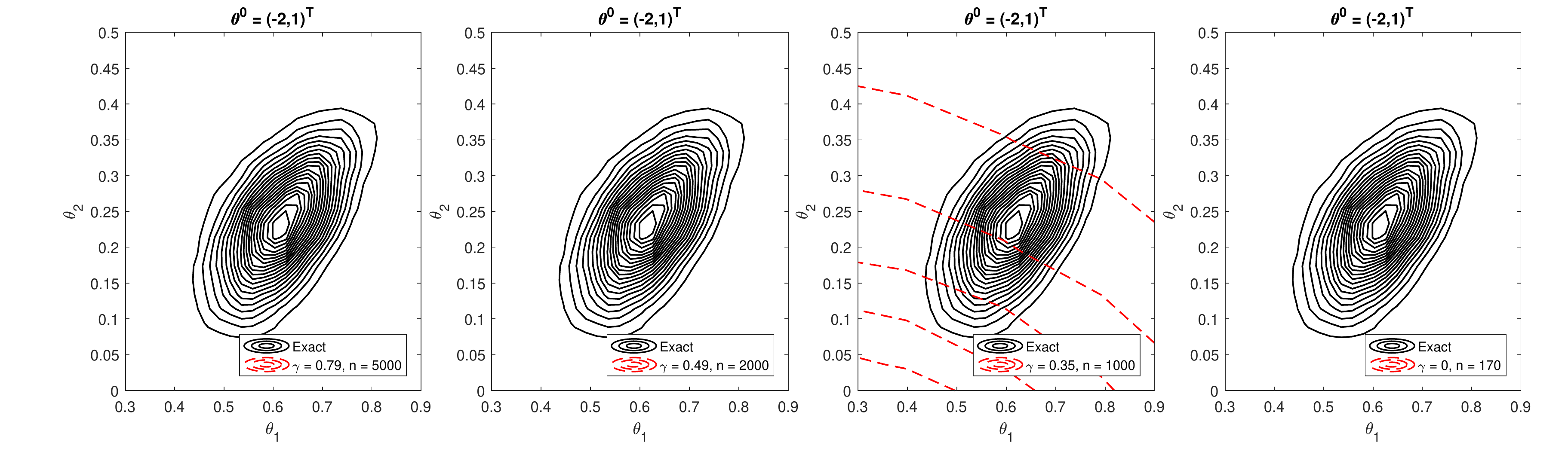}
\end{subfigure}

\begin{subfigure}
\centering\includegraphics[width = 11cm]{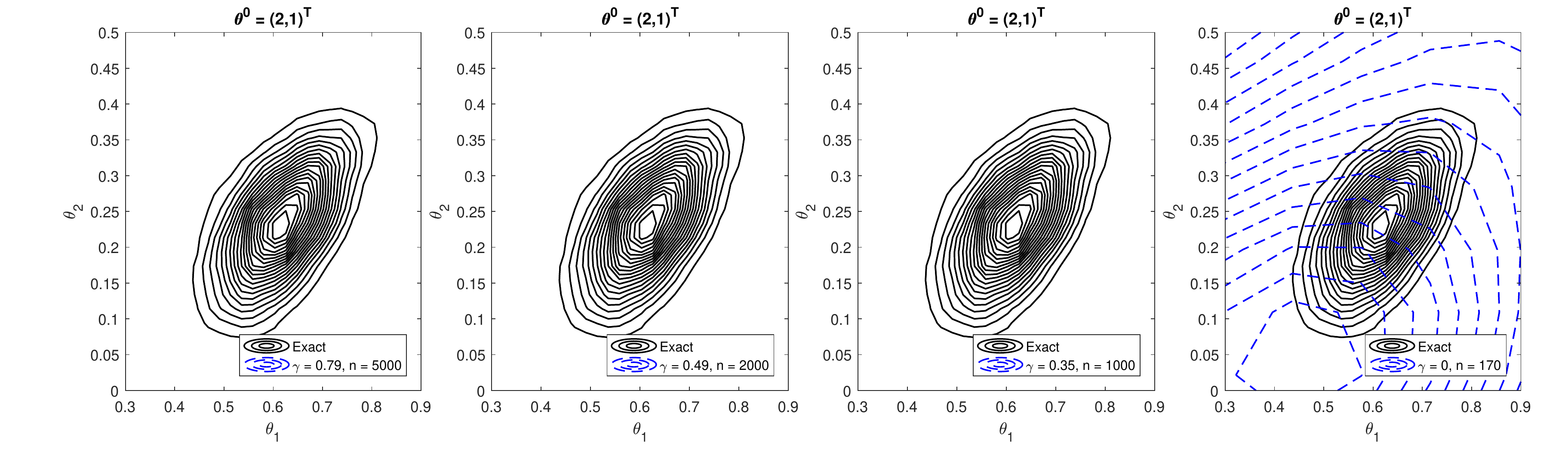}
\end{subfigure}

\begin{subfigure}
\centering\includegraphics[width = 11cm]{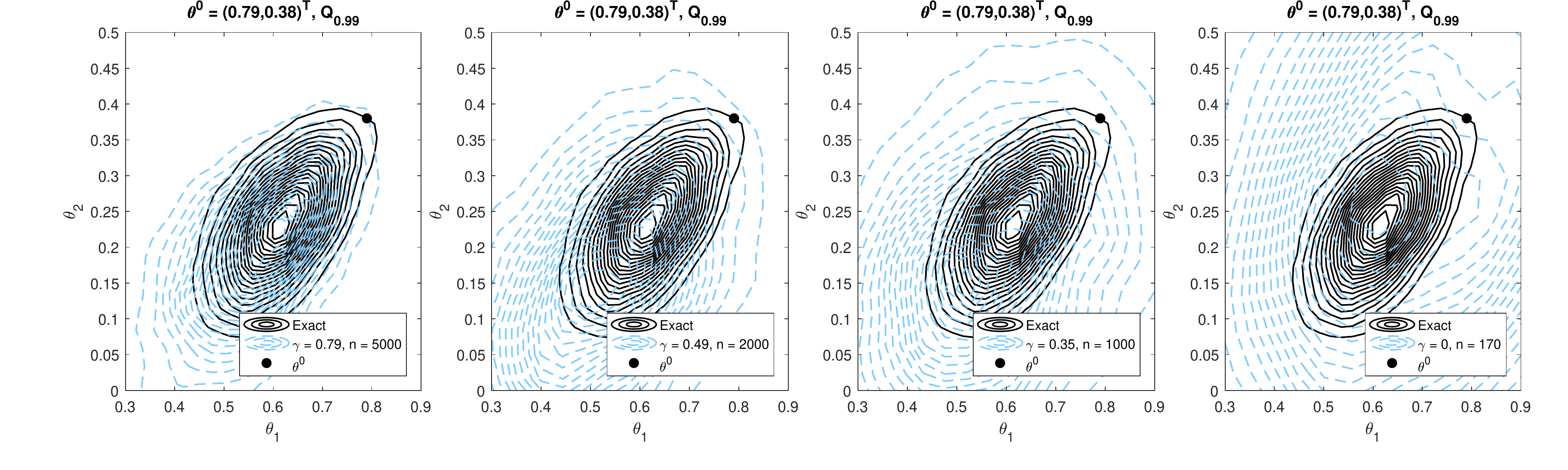}
\end{subfigure}

\begin{subfigure}
\centering\includegraphics[width = 11cm]{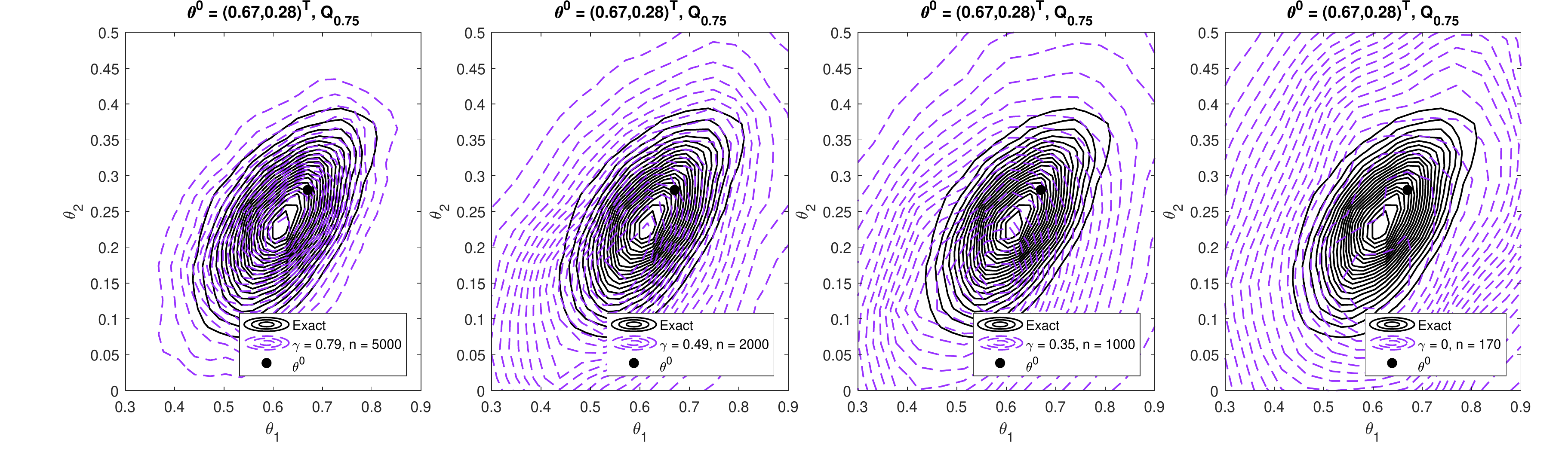}
\end{subfigure}

\caption{\small As Figure 5 in the main text, but for Cholesky whitening. Note some panels do not show any posterior support within the plotted bounds.
%Sensitivity of wBSL posterior with Cholesky whitening to point estimate for MA(2) example. Note some results do not show any posterior support within the plotted bounds.
}
\label{fig:sensMA2pcacor5}
\end{figure}

\clearpage
\subsubsection{An AR(1) model}

\begin{figure}[h!]
    \centering
    \includegraphics[width = 15cm]{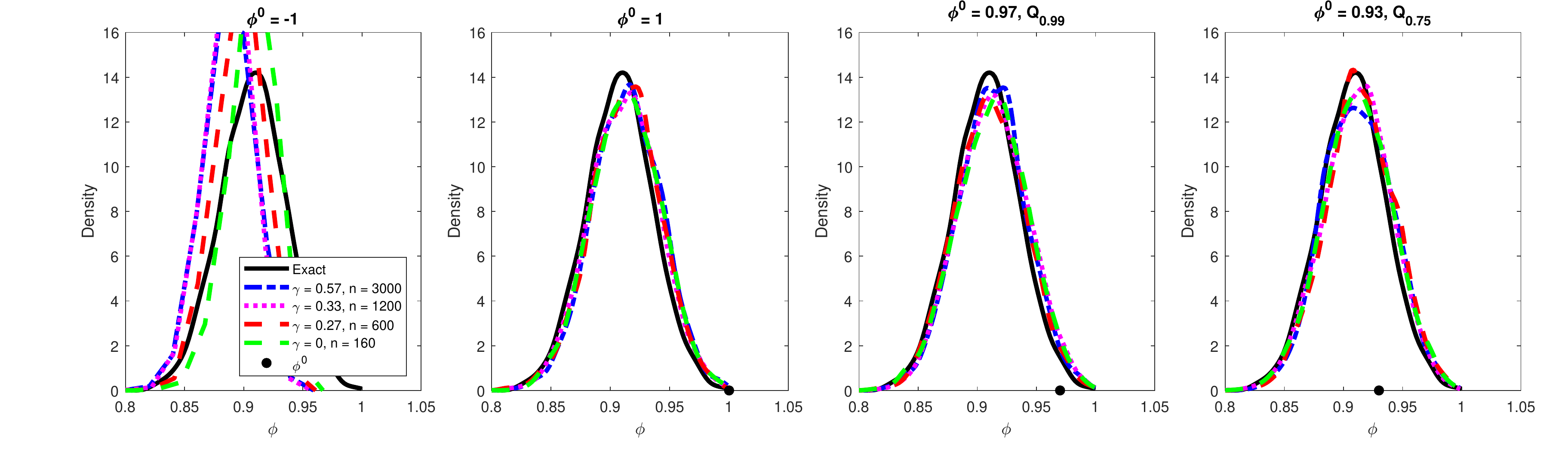}
    \caption{\small
    Sensitivity of the wBSL posterior approximation with PCA whitening (dashed lines) to the point estimate for $\phi^0$ for the AR(1) model. Panels correspond to different point estimates for $\phi^0$, and line types to different $\gamma,n$ combinations.
Estimated true posterior is shown by solid lines, $\phi^0$ is illustrated by a dot.
 %   AR(1) example sensitivity of wBSL posterior with PCA whitening to initial point estimate. Each panel represents a different point estimate, and each color is a different level of shrinkage. The exact MCMC posterior obtained using the likelihood is shown in black.
    }
    \label{fig:sensar}
\end{figure}

Under the same sampler settings we also consider the sensitivity of the wBSL posterior approximation for the AR(1) model, for which $|\phi|<1$. As for the MA(2) model analysis,  we consider boundary specifications $\phi^0=-1$ and $\phi^0=1$, as well as setting $\phi^0$ to be the 0.75 and 0.99 quantiles of the true posterior as proxy estimates of the location of the posterior high density region. The results are shown in Figure \ref{fig:sensar}.
As expected, when $\phi^0$ is close to regions of high posterior density ($\phi_{\text{true}}=0.9$) the wBSL posterior approximation is accurate for all levels of shrinkage. When $\phi^0=-1$ is very far from the high posterior density region, then the wBSL approximation becomes poorer.
%We also consider the sensitivity of the AR$(1)$ example to the initial point estimate. Again we use $T = 100000$, $n_{\text{cov}} = 20000$ and discard a burn-in of 1000 iterations from each run. We use the same prior as in Section 6.2. Recall the parameter of the AR$(1)$ model is $\phi$, subject to the constraint $|\phi| < 1$. We again consider the theoretical worse possible parameter choices, $\phi = -1$ and $\phi = 1$, as well as the $0.75$ and $0.99$ quantiles. \\

%Recall the `true' parameter value for the AR$(1)$ example is $\phi = 0.9$. Thus, the boundary value of $\phi = 1$ is not far from parameter values of high posterior support, so should perform well. From Figure \ref{fig:sensar}, we observe similar behaviour to the MA$(2)$ example. For point estimates at $\phi = 1$, $\phi = 0.97$ and $\phi = 0.93$, the wBSL posteriors are very close to the BSL posterior for all levels of shrinkage. There is some deviation for the BSL posterior on the far boundary at $\phi = -1$. This is a very naive parameter choice which would probably not be made in practice; it is merely shown for the purpose of demonstration.\\

\begin{figure}[h!]
    \centering
    \includegraphics[width = 15cm]{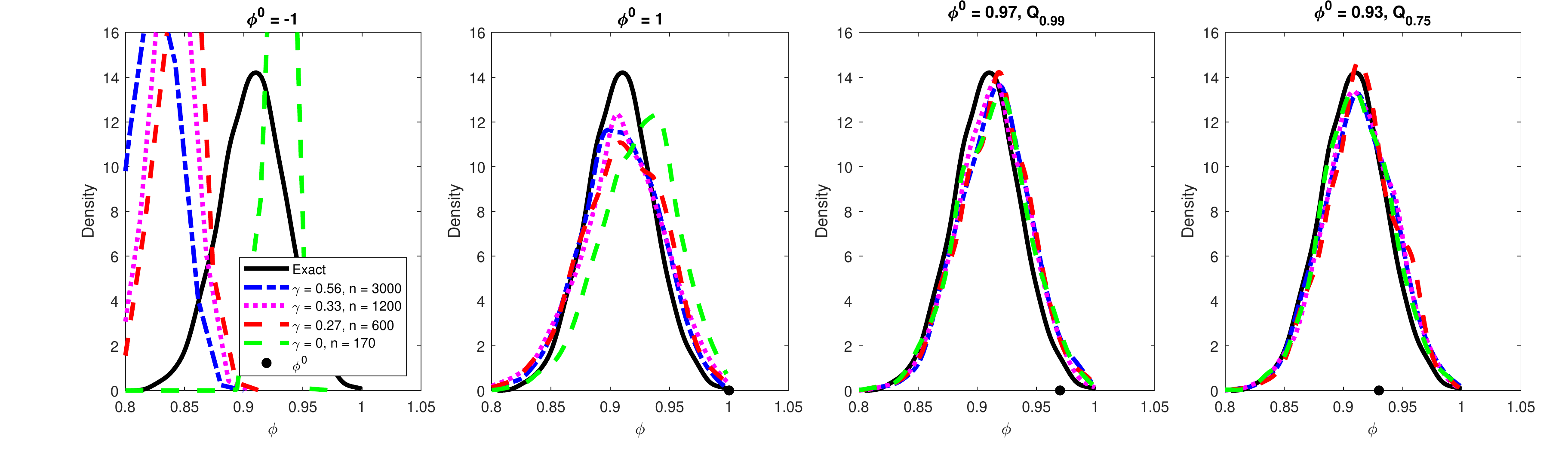}
    \caption{\small As Figure \ref{fig:sensar}, but for PCA-cor whitening. 
    %AR(1) example sensitivity of wBSL posterior with PCA-cor whitening to initial point estimate.
    }
    \label{fig:sensar2}
\end{figure}

\begin{figure}[h!]
    \centering
    \includegraphics[width = 15cm]{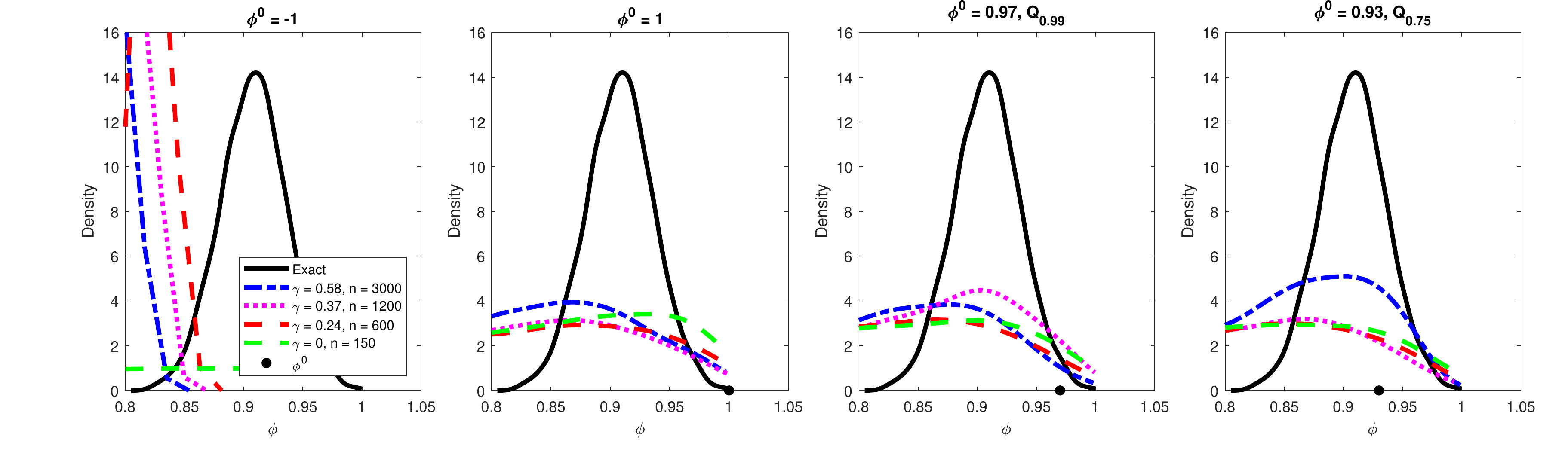}
    \caption{\small As Figure \ref{fig:sensar}, but for ZCA whitening. 
    %AR(1) example sensitivity of wBSL posterior with ZCA whitening to initial point estimate.
    }
    \label{fig:sensar3}
\end{figure}

\begin{figure}[h!]
    \centering
    \includegraphics[width = 15cm]{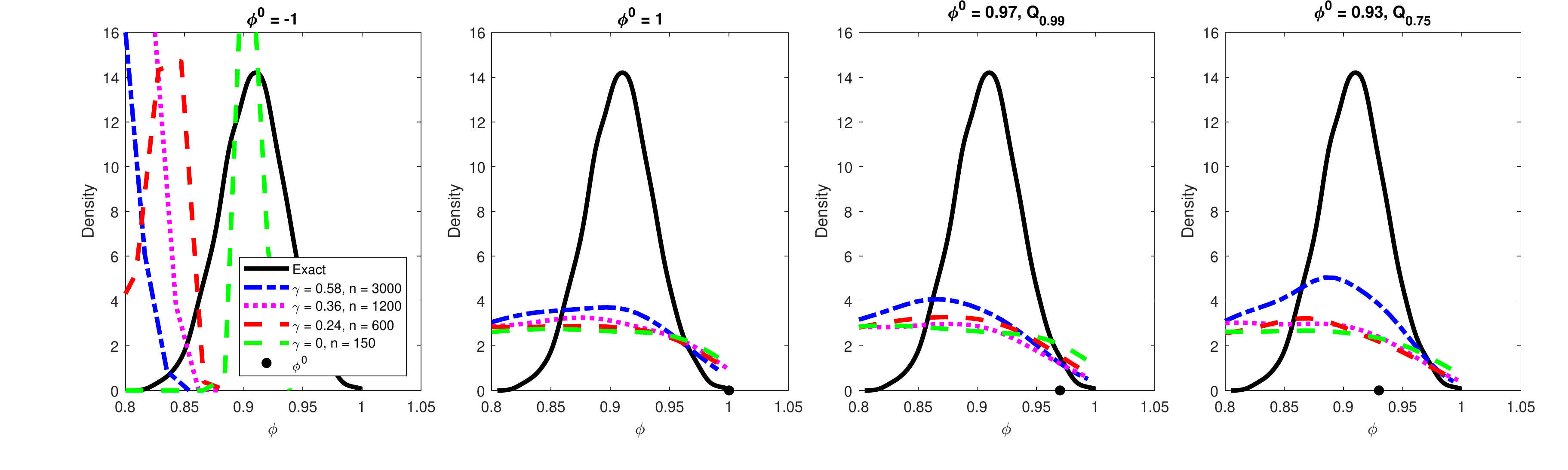}
    \caption{\small As Figure \ref{fig:sensar}, but for ZCA-cor whitening. 
    %AR(1) example sensitivity of wBSL posterior with ZCA-cor whitening to initial point estimate.
    }
    \label{fig:sensar4}
\end{figure}

\begin{figure}[h!]
    \centering
    \includegraphics[width = 15cm]{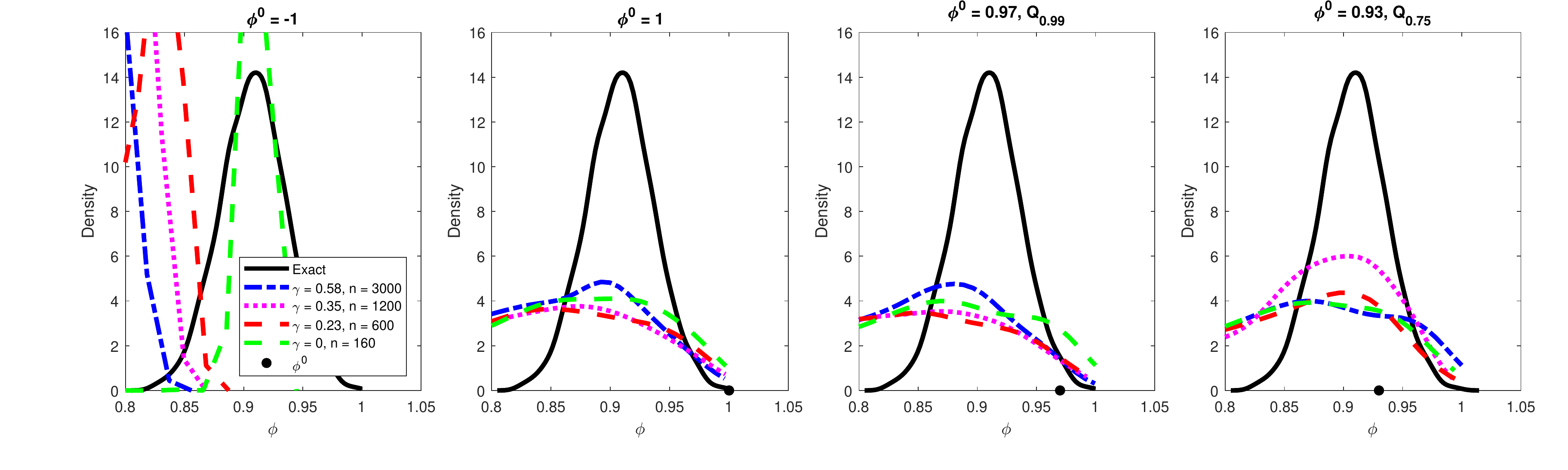}
    \caption{
    \small As Figure \ref{fig:sensar}, but for Cholesky whitening. 
    %AR(1) example sensitivity of wBSL posterior with Cholesky whitening to initial point estimate.
    }
    \label{fig:sensar5}
\end{figure}

\end{document}